%% file: main.tex
\documentclass[11pt]{article} 
\usepackage{amsfonts,amssymb,amsmath,amsthm,dsfont,graphicx,hyperref,mathrsfs,euscript,enumitem,bm}
\usepackage[onehalfspacing]{setspace}
\usepackage{caption,subcaption,lscape}
\hypersetup{pdfborder = {0 0 0},colorlinks=true,linkcolor=blue,urlcolor=blue,citecolor=blue}
\usepackage{natbib}
\usepackage[title]{appendix}
\usepackage[top=1in,bottom=1in,left=.95in,right=.95in]{geometry}
\usepackage{placeins}
\usepackage{threeparttable}

\newtheorem{thm}{Theorem}

\newtheorem{lem}{Lemma}

\numberwithin{equation}{section}

\theoremstyle{definition}
\newtheorem{remark_tmp}{Remark}
\newenvironment{remark}
	{ \begin{remark_tmp} 	}
	{ 
		\medskip\hfill{\LARGE$\lrcorner$}
		\end{remark_tmp} 
	}
\theoremstyle{definition}

\allowdisplaybreaks[1]

\renewcommand*{\arraystretch}{.6}

	\DeclareMathOperator*{\argmin}{arg\,min}
	
	\DeclareMathOperator*{\diag}{diag}
		
\renewcommand{\P}{\mathbb{P}}
\newcommand{\E}{\mathbb{E}}
\newcommand{\V}{\mathbb{V}}

\newcommand{\I}{\mathds{1}}

\newcommand{\bA}{\mathbf{A}}
\newcommand{\bB}{\mathbf{B}}
\newcommand{\bC}{\mathbf{C}}
\newcommand{\bD}{\mathbf{D}}

\newcommand{\bG}{\mathbf{G}}
\newcommand{\bH}{\mathbf{H}}
\newcommand{\bI}{\mathbf{I}}

\newcommand{\bM}{\mathbf{M}}

\newcommand{\bQ}{\mathbf{Q}}

\newcommand{\bS}{\mathbf{S}}
\newcommand{\bU}{\mathbf{U}}

\newcommand{\bZ}{\mathbf{Z}}

\newcommand{\bb}{\mathbf{b}}
\newcommand{\bc}{\mathbf{c}}
\newcommand{\bg}{\mathbf{g}}

\newcommand{\bp}{\mathbf{p}}
\newcommand{\bq}{\mathbf{q}}
\newcommand{\br}{\mathbf{r}}
\newcommand{\bs}{\mathbf{s}}

\newcommand{\bu}{\mathbf{u}}
\newcommand{\bv}{\mathbf{v}}
\newcommand{\bx}{\mathbf{x}}

\newcommand{\bw}{\mathbf{w}}
\newcommand{\bz}{\mathbf{z}}
\newcommand{\ttb}{\star}

\newcommand{\bbeta}{\boldsymbol{\beta}}

\newcommand{\bgamma}{\boldsymbol{\gamma}}

\newcommand{\bxi}{\boldsymbol{\xi}}

\newcommand{\bSigma}{\boldsymbol{\Sigma}}

\newcommand{\bdelta}{\boldsymbol{\delta}}

\begin{document}

\title{\vspace{-0.0in}Prediction Intervals for Synthetic Control Methods\thanks{We thank Alberto Abadie, Amir Ali Ahmadi, Michael Jansson, Filippo Palomba and Mykhaylo Shkolnikov for many insightful discussions, and Kaspar W\"uthrich for sharing replication codes used in the simulations. We also thank three reviewers for their critical comments and suggestions that helped improve our manuscript. Cattaneo and Titiunik gratefully acknowledges financial support from the National Science Foundation (SES-2019432), and Cattaneo gratefully acknowledges financial support from the National Institutes of Health (R01 GM072611-16).}
\bigskip }
\author{Matias D. Cattaneo\thanks{Department of Operations Research and Financial Engineering, Princeton University.} \and
	    Yingjie Feng\thanks{School of Economics and Management, Tsinghua University.}  \and
	    Rocio Titiunik\thanks{Department of Politics, Princeton University.}}
\maketitle

\begin{abstract}
	Uncertainty quantification is a fundamental problem in the analysis and interpretation of synthetic control (SC) methods. We develop conditional prediction intervals in the SC framework, and provide conditions under which these intervals offer finite-sample probability guarantees. Our method allows for covariate adjustment and non-stationary data. The construction begins by noting that the statistical uncertainty of the SC prediction is governed by two distinct sources of randomness: one coming from the construction of the (likely misspecified) SC weights in the pre-treatment period, and the other coming from the unobservable stochastic error in the post-treatment period when the treatment effect is analyzed. Accordingly, our proposed prediction intervals are constructed taking into account both sources of randomness. For implementation, we propose a simulation-based approach along with finite-sample-based probability bound arguments, naturally leading to principled sensitivity analysis methods. We illustrate the numerical performance of our methods using empirical applications and a small simulation study. \texttt{Python}, \texttt{R} and \texttt{Stata} software packages implementing our methodology are available. 
\end{abstract}


\thispagestyle{empty}
\clearpage

\doublespacing
\setcounter{page}{1}
\pagestyle{plain}

\pagestyle{plain}

\section{Introduction}

The synthetic control (SC) method was first introduced by \cite{Abadie-Gardeazabal_2003_AER} as an approach to study the causal effect of a treatment affecting a single aggregate unit that is observed both before and after the treatment occurs.  The authors originally motivated the method with a study of the effect of terrorism in the Basque Country on its GDP per capita. The Basque Country was one of the three richest regions in Spain before the outset of terrorism around the mid-1970s, but the region became relatively poorer in the decades that followed. The question is whether this relative decline can be attributed to terrorism. Their analysis covers the 1955-2000 period and places the beginning of intense terrorism in 1975, thus defining a ``pre-treatment'' period when terrorism is not salient (roughly 1955-1975), and a ``post-treatment'' period that starts when terrorism intensifies (roughly 1975 onward). The time series data allows for a comparison of Basque GDP before and after the onset of terrorism, but to interpret this change as the causal effect of terrorism would require assuming the absence of time trends. Instead, \cite{Abadie-Gardeazabal_2003_AER} proposed to use other regions in Spain, whose GDP is also observed before and after the onset of terrorism in the Basque Country, to build an aggregate or ``synthetic'' control unit that captures the GDP trajectory that would have occurred in the Basque Country if terrorism had never occurred. The synthetic control is built as a weighted average of all units in the control group (the ``donor pool''), where the weights are chosen so that the synthetic control's outcome in the pre-treatment period closely matches the treated unit's trajectory while also satisfying some constraints such as being non-negative, adding up to one, and accounting for other pre-treatment covariates. For a contemporaneous review of this literature, see \citet{Abadie_2021_JEL} and the references therein.

The SC method has received increasing attention since its introduction, and is now a popular component of the methodological toolkit for causal inference and program evaluation \citep{Abadie-Cattaneo_2018_ARE}. Methodological and theoretical research concerning SC methods has mostly focused either on expanding the SC causal framework (e.g., to dissagregated data or staggered treatment adoption settings) or on developing new implementations of the SC prediction (e.g., via different penalization constraints or matrix completion methods). Recent examples include \citet{Abadie-LHour_2021_JASA}, \citet{Agarwal-Shah-Shen-Song_2021_JASA}, \citet{Athey-et-al_2021_JASA}, \citet{Bai-Ng_2021_JASA}, \citet{BenMichael-Feller-Rothstein_2021_JASA}, \citet{Chernozhukov-Wuthrich-Zhu_2021_ttest}, \citet{Ferman_2021_JASA}, \citet{Kellogg-Mogstad-Pouliot-Torgovitsky_2021_JASA}, and \citet{Masini-Medeiros_2021_JASA}; see their references for many more. In contrast, considerably less effort has been devoted to develop principled statistical inference procedures for uncertainty quantification within the SC framework. In particular, \citet{Abadie-Diamond-Hainmueller_2010_JASA} propose a design-based permutation approach under additional assumptions, \citet{Li_2020_JASA} relies on large-sample approximations for disaggregated data under correct specification, \citet{Chernozhukov-Wuthrich-Zhu_2021_JASA} develop time-series permutation-based inference methods, and \citet{Shaikh-Toulis_2021_JASA} discuss cross-sectional permutation-based inference methods in semiparametric duration-type settings. See also \citet{Feng_2021_wp}, and references therein, for related large sample inference methods employing local principal component analysis based on nearest-neighbor approximations in possibly non-linear factor model settings. 

We develop conditional prediction intervals for the SC framework, offering an alternative (conditional) inference method to assess statistical uncertainty. Our proposed approach builds on ideas from the literature on conditional prediction intervals \citep{Vovk_2012_ACML,Chernozhukov-Wuthrich-Zhu_2021_distributional} and non-asymptotic concentration \citep{Vershynin_2018_Book,Wainwright_2019_Book} in probability and statistics. As a consequence, the resulting (conditional) prediction intervals are conservative but formally shown to offer probability guarantees. We focus on uncertainty quantification via (conditional) prediction intervals because, in the SC framework, the treatment effect estimator is a random variable emerging from an out-of-sample prediction problem, based on the estimated SC weights constructed using pre-treatment data. Our inference procedures are not confidence intervals in the usual sense (i.e., giving a region in the parameter space for a non-random parameter of interest), but rather intervals describing a region on the support of a random variable where a new realization is likely to be observed.

Our construction begins by noting that the statistical uncertainty of the SC prediction is governed by two distinct sources of randomness: one due to the construction of the  (likely misspecified) SC weights in the pre-treatment period, and the other due to the unobservable stochastic error in the post-treatment period when the treatment effect is analyzed. Accordingly, our proposed prediction intervals are constructed taking into account both sources of randomness. For the first source of uncertainty, we propose a simulation-based approach that is justified via non-asymptotic probability concentration and hence enjoys probability guarantees. This approach takes into account the specific construction of the SC weights. For the second source of uncertainty, which comes from out-of-sample prediction due to the unobservable error in the post-treatment period, we discuss several approaches based on nonparametric and parametric probability approximations as a framework for principled sensitivity analysis. This second uncertainty source is harder to handle nonparametrically, and hence its contribution to the overall prediction intervals should be considered with care. Our approach in this paper is to employ an agnostic sensitivity analysis, but future work will consider other approaches.

Our results are obtained under high-level conditions, but we provide primitive conditions for three examples: an outcomes-only setting with i.i.d. data, a multi-equation setting allowing for stationary weakly dependent data where the weights are obtained by not only matching the pre-treatment trends of the outcome of interest but also approximating the trajectories of additional variables such as important covariates or secondary outcomes; and a non-stationary cointegration setting. All three settings allow the weights to be covariate-adjusted in each equation. We also showcase our methods numerically, using both simulated and real data. The methods perform well in finite samples.

The rest of the paper proceeds as follows. Section \ref{sec:setup} provides a formal introduction to the SC framework and defines the basic quantities of interest. Section \ref{sec:predictint} introduces the prediction intervals we focus on, and provides basic intuition for their decomposition in terms of the SC weights estimation error and the unobservable post-treatment error. Section \ref{sec:bound w.hat} develops a simulation-based method to account for the first source of uncertainty, and Section \ref{sec:bound e.T} discusses how to (model and) account for the second source of uncertainty. Section \ref{sec:numerical} illustrates the performance of our proposed prediction intervals with a Monte Carlo experiment and two empirical examples from the SC literature. Section \ref{sec:conclusion} concludes. Appendix \ref{sec:extension weakly dependent data} provides an extension of our main in-sample uncertainty quantification approach to the case of weakly dependent ($\beta$-mixing) stationary time series data. All the proofs of our technical results, as well as additional numerical evidence, are collected in the online supplemental appendix. We provide companion replication codes in \texttt{R}, and a general-purpose software package is underway \citep*{Cattaneo-Feng-Palomba-Titiunik_2021_scpi}.

\section{Setup}\label{sec:setup}

We consider the standard synthetic control framework with a single treated unit and several control units, allowing for both stationary and non-stationary data. The data may include only the outcome of interest, or the outcome of interest plus other variables. The researcher observes $N+1$ units for $T_0+T_1$ periods of time. Units are indexed by $i = 1,2,\ldots N, N+1$, and time periods are indexed by $t=1,2, \ldots,T_0, T_0+1, \ldots, T_0+T_1$. During the first $T_0$ periods, all units are untreated. Starting at $T_0 + 1$, unit $1$ receives treatment but the other units remain untreated. Once the treatment is assigned at $T_0+1$, there is no change in treatment status: the treated unit continues to be treated  and the untreated units remain untreated until the end of the series, $T_1$ periods later.

Each unit $i$ at period $t$ has two potential outcomes, $Y_{it}(1)$ and $Y_{it}(0)$, respectively denoting the outcome under treatment and the outcome in the absence of treatment (which we call the \textit{control} or the \textit{untreated} condition). This notation imposes two additional implicit assumptions that are standard in this setting: no spillovers (the potential outcomes of unit $i$ depend only on $i$'s treatment status) and no anticipation (the potential outcomes at $t$ depend only on the treatment status of the same period).

Attention is restricted to the impact of the treatment on the treated unit. By treatment impact, we mean the difference between the outcome path taken by the treated unit, and the path it would have taken in the absence of the treatment. The quantity of interest is 
\begin{equation}\label{eq: parameter tau}
\tau_{t}=Y_{1t}(1)-Y_{1t}(0), \qquad t> T_0,
\end{equation}
where $\tau_{t}$ may be regarded as random or non-random depending on the framework considered. In this paper, we view $\tau_{t}$ as a random variable.

For each unit, we only observe the potential outcome corresponding to the treatment status actually received by the unit. We denote the observed outcome by $Y_{it}$, which is defined as
\[
Y_{it}=\begin{cases}
Y_{it}(0) & \text{if}\; i = 2, \ldots N+1   \\
Y_{it}(0) & \text{if}\; i=1 \; \text{and}\; t \in \left\{1, 2, \ldots, T_0 \right\}\\
Y_{it}(1) & \text{if}\; i=1 \; \text{and}\; t \in \left\{T_0+1, \ldots, T_0+T_1 \right\}
\end{cases}.
\]

This means that, in $\tau_t$, the treated unit's potential outcome  $Y_{1t}(0)$ is unobservable for all $t > T_0$. The idea of the synthetic control method is to use an appropriate combination of the post-treatment observed outcomes of the untreated units to approximate the treated unit's counterfactual post-treatment outcome, $Y_{1t}(0)$ for $t>T_0$. This idea has been formalized in different ways since it was originally proposed by \citet{Abadie-Gardeazabal_2003_AER}.

In all SC frameworks, the formalization chooses a set of weights $\bw = (w_2,w_3,\dots,w_{N+1})'$ such that a given loss function is minimized under constraints. Given a set of estimated weights $\widehat{\bw}$, the treated unit's counterfactual predicted outcome is then calculated as $\widehat{Y}_{1t}(0) = \sum_{i=2}^{N+1} \widehat{w}_{i}Y_{it}(0)  $ for $t>T_0$. The weighted average $\widehat{Y}_{1t}(0)$ is often referred to as the \textit{synthetic control} of the treated unit, as it represents how the untreated units can be combined to provide the best counterfactual for the treated unit in the post-treatment period.

When the data contains only information on the outcome of interest, $\bw$ is chosen such that the weighted average of the outcomes of the untreated units approximates well the outcome trajectory of the treated unit in the period before the treatment. That is, the weights $\bw$ are chosen so that
\[\sum_{i=2}^{N+1} w_{i}Y_{it}(0) \approx Y_{1t}(0), \qquad \text{for}\quad t=1,2,\dots,T_0,\]
where the meaning of the symbol ``$\approx$'' varies depending on the specific framework considered. A leading example constrains the weights to be non-negative and sum to one, and estimates $\bw$ by constrained least squares:
\begin{equation}\label{eq:SC-outcomeonly-estimate}
    (\widehat{\bw}',\widehat{r})' \in \underset{\bw\in\mathcal{W},\, r\in\mathcal{R}}{\argmin}\; \sum_{t=1}^{T_0}(Y_{1t} - Y_{2t}w_{2}- \cdots - Y_{(N+1)t}w_{N+1} - r)^2,
\end{equation}
where $r$ denotes the intercept, and $\mathcal{W}$ and $\mathcal{R}$ denote the corresponding constraint (or feasibility) sets---we give formal definitions in the next subsection.

When the weights are chosen according to (\ref{eq:SC-outcomeonly-estimate}), the resulting synthetic control will reproduce as closely as possible the outcome trajectory of the treated unit in the pre-treatment period. For example, in the Basque terrorism application, this procedure would lead to a synthetic Basque Country that would have a similar per capita GDP to the Basque Country's per capita GDP in the 1955-1975 period when terrorism is not salient. 

This outcomes-only version of the SC method, however, cannot guarantee that the resulting synthetic control unit will be similar to the treated unit in any characteristics other than the (pre-treatment) outcome. In some applications, this feature may be undesirable, as researchers may have access to additional characteristics such as baseline covariates or secondary outcomes and may want to also ensure that the synthetic control approximates the treated unit in terms of these additional characteristics. The SC framework can handle this case by including additional equations for these additional characteristics and minimizing the combined loss. In this case, letting $l=1,2,\ldots, M$ index the variables that will be ``matched'' to produce the weights, the minimization problem above can be generalized as 
\begin{equation}\label{eq:SC-outcome+covs-estimate}
    (\widehat{\bw}',\widehat{\br})' \in \underset{\bw\in\mathcal{W},\, \br\in\mathcal{R}}{\argmin}\; \sum_{l=1}^{M}\sum_{t=1}^{T_0}\upsilon_{t,l}(Y_{1t,l} - Y_{2t,l}w_{2}- \cdots - Y_{(N+1)t,l}w_{N+1} - r_l)^2,
\end{equation}
where $\widehat{\br}=(\widehat{r}_1,\dots,\widehat{r}_M)'$ and $\{\upsilon_{t,l}\}_{1\leq t\leq T_0, 1\leq l\leq M}$ are positive constants reflecting the relative importance of different equations and periods.

For example, in the original Basque terrorism example, \cite{Abadie-Gardeazabal_2003_AER} show that the Basque country differs from the rest of Spain in terms of population density, and they are concerned that pre-terrorism differences in population density may affect economic growth in the post-treatment period. In this case, we can choose the weights $\widehat{\bw}$ to ensure not only that the per capita GDP trajectory is similar between the treated unit and the synthetic control unit, but also to ensure that the synthetic control is similar to the treated unit in terms of population density. To implement this multi-equation SC method, we fit equation (\ref{eq:SC-outcome+covs-estimate}) with two variables ($M=2$) where $Y_{it,1}$ ($l=1$) is per capita GDP for region $i$ in year $t$ and $Y_{it,2}$ ($l=2$) is population density for region $i$ in year $t$. When $\widehat{\bw}$ is chosen this way, the resulting synthetic control will resemble (to the extent that the data allows) the treated unit in terms of both per capita GDP and population density.

Equation \eqref{eq:SC-outcome+covs-estimate} can be viewed as a (weighted) combination of $M$ optimization problems in \eqref{eq:SC-outcomeonly-estimate}, satisfying an additional constraint that the weights $\bw$ must be the same across the $M$ equations. For simplicity, we let $\upsilon_{t,l}=1$ for all $t$ and $l$, but the analysis below can be applied to the more general case if additional regularity conditions are imposed on $\{\upsilon_{t,l}\}_{1\leq t\leq T_0, 1\leq l\leq M}$.

The two cases just discussed (outcomes-only and multi-equation SC frameworks) allow for weakly dependent and cointegrated data, and they also can be generalized further by including covariates in a linear and additive way in  (\ref{eq:SC-outcomeonly-estimate}) or (\ref{eq:SC-outcome+covs-estimate}). This covariate adjustment would introduce additional parameters to the fit that would not be of primary interest; rather, they would be included to ``partial out'' the effect of additional covariates.

\subsection{General Framework}

We now introduce a general framework and further notation that encompass and formalize the two particular examples discussed above as well as other synthetic control approaches in the literature. Our general framework includes the outcomes-only fit and the multi-equation fit (i.e., outcome plus other variables) as particular cases, allowing for covariate adjustment and non-stationary data in a unified way. 

Consider synthetic control weights constructed simultaneously for $M$ features of the treated unit, denoted by $\bA_l=(a_{1,l}, \cdots, a_{T_0,l})'\in\mathbb{R}^{T_0}$, with index $l=1,\cdots, M$. For each feature $l$, there exist $J+K$ variables that can be used to predict or ``match'' the $T_0$-dimensional vector $\bA_l$. These $J+K$ variables are separated into two groups denoted by $\bB_l=(\bB_{1,l}, \bB_{2,l}, \cdots, \bB_{J, l})\in\mathbb{R}^{T_0\times J}$ and $\bC_l=(\bC_{1,l}, \cdots, \bC_{K,l})\in\mathbb{R}^{T_0\times K}$, respectively. More precisely, for each $j$, $\bB_{j,l}=(b_{j1,l}, \cdots, b_{jT_0,l})'$ corresponds to the $l$th feature of the $j$th unit observed in $T_0$ pre-treatment periods and, for each $k$, $\bC_{k,l}=(c_{k1,l}, \cdots, c_{kT_0,l})'$ is another vector of control variables also possibly used to predict $\bA_l$ over the same pre-intervention time span. For ease of notation, we let $d=J+KM$. 

The goal of the synthetic control method is to search for a vector of common weights $\bw\in\mathcal{W}\subseteq\mathbb{R}^{J}$ across the $M$ features and a vector of coefficients $\br\in\mathcal{R}\subseteq\mathbb{R}^{KM}$, such that the linear combination of $\bB_l$ and $\bC_l$ ``matches'' $\bA_l$ as close as possible, for all $1\leq l\leq M$. This goal is typically achieved via the following optimization problem:
\begin{equation}\label{eq: estimated weight}
\widehat{\bbeta} := (\widehat{\bw}', \widehat{\br}')' \in\underset{\bw\in\mathcal{W},\, \br\in\mathcal{R}}{\argmin}\;
(\bA-\bB\bw-\bC\br)'(\bA-\bB\bw-\bC\br)
\end{equation}
where
\[
\bA=\begin{bmatrix}
\bA_1 \\
\bA_2 \\
\vdots\\
\bA_M
\end{bmatrix},\quad
\bB=\begin{bmatrix}
\bB_1 \\
\bB_2 \\
\vdots\\
\bB_M
\end{bmatrix}, \quad
\bC=\begin{bmatrix}
\bC_1 & \bm{0} &\cdots & \bm{0}\\
\bm{0}& \bC_2  &\cdots & \bm{0}\\
\vdots& \vdots &\ddots & \vdots\\
\bm{0}& \bm{0} &\cdots & \bC_M
\end{bmatrix},
\]
and where the feasibility sets $\mathcal{W}$ and $\mathcal{R}$ capture the restrictions imposed. (For simplicity we do not introduce an explicit re-weighting of the $M$ equations, but recall that this extension is possible.) This framework encompasses multiple prior synthetic control formalizations in the literature, which differ in whether they include additional covariates, whether the data is assumed to be stationary, and the particular choice of constraint sets $\mathcal{W}$  and $\mathcal{R}$ used, among other possibilities.

The following list provides some examples of different constraint sets used in practice, where $\|\cdot\|_p$ denotes the $L_p$ vector norm and $Q$ and $\alpha$ are tuning parameters.
\begin{itemize}[noitemsep, leftmargin=*]
	\item \cite{Abadie-Diamond-Hainmueller_2010_JASA}: $\mathcal{W}=\{\bw \in \mathbb{R}^{N}_+: \|\bw\|_1=1 \}$ and $\mathcal{R}=\{0\}$.
	\item \cite{Hsiao-et-al_2012_JAE}: $\mathcal{W}=\mathbb{R}^{N}$ and $\mathcal{R}=\mathbb{R}$.
	\item \cite{Ferman-Pinto_2021_wp}: $\mathcal{W}=\{\bw \in \mathbb{R}^{N}_+: \|\bw\|_1=1\}$ and $\mathcal{R}=\mathbb{R}$.
	\item \cite{Chernozhukov-Wuthrich-Zhu_2021_JASA}: $\mathcal{W}=\{\bw \in \mathbb{R}^{N}: \|\bw\|_1 \leq 1\}$ and $\mathcal{R}=\mathbb{R}$.
	\item \cite{Amjad-Shah-Shen_2018_JMLR}: $\mathcal{W}=\{\bw \in \mathbb{R}^{N}: \|\bw\|_2 \leq Q\}$ and $\mathcal{R}=\{0\}$.
	\item \cite{Arkhangelsky-et-al_2021_wp}: $\mathcal{W}=\{\bw \in \mathbb{R}^{N}: \|\bw\|_2 \leq Q,  \|\bw\|_1=1\}$ and  $\mathcal{R}=\mathbb{R}$. 
	\item \cite{Doudchenko-Imbens_2016_wp}: $\mathcal{W}=\{\bw \in \mathbb{R}^{N}: \frac{1-\alpha}{2}\|\bw\|_2^2+\alpha\|\bw\|_1 \leq Q\}$ and $\mathcal{R}=\mathbb{R}$.
\end{itemize}
In some applications the intercept in \eqref{eq: estimated weight} is removed by demeaning the data before the analysis. Section \ref{sec:examples} discusses in detail the outcomes-only case, as well as the multi-equation case where the researcher ``matches'' on pre-treatment characteristics and pre-intervention outcomes simultaneously. That section also deals with stationary weakly dependant data, and non-stationary data (i.e., cointegration system).

For example, the outcomes-only setup can be obtained as a particular case of \eqref{eq: estimated weight} with $M=1$ (there is only one feature to match on), $J=N$ (there are $N$ units in the donor pool), and $K=1$ (there is an intercept). Then, $\bA_1=(Y_{11}, Y_{12}, \cdots, Y_{1T_0})'$, $\bB_{j,1}=(Y_{(j+1)1}, Y_{(j+1)2},\cdots,Y_{(j+1)T_0})'$, $\bC_{j,1}=(1, 1,\cdots,1)'$, and \eqref{eq: estimated weight} reduces to the (possibly constrained) optimization problem \eqref{eq:SC-outcomeonly-estimate}. The multi-equation setup with one outcome and one covariate can be obtained similarly by setting $M=2$ (there are two features to match on), $J=N$ ( $N$ units in the donor pool), and $K=1$ (there is an intercept), which reduces to (\ref{eq:SC-outcome+covs-estimate}). 

To further understand our proposed inference approach, we define the pseudo-true values $\bw_0$ and $\br_0$ relative to a sigma field $\mathscr{H}$:
\begin{equation}\label{eq: pseudo true value}
\bbeta_0 := (\bw_0', \br_0')' = \underset{\bw\in\mathcal{W},\, \br\in\mathcal{R}}{\arg\min}\; \E[(\bA-\bB\bw-\bC\br)'(\bA-\bB\bw-\bC\br)|\mathscr{H}],
\end{equation}
and thus write
\begin{equation}\label{eq: vertical regression}
\bA=\bB\bw_0+\bC\br_0+\bU, \qquad \bw_0\in\mathcal{W}, \qquad \br_0\in\mathcal{R},
\end{equation}
where $\bU=(u_{1,1}, \cdots, u_{T_0,1}, \cdots, u_{1,M}, \cdots, u_{T_0, M})'\in\mathbb{R}^{T_0M}$ is the corresponding pseudo-true residual relative to a sigma field $\mathscr{H}$. That is,
$\bw_0$ and $\br_0$ are the mean square error estimands associated with the (possibly constrained) best linear prediction coefficients $\widehat{\bw}$ and $\widehat{\br}$ conditional on $\mathscr{H}$. Importantly, we \textit{do not} attach any structural meaning to equation \eqref{eq: vertical regression}. The population vectors $\bw_0$ and $\br_0$ are (conditional) pseudo-true values whose meaning should be understood in context, and are determined by the assumptions imposed on the data generating process. In particular, with strong parametric functional form assumptions or rich enough nonparametric basis expansions, equation \eqref{eq: vertical regression} may be viewed as a representation (or approximation) of $\E[\bA|\bB,\bC,\mathscr{H}]$. In such cases, $\E[\bU|\bB,\bC,\mathscr{H}]=\bm{0}$ or, at least, $\E[\bU|\bB,\bC,\mathscr{H}]$ is taken to be ``small''. Alternatively, if the (population, conditional) linear projection coefficients lie on $\mathcal{W}\times\mathcal{R}$, that is, the constraints imposed by $\mathcal{W}$ and $\mathcal{R}$ in \eqref{eq: pseudo true value} are not binding, then \eqref{eq: vertical regression} represents the best linear approximation of $\bA$ based on $(\bB,\bC)$, conditional on $\mathscr{H}$. In this scenario, $\bU$ is uncorrelated with $(\bB,\bC)$, conditional on $\mathscr{H}$. Most importantly, in general, $\bU$ may not be mean zero due to the (binding) constraints imposed in the construction of $\widehat{\bw}$ and $\widehat{\br}$.

Given estimated weights $\widehat{\bw}$ and coefficients $\widehat{\br}$, the post-treatment counterfactual outcome for the treated unit is predicted by
\[\widehat{Y}_{1T}(0) = \bx_T'\widehat{\bw}+\bg_T'\widehat{\br} = \bp_T'\widehat{\bbeta}, \qquad \bp_T := (\bx_T', \bg_T')', \qquad T>T_0, \]
where $\bx_T\in\mathbb{R}^{N}$ is a vector of predictors for control units observed in time $T$ and $\bg_T\in\mathbb{R}^{KM}$ is another set of user-specified predictors observed at time $T$. Variables included in $\bx_T$ and $\bg_T$ need not be the same as those in $\bB$ and $\bC$, but will be part of the sigma field $\mathscr{H}$, as explained in more detail in the next section. Therefore, from this perspective, our focus is on conditional inference. We decompose the potential outcome of the treated unit accordingly:
\begin{equation}\label{eq: SC Y1T(0)}
    Y_{1T}(0) \equiv \bx_T'\bw_0+\bg_T'\br_0+e_T = \bp_T'\bbeta_0 + e_T, \qquad T>T_0,
\end{equation}
where $e_T$ is defined by construction. In our analysis, $\bw_0$ and $\br_0$ are assumed to be possibly random elements around which $\widehat{\bw}$ and $\widehat{\br}$ are concentrating in probability, respectively, which is why we called them pseudo-true values.

The distance between the estimated treatment effect on the treated and the target population one is
\begin{equation}\label{eq: tauhat - tau}
    \widehat{\tau}_{T} - \tau_{T}  = \Big( Y_{1T}(1)-\widehat{Y}_{1T}(0)  \Big) - \Big(Y_{1T}(1)-Y_{1T}(0) \Big) = Y_{1T}(0)  - \widehat{Y}_{1T}(0).
\end{equation}

Within the synthetic control framework, we view the quantity of interest $\tau_{T}$ as a random variable, and hence we refrain from calling it a ``parameter''. Consequently, we call $\widehat{\tau}_{T}$ a prediction of $\tau_{T}$ rather than an ``estimator'' of it, and focus on building prediction intervals rather than confidence intervals.

\section{Prediction Intervals}\label{sec:predictint}

Given the generic framework introduced in the previous section, we now present our proposed prediction intervals for $\tau_T$. See \citet{Vovk_2012_ACML}, \citet{Chernozhukov-Wuthrich-Zhu_2021_distributional,Chernozhukov-Wuthrich-Zhu_2021_JASA}, and references therein, for recent papers on (conditional) prediction intervals and related methods. Let $\bA$, $\bB$ and $\bC$ be random quantities defined on a probability space $(\Omega, \mathscr{F}, \P)$, and $\mathscr{H}\subseteq\mathscr{F}$ be a sub-$\sigma$-field. For some $\alpha, \pi\in(0,1)$, we say a random interval $\mathcal{I}$ is an $(\alpha,\pi)$-valid $\mathscr{H}$-conditional prediction interval for $\tau_T$ if
\begin{equation}\label{eq: PI definition}
    \P\Big\{\P\big[\tau_T\in\mathcal{I} \,\big|\, \mathscr{H}\big]\geq 1-\alpha\Big\}\geq 1-\pi.
\end{equation}

If $\mathscr{H}$ is the trivial $\sigma$-field over $\Omega$, then $\mathcal{I}$ reduces to an unconditional prediction interval for $\tau_T$. In the general case, $\mathcal{I}$ is an $\mathscr{H}$-conditionally $(\alpha,\pi)$-valid prediction interval: the conditional coverage probability of $\mathcal{I}$ is at least $(1-\alpha)$, which holds with probability over $\mathscr{H}$ at least $ (1-\pi)$. In practice, $(1-\alpha)$ is a desired confidence level chosen by users, say $95\%$, and $\pi$ is a ``small'' number that depends on the sample size and typically goes to zero in some asymptotic sense. In this paper, all the results are valid for all $T_0$ large enough, with the associated probability loss $\pi$ characterized precisely. Thus, we say that the conditional coverage of the prediction interval $\mathcal{I}$ is at least $(1-\alpha)$ with high probability, or that the conditional prediction interval offers finite-sample probability guarantees. Our results imply $\pi\to 0$ as $T_0\to\infty$, but no limits or asymptotic arguments are used in this paper.

An asymptotic analogue to the above definition \eqref{eq: PI definition} would be $\P(\tau_T\in\mathcal{I}|\mathscr{H})\geq 1-\alpha-o_\P(1)$ or, perhaps, $\P(\tau_T\in\mathcal{I}|\mathscr{H})\to_\P 1-\alpha$, where the probability limit is taken as the sample size grows to infinity (e.g., as $T_0\to\infty$). In this case, we say $\mathcal{I}$ is an $\mathscr{H}$-conditional prediction interval for $\tau_T$ that is asymptotically valid with coverage probability (at least) $(1-\alpha)$. This is a weaker property because it does not offer any finite-sample probability guarantees for the (conditional) coverage of the prediction interval.

We employ the following lemma to construct valid, conditional prediction intervals in the sense of \eqref{eq: PI definition}. This lemma follows from the union bound applied to $\widehat{\tau}_T - \tau_{T} = \bp_T'(\bbeta_0-\widehat{\bbeta}) + e_T$.

\begin{lem}[Prediction Interval] \label{lem: PI}
	Suppose that there exist $M_{1,\mathtt{L}}$, $M_{1,\mathtt{U}}$, $M_{2,\mathtt{L}}$ and $M_{2,\mathtt{U}}$, possibly depending on $\alpha_1, \alpha_2, \pi_1, \pi_2\in(0,1)$ and the conditioning $\sigma$-field $\mathscr{H}$, such that
	\begin{alignat*}{4}
	&\P\Big\{\P\big[M_{1,\mathtt{L}}\leq\bp_T'(&&\bbeta_0-\widehat{\bbeta})&&\leq M_{1,\mathtt{U}} \; &&\big| \; \mathscr{H} \big]\geq 1-\alpha_1\Big\}\geq 1-\pi_1, \qquad \text{and}\\
	&\P\Big\{\P\big[M_{2,\mathtt{L}}\leq &&e_T&&\leq M_{2,\mathtt{U}} \; &&\big| \; \mathscr{H} \big] \geq 1-\alpha_2\Big\}\geq 1-\pi_2.
	\end{alignat*} 
	Then, $\P\Big\{\P\big[\widehat{\tau}_T-M_{1,\mathtt{U}}-M_{2,\mathtt{U}} \leq \tau_T\leq \widehat{\tau}_T-M_{1,\mathtt{L}}-M_{2,\mathtt{L}} \big|  \mathscr{H}  \big]
	\geq 1-\alpha_1-\alpha_2\Big\}\geq 1-\pi_1-\pi_2$.
\end{lem}
This lemma provides a simple way to construct an $\mathscr{H}$-conditional prediction interval enjoying $(\alpha,\pi)$-validity with $\alpha=\alpha_1+\alpha_2$ and $\pi=\pi_1+\pi_2$:
\[\mathcal{I} = \Big[\widehat{\tau}_T-M_{1,\mathtt{U}}-M_{2,\mathtt{U}} \, , \, \widehat{\tau}_T-M_{1,\mathtt{L}}-M_{2,\mathtt{L}}\Big],\]
for appropriate choices of $M_{1,\mathtt{L}}$, $M_{1,\mathtt{U}}$, $M_{2,\mathtt{L}}$ and $M_{2,\mathtt{U}}$ and conditioning sigma field. In this paper, we consider conditional prediction intervals with $\mathscr{H}= \sigma(\bB, \bC, \bx_T, \bg_T)$ and focus on building a probability bound for each of the two terms, $\bp_T'(\bbeta_0-\widehat{\bbeta})$ and $e_T$, separately, and then combine them to build an overall probability bound via Lemma \ref{lem: PI}. In the decomposition leading to the prediction interval construction, we interpret $\bp_T'(\bbeta_0-\widehat{\bbeta})$ as capturing the in-sample uncertainty coming from constructing the SC weights using pre-treatment information, and $e_T$ the out-of-sample uncertainty coming from misspecification along with any additional noise occurring at the post-treatment period $T>T_0$. The next two subsections are devoted to handle each of these terms, respectively.

\begin{remark}[Prediction Interval for $Y_{1T}(0)$]\label{rmk:PI for Y_1T(0)}
Once the $\mathscr{H}$-conditionally $(\alpha,\pi)$-valid prediction interval $\mathcal{I}$ for $\tau_T$ is constructed, an analogous prediction interval for the counterfactual outcome of the treated unit in the post-treatment period $T$, $Y_{1T}(0)$, is also readily available. To be precise, using \eqref{eq: parameter tau}, it follows that
\[\P\Big\{\P\big[Y_{1T}(1) - Y_{1T}(0) \in \mathcal{I} \big|  \mathscr{H}  \big] \geq 1-\alpha\Big\}\geq 1-\pi,\]
that is, $\big[M_{1,\mathtt{L}}+M_{2,\mathtt{L}}+\widehat{Y}_{1T}(0) \, , \, M_{1,\mathtt{U}}+M_{2,\mathtt{U}}+\widehat{Y}_{1T}(0) \big]$ is a conditionally valid prediction interval for $Y_{1T}(0)$.
\end{remark}

\section{In-Sample Uncertainty} \label{sec:bound w.hat}

We first quantify the in-sample uncertainty coming from $\bp_T'(\bbeta_0-\widehat{\bbeta})$, thereby providing methods to determine $(M_{1,\mathtt{L}},M_{1,\mathtt{U}})$ and their probability guarantees $(\alpha_1,\pi_1)$ in Lemma \ref{lem: PI}. Let $\bZ=(\bB, \bC)$, $\bD$ be a non-negative diagonal (scaling) matrix of size $d$, possibly depending on the pre-treatment sample size $T_0$, and recall that $\mathscr{H} = \sigma(\bB, \bC, \bx_T, \bg_T)$. Because $\widehat{\bbeta}$ solves \eqref{eq: estimated weight}, we can define $\widehat{\bdelta}:=\bD(\widehat{\bbeta}-\bbeta_0)$ as the optimizer of the centered criterion function:
\[
\widehat{\bdelta}=\underset{\bdelta\in\Delta}{\argmin}\; \big\{\bdelta'\widehat{\bQ}\bdelta-2\widehat{\bgamma}'\bdelta\big\},
\]
where $\widehat{\bQ}=\bD^{-1}\bZ'\bZ\bD^{-1}$,  
$\widehat{\bgamma}'=\bU'\bZ\bD^{-1}$, and $\Delta=\{\bm{h}\in\mathbb{R}^d: \bm{h}=\bD(\bbeta-\bbeta_0),\, \bbeta\in\mathcal{W}\times \mathcal{R}\}$.

The following lemma, which holds whether or not $\bgamma:=\E[\widehat{\bgamma}|\mathscr{H}] = \mathbf{0}$, is a key building block for our prediction interval construction.

\begin{lem}[Optimization Bounds]\label{lem: Opt Bounds}
	\label{lem: concentration of weights}
	Fix $\widehat{\bQ}$ and $\bp_T$. Assume $\mathcal{W}$ and $\mathcal{R}$ are convex, and let $\widehat{\bbeta}$ in \eqref{eq: estimated weight} and $\bbeta_0$ in \eqref{eq: pseudo true value} exist. Then,
	\[
	\varsigma_{\mathtt{L}}:=\inf_{\bdelta\in\mathcal{M}_{\widehat{\bgamma}-\bgamma}}\bp_T'\bD^{-1}\bdelta \,\leq \, 
	\bp_T'\bD^{-1}\widehat{\bdelta} \, \leq \, \sup_{\bdelta\in\mathcal{M}_{\widehat{\bgamma}-\bgamma}} \bp_T'\bD^{-1}\bdelta =:\varsigma_{\mathtt{U}},
	\]
	where $\mathcal{M}_{\bxi}=\{\bdelta\in\Delta:\bdelta'\widehat{\bQ}\bdelta-2\bxi'\bdelta\leq 0\}$. Furthermore, for any $\kappa\in\mathbb{R}$, \[\Big\{\bxi\in\mathbb{R}^{d}:\inf_{\bdelta\in\mathcal{M}_{\bxi}}\;\bp_T'\bD^{-1}\bdelta \geq \kappa\Big\} \qquad \text{and} \qquad
	  \Big\{\bxi\in\mathbb{R}^{d}:\sup_{\bdelta\in\mathcal{M}_{\bxi}}\;\bp_T'\bD^{-1}\bdelta \leq \kappa\Big\}
	\]
	are convex sets.
\end{lem}

This lemma does not involve probabilistic statements, but rather follows from basic features of constrained least squares optimization. In particular, simple bounds on $\bp_T'\bD^{-1}\widehat{\bdelta}$ can be deduced based on the basic inequality from optimization $\widehat{\bdelta}'\widehat{\bQ}\widehat{\bdelta}-2(\widehat{\bgamma}-\bgamma)'\widehat{\bdelta}\leq 0$, and the fact that any solution must satisfy the constraints imposed in \eqref{eq: estimated weight}, i.e., $\widehat{\bdelta}\in\Delta$. The second part of the lemma establishes that the set of possible localization values ($\bxi$) determining the feasibility set ($\mathcal{M}_{\bxi}$) of the bounding (random) quantities $\varsigma_{\mathtt{L}}$ and $\varsigma_{\mathtt{U}}$ in Lemma \ref{lem: Opt Bounds} (when $\bxi=\widehat{\bgamma}-\bgamma$) form (random) convex sets.

Conditional on $\mathscr{H}$, the set $\mathcal{M}_{\bxi}$ is not random due to $\widehat{\bQ}$, which is random only unconditionally. As a consequence, conditional on $\mathscr{H}$, both $\mathcal{M}_{\widehat{\bgamma}-\bgamma}$ and $\{\bp_T'\bD^{-1}\bdelta:\bdelta\in\mathcal{M}_{\widehat{\bgamma}-\bgamma}\}$ are random sets only because $\widehat{\bgamma}$ is a random quantity and, accordingly, $\varsigma_{\mathtt{L}}$ and $\varsigma_{\mathtt{U}}$ are random variables defined by a random set. If the conditional distributions of $\varsigma_{\mathtt{L}}$ and $\varsigma_{\mathtt{U}}$ were known, we could take their quantiles as lower and upper bounds for the quantiles of the conditional distribution of $\bp_T'\bD^{-1}\widehat{\bdelta}=\bp_T'(\widehat{\bbeta}-\bbeta_0)$, thereby transforming the first conclusion of Lemma \ref{lem: concentration of weights} into a probabilistic statement. However, this approach requires knowledge of the conditional (on $\mathscr{H}$) distribution of the bounding random variables $\varsigma_{\mathtt{L}}$ and $\varsigma_{\mathtt{U}}$. The convexity properties also established in Lemma \ref{lem: Opt Bounds} allow us to provide precise bounds on the desired conditional distribution of the bounding random variables using Berry-Esseen bounds for convex sets \citep{Raivc_2019_Bernoulli}.

The following theorem formalizes our first main result based on Lemma \ref{lem: Opt Bounds}. We only present the result for the upper bound to conserve space, but the analogous result holds for the lower bound. See Remarks SA-2.1, SA-2.2 and SA-2.3 in the supplemental appendix for more details. Let $\bSigma=\V[\widehat{\bgamma}|\mathscr{H}]$ and $\bSigma^{-1/2}\bD^{-1}\bZ'=(\tilde{\bz}_{1,1}, \cdots, \tilde{\bz}_{T_0,1}, \cdots, \tilde{\bz}_{1,M}, \cdots, \tilde{\bz}_{T_0,M})$. In addition, let $\|\cdot\|$ denote the spectral matrix norm (so that $\|\cdot\|=\|\cdot\|_2$ for vectors).

\begin{thm}[Distributional Approximation, Independent Case]
	\label{thm: coverage error approximation}
	Assume $\mathcal{W}$ and $\mathcal{R}$ are convex, $\widehat{\bbeta}$ in \eqref{eq: estimated weight} and $\bbeta_0$ in \eqref{eq: pseudo true value} exist, and $\mathscr{H} = \sigma(\bB, \bC, \bx_T, \bg_T)$. In addition, for some finite non-negative constants $\epsilon_{\gamma}$ and $\pi_{\gamma}$, the following conditions hold:
	\begin{enumerate}[label=\normalfont(T\thethm.\roman*),noitemsep]
		\item $\bu_t=(u_{t,1}, \cdots, u_{t,M})'$ is independent over $t$ conditional on $\mathscr{H}$;
		\item $\P\{\sum_{t=1}^{T_0}\E[\|\sum_{l=1}^{M}\tilde{\bz}_{t,l}(u_{t,l}-\E[u_{t,l}|\mathscr{H}])\|^3|\mathscr{H}]\leq \epsilon_{\gamma}(42(d^{1/4}+16))^{-1}\}\geq 1-\pi_{\gamma}$.
	\end{enumerate}
	Then,
	\[\P\Big[\P\big(\bp_T'\bD^{-1}\widehat{\bdelta}\leq \mathfrak{c}^\dagger(1-\alpha)\big|\mathscr{H}\big)\geq 1-(\alpha+\epsilon_\gamma)\Big]\geq 1-\pi_\gamma,\]
	where $\mathfrak{c}^\dagger(1-\alpha)$ denotes the $(1-\alpha)$-quantile of $\varsigma_\mathtt{U}^\dagger=\sup\big\{\bp_T'\bD^{-1}\bdelta:\bdelta\in\mathcal{M}_{\bG}\big\}$ conditional on $\mathscr{H}$, with $\mathcal{M}_{\bG}=\{\bdelta\in\Delta: \ell^\dagger(\bdelta)\leq 0\}$, $\ell^\dagger(\bdelta):=\bdelta'\widehat{\bQ}\bdelta-2\bG'\bdelta$, and $\bG|\mathscr{H}\thicksim\mathsf{N}(\mathbf{0}, \bSigma)$.
\end{thm}

This theorem is established under two high-level conditions. Condition (T1.i) imposes independence across time for the pseudo-residuals underlying the population analogue construction of the synthetic control weights in \eqref{eq: vertical regression}. In Appendix \ref{sec:extension weakly dependent data} we relax this requirement by allowing for weak dependence across time via a $\beta$-mixing condition \citep{Doukhan_2012_Book}, but to avoid untidy conditions we focus on the independent case here. Importantly, even in this case, Theorem \ref{thm: coverage error approximation} covers non-stationarity in the outcome variable (via a cointegration relationship). See Section \ref{sec:examples} for different examples with independent, weakly stationary, and non-stationary data.

The second high-level requirement in Theorem \ref{thm: coverage error approximation} helps control the (distributional) distance between $\widehat{\bgamma}-\bgamma$ and the Gaussian random vector $\mathsf{N}(\mathbf{0}, \bSigma)$, conditionally on $\mathscr{H}$, as well as the unconditional probability loss $\pi_\gamma$. Condition (T1.ii) can be verified in a variety of ways depending on the dependence structure imposed on the data and other regularity conditions, as we illustrate in Section \ref{sec:examples}. For instance, two sufficient conditions are: $\max_{1\leq l\leq M}\max_{1\leq t\leq T_0} \E[|u_{t,l}-\E[u_{t,l}|\mathscr{H}]|^3|\mathscr{H}]\leq \eta$, a.s. on $\mathscr{H}$ for some constant $\eta>0$, and $\P\{\sum_{t=1}^{T_0}\sum_{l=1}^{M}\|\tilde{\bz}_{t,l}\|^3\leq\epsilon_\gamma(42(d^{1/4}+16)\eta M^2)^{-1}\}\geq 1-\pi_\gamma$. 

Since $\bp_T'\bD^{-1}\widehat{\bdelta}=\bp_T'(\widehat{\bbeta}-\bbeta_0)$, Theorem \ref{thm: coverage error approximation} could immediately be applied to construct valid $M_{1,\mathtt{L}}$ and $M_{1,\mathtt{U}}$ in Lemma \ref{lem: PI} if $\bSigma=\V[\widehat{\bgamma}|\mathscr{H}]$ was known. Thus, to finalize the in-sample uncertainty quantification we discuss a feasible simulation-based approximation for the critical value $\mathfrak{c}^\dagger(1-\alpha)$. To describe such approach, define a simulation-based criterion function conditional on the data
\[\ell^{\ttb}(\bdelta) = \bdelta'\widehat{\bQ}\bdelta-2(\bG^{\ttb})'\bdelta,\qquad \bG^{\ttb}\thicksim\mathsf{N}(0,\widehat{\bSigma}),\]
where $\widehat{\bSigma}$ is some estimate of $\bSigma$. The form of $\widehat{\bSigma}$ depends on the specific dependence structure underlying the data and related regularity conditions, as we illustrate in Section \ref{sec:examples}. Naturally, the important high-level requirement is that $\widehat{\bSigma}$ should concentrate around $\bSigma$ with known probability; see Theorem \ref{thm: coverage error approximation, plug-in} below for the precise statement. In addition, the constraint set used in the simulation has to be properly defined to account for the parameters being possibly near or at the boundary, so that it mimics the local geometry of $\Delta$. Specifically, let $\Delta^{\ttb}$ denote the constraint set used in simulation. We require that
\begin{equation} \label{eq: locally equal}
\Delta^{\ttb}\cap\mathcal{B}(\mathbf{0}, \varepsilon)= \Delta\cap \mathcal{B}(\mathbf{0}, \varepsilon), \qquad \text{for some} \;\varepsilon>0,
\end{equation}
where $\mathcal{B}(\mathbf{0}, \varepsilon)$ is an $\varepsilon$-neighborhood around zero. We say  $\Delta^{\ttb}$ is \textit{locally equal to} $\Delta$ if \eqref{eq: locally equal} is satisfied. Consequently, searching for the desired region under constraints in $\Delta^{\ttb}$ is almost equivalent to doing so under constraints in $\Delta$. We discuss below more implementation details. 

The next theorem establishes the validity of our proposed simulation-based inference method and provides the associated probability guarantees, under high-level conditions. Let $\|\cdot\|_\mathtt{F}$ denote the Frobenius matrix norm (so that $\|\cdot\|_\mathtt{F}=\|\cdot\|=\|\cdot\|_2$ for vectors), and $\bI_q$ the identity matrix of size $q$ for an integer $q>0$.

\begin{thm}[Plug-in Approximation]
	\label{thm: coverage error approximation, plug-in}
	Assume $\mathcal{W}$ and $\mathcal{R}$ are convex, $\widehat{\bbeta}$ in \eqref{eq: estimated weight} and $\bbeta_0$ in \eqref{eq: pseudo true value} exist, and $\mathscr{H} = \sigma(\bB, \bC, \bx_T, \bg_T)$. In addition, for some finite non-negative constants $\epsilon_{\gamma}$, $\pi_{\gamma}$, $\varpi_\delta^{\ttb}$, $\epsilon_{\delta}^{\ttb}$, $\pi_\delta^{\ttb}$, $\epsilon_{\Delta}^{\ttb}$, $\pi_{\Delta}^{\ttb}$, $\epsilon_{\gamma,1}^{\ttb}$, $\epsilon_{\gamma,2}^{\ttb}$ and $\pi_{\gamma}^{\ttb}$, the following conditions hold:
	\begin{enumerate}[label=\normalfont(T\thethm.\roman*),noitemsep]
		\item $\P[\P(\bp_T'\bD^{-1}\widehat{\bdelta}\leq \mathfrak{c}^\dagger(1-\alpha)|\mathscr{H})\geq 1-\alpha-\epsilon_\gamma]\geq 1-\pi_\gamma$;
		\item $\P[\P(\sup\{\|\bdelta\|:	\bdelta\in\mathcal{M}_\bG\} \leq \varpi_\delta^{\ttb}|\mathscr{H})\geq 1-\epsilon_{\delta}^{\ttb}]\geq 1-\pi_\delta^{\ttb}$;
		\item $\P[\P(\Delta^{\ttb} \text{ is locally equal to } \Delta \;|\mathscr{H})\geq 1-\epsilon_{\Delta}^{\ttb}]\geq 1-\pi_{\Delta}^{\ttb}$ for $\varepsilon=\varpi_\delta^{\ttb}$ in \eqref{eq: locally equal};
		\item $\P[\P(\|\bSigma^{-1/2}\widehat{\bSigma}\bSigma^{-1/2}-\bI_d\|_\mathtt{F}\leq 2\epsilon_{\gamma,1}^{\ttb}|\mathscr{H})\geq 1-\epsilon_{\gamma,2}^{\ttb}]\geq 1-\pi_{\gamma}^{\ttb}$.
	\end{enumerate}
    Then, for $\epsilon_{\gamma,1}^{\ttb}\in[0,1/4]$,
	\[
	\P\Big[\P\big(\bp_T'\bD^{-1}\widehat{\bdelta}\leq \mathfrak{c}^{\ttb}(1-\alpha)\big|\mathscr{H}\big)\geq 1-\alpha-\epsilon\Big]\geq 1-\pi,
	\]
	where $\epsilon=\epsilon_{\gamma}+\epsilon_{\gamma,1}^{\ttb}+\epsilon_{\gamma,2}^{\ttb}+\epsilon_{\delta}^{\ttb}+\epsilon_{\Delta}^{\ttb}$,
	$\pi=\pi_{\gamma}+\pi_\gamma^{\ttb}+\pi_{\delta}^{\ttb}+\pi_{\Delta}^{\ttb}$, and $\mathfrak{c}^{\ttb}(1-\alpha)$ denotes the $(1-\alpha)$-quantile of $\varsigma_\mathtt{U}^{\ttb}:=\sup\{\bp_T'\bD^{-1}\bdelta: \bdelta\in\Delta^{\ttb},\,\ell^{\ttb}(\bdelta)\leq 0\}$, conditional on the data.
\end{thm}

This theorem gives a feasible, simulation-based approach to determine valid $M_{1,\mathtt{L}}$ and $M_{1,\mathtt{U}}$ in Lemma \ref{lem: PI}, with precise coverage probability guarantees. The first high-level Condition (T2.i) in Theorem \ref{thm: coverage error approximation, plug-in} takes as a starting point the conclusion of Theorem \ref{thm: coverage error approximation} or, alternatively, the conclusion of Theorem \ref{thm: coverage error approximation, mixing} in the appendix when the data is assumed to exhibit weak dependence via a $\beta$-mixing condition. The other three high-level conditions in Theorem \ref{thm: coverage error approximation, plug-in} are intuitive. Conditions (T2.ii) and (T2.iii) control the local geometry of the simulation feasibility set, as discussed earlier, while Condition (T2.iv) requires $\widehat{\bSigma}$ to be a ``good'' approximation of $\bSigma$, in the sense that $\widehat{\bSigma}$ concentrates in probability around $\bSigma$ with well-controlled errors. Importantly, Theorem \ref{thm: coverage error approximation, plug-in} is carefully crafted to accommodate both Theorem \ref{thm: coverage error approximation} (independent data) and Theorem \ref{thm: coverage error approximation, mixing} in the appendix (weakly dependent time series data) in a unified way. The next section illustrates different cases with practically relevant examples, and gives precise primitive conditions.

\subsection{Examples} \label{sec:examples}

We consider the standard synthetic control constraints $\mathcal{W}=\{\bw\in\mathbb{R}^N_+: \|\bw\|_1=1\}$ and $\mathcal{R}=\mathbb{R}^{KM}$. For simulation-based inference, we define explicitly a relaxed constraint set based on the original estimated coefficients $\widehat{\bbeta}$: $\Delta^{\star}= \{\bD(\bbeta-\widehat{\bbeta}^{\ttb}):\bbeta=(\bw', \br')', \bw\in\mathbb{R}^N_+,\; \|\bw\|_1=\|\widehat{\bw}^{\ttb}\|_1\}$, where $\widehat{\bbeta}^{\ttb}=(\widehat{\bw}^{\ttb '},\widehat{\br}')'$, $\widehat{\bw}^{\ttb}=(\widehat{\omega}_2^{\ttb}, \cdots, \widehat{\omega}_{N+1}^{\ttb})'$, $\widehat{\omega}_j^{\ttb}=\widehat{\omega}_j\I(|\widehat{\omega}_j|>\varrho)$, and $\varrho$ is a tuning parameter that ensures the constraint set in the simulation world preserves the local geometry of $\Delta$. Moreover, we set $\bx_T=(Y_{2T}(0), \cdots, Y_{(N+1)T}(0))'$ as it is common in the SC literature. Other synthetic control methods that vary these choices, including the other constraint sets $\mathcal{W}$ discussed previously, can be handled analogously, but we do not discuss them in this paper due to space limitations. Finally, in the remaining of this paper, we let $\mathfrak{C}$, $\mathfrak{C}^{\ttb}$ and $\mathfrak{c}$, with various sub-indexes, denote non-negative finite constants not depending on $T_0$. In simple cases, we give the exact expression of these constants, while in other cases they can be characterized from the proofs of the results. Let $\lambda_{\min}(\bM)$ and $\lambda_{\max}(\bM)$ be the minimum and the maximum eigenvalues of a generic square matrix $\bM$.

\subsubsection{Outcomes-only}

We start with the simplest possible example already introduced in Section \ref{sec:setup}. The SC weights are constructed based on past outcomes only, and the model allows for an intercept. Thus, the working model simplifies to 
\[a_{t}=\bb_{t}'\bw_{0} + r_0 + u_t, \qquad t=1, \cdots, T_0,\]
where $a_{t}:=Y_{1t}(0)$, $\bb_{t} :=(Y_{2t}(0), Y_{3t}(0), \ldots, Y_{(N+1)t}(0))'$, and with $M=1$, $K=1$, and $d=N+1$. Recall that $\bw_0=(w_{0,1},w_{0,2},\dots,w_{0,J})'$ is defined in \eqref{eq: pseudo true value}, and let $\bz_t=(\bb_t', 1)'$, $\bbeta_0=(\bw_0', r_0)'$. We further assume independent sampling across time, and thus set $\bD=T_0^{1/2}\bI_d$. A natural variance estimator is
\[\widehat{\bSigma}=\frac{1}{T_0}\sum_{t=1}^{T_0}\bz_t\bz_t' (\widehat{u}_{t}-\widehat{\E}[u_{t}|\bb_t])^2,\]
where $\widehat{u}_t=a_t-\bz_t'\widehat{\bbeta}$, and $\widehat{\E}[u_{t}|\bb_t]$ denotes some estimate of the conditional mean of the pseudo-residuals.

Theorem SA-1 in the supplemental appendix gives precise primitive conditions to verify the high-level conditions of Theorems \ref{thm: coverage error approximation} and \ref{thm: coverage error approximation, plug-in}. In particular, assuming that $\{\bz_t, u_t\}_{t=1}^T$ is i.i.d. over $t=1, \cdots, T_0$, and that $\max_{1\leq t\leq T_0}\E[|u_t|^3|\bB]\leq\bar{\eta}_1$ a.s. on $\sigma(\bB)$ and $\E[\|\bz_t\|^6]\leq \bar{\eta}_2$, $\min_{1\leq t\leq T_0}\V[u_{t}|\bB]\geq\underline{\eta}_1$ a.s. on $\sigma(\bB)$, and $\lambda_{\min}(\E[\bz_t\bz_t'])\geq \underline{\eta}_2$, for finite non-negative constants $\bar{\eta}_1$, $\bar{\eta}_2$, $\underline{\eta}_1$ and $\underline{\eta}_2$, we show that the conditions of Theorem \ref{thm: coverage error approximation} hold with $\pi_{\gamma}=\mathfrak{C}_{\pi}T_0^{-1}$ and $\epsilon_{\gamma}=\mathfrak{C}_{\epsilon}T_0^{-1/2}$, where $\mathfrak{C}_{\pi}=\frac{d}{\bar{\eta}_2}+\frac{4d^4\bar{\eta}_2}{\underline{\eta}_2^2}$ and $\mathfrak{C}_{\epsilon}=42(d^{1/4}+16) \frac{2^{5/2}d^{3/2}\bar{\eta}_1\bar{\eta}_2}{(\underline{\eta}_1\underline{\eta}_2)^{3/2}}$. Furthermore, under additional primitive conditions, we also show that the conditions of Theorem \ref{thm: coverage error approximation, plug-in} hold with precise non-asymptotic probability bounds characterized in the proof.

Theorem SA-1 characterizes precisely the probability guarantees for the in-sample prediction---that is, the precise values of $\alpha_1$ and $\pi_1$ in Lemma \ref{lem: PI} obtained via Theorems \ref{thm: coverage error approximation} and \ref{thm: coverage error approximation, plug-in}. The conditions imposed are primitive (e.g., moment bounds and rank conditions), with perhaps the exception of conditions (SA-1.iii) and (SA-1.v) in Theorem SA-1 in the supplemental appendix. Specifically, Condition (SA-1.iii) requires $\varrho=\varpi_\delta^{\ttb}/\sqrt{T_0}$ and $\P(\min\{|w_{0,j}|:w_{0,j}\neq 0\}\geq\varrho)\geq 1-\pi_{w}^{\ttb}$, for non-negative constants $\varpi_\delta^{\ttb}$ and $\pi_{w}^{\ttb}$, which is also primitive insofar it relates to the separation from zero of the non-zero (possibly random) coefficients $\bw_0$ entering the best linear approximation \eqref{eq: pseudo true value}, which is a standard (sparsity-type) assumption in the literature of constrained least squares estimation. On the other hand, Condition (SA-1.v) requires $\P[\P(\max_{1\leq t\leq T_0}|\widehat{\E}[u_t|\bb_t]-\E[u_t|\bb_t]|\leq \varpi_u^{\ttb}|\mathscr{H})\geq 1-\epsilon_u^{\ttb}]\geq 1-\pi_u^{\ttb}$, for non-negative constants $\varpi_u^{\ttb}$, $\epsilon_u^{\ttb}$ and $\pi_u^{\ttb}$, which is purposely not as primitive (but still easily interpretable) because it is meant to cover many different approximation approaches for $\E[u_t|\bb_t]$. In practice, researchers may assume $\E[u_t|\bb_t]=0$ or, alternatively, employ flexible-parametric/nonparametric approaches to form the estimator $\widehat{\E}[u_t|\bb_t]$. Since the latter approaches are setting-specific and technically well-understood, we chose to present our results using the generic condition (SA-1.v) rather than providing primitive conditions for a specific example of $\widehat{\E}[u_t|\bb_t]$.

\subsubsection{Multi-equation with Weakly Dependent Data}\label{sec: cov. dep. case}

The second example is the multi-equation setup introduced in Section \ref{sec:setup}, where we incorporate pre-intervention covariates in the construction of the SC weights and allow for stationary weakly dependent time series data. See \citet{Kilian-Lutkepohl_2017_Book} and references therein for an introduction to time series analysis. We let $M=2$ (two features) and $K=0$ (no additional controls) for simplicity, which gives the working model
\begin{align*}
   a_{t,1} &=\sum_{j=1}^J b_{jt,1}w_{0,j} + u_{t,1}, \\
   a_{t,2} &=\sum_{j=1}^J b_{jt,2}w_{0,j} + u_{t,2},
\end{align*}
$t=1, \cdots, T_0$. The first equation could naturally correspond to pre-intervention outcomes as in the previous example, i.e., $a_{t,1}:=Y_{1t}(0)$ and $\bb_{t,1} :=(Y_{2t}(0), Y_{3t}(0), \ldots, Y_{(N+1)t}(0))'$, while the second equation could correspond to some other covariate (such as population density in the Basque terrorism application) also used to construct $\widehat{\bw}$ in \eqref{eq: estimated weight}. Let $\bb_{t,l}=(b_{1t,l}, \cdots, b_{Jt,l})'$, for $l=1,2$. To provide interpretable primitive conditions, we also assume $\bu_t=(u_{t,1},u_{t,2})'$ and $\bb_t=(\bb_{t,1}',\bb_{t,2}')'$ follow independent first-order stationary autoregressive (AR) processes:
\begin{alignat*}{2}
    \bu_t &= \bH_u\bu_{t-1}+\bm{\zeta}_{t,u}, \qquad &&\bH_u=\diag(\rho_{1,u}, \rho_{2,u}),\\
    \bb_t &= \bH_b\bb_{t-1}+\bm{\zeta}_{t,b}, \qquad &&\bH_b=\diag(\rho_{1,b}, \rho_{2,b}, \cdots, \rho_{J,b}),
\end{alignat*}
where $\bm{\zeta}_{t,u}$ and $\bm{\zeta}_{t,b}$ are i.i.d. over $t$, independent of each other, and $\diag(\cdot)$ denotes a diagonal matrix with the function arguments as the corresponding diagonal elements. Let $\bD=T_0^{1/2}\bI_d$, and note that $\bU=(u_{1,1}, \cdots, u_{T_0,1}, u_{1,2}, \cdots, u_{T_0, 2})'$ in this case. A natural, generic variance estimator is
\[\widehat{\bSigma}=\frac{1}{T_0}\bZ'\widehat{\V}[\bU|\mathscr{H}]\bZ,\]
where $\widehat{\V}[\bU|\mathscr{H}]$ is an estimate of $\V[\bU|\mathscr{H}]$. In this example, $\bSigma$ corresponds to the (conditional) long-run variance, and naturally $\widehat{\bSigma}$ can be chosen to be any standard estimator thereof.

Theorem  SA-2 in the supplemental appendix gives primitive conditions that verify the high-level conditions of Theorem \ref{thm: coverage error approximation, mixing} in the appendix, and the high-level conditions of Theorem \ref{thm: coverage error approximation, plug-in} for implementation. Note that because of the time dependence in this example, the primitive conditions are for Theorem \ref{thm: coverage error approximation, mixing}  instead of Theorem \ref{thm: coverage error approximation}. In particular, we show that under standard conditions guaranteeing $\beta$-mixing and moment and rank conditions (similar to those imposed in the previous example), the conditions of Theorem \ref{thm: coverage error approximation, mixing} hold with $\pi_{\gamma}=\mathfrak{C}_\pi T_0^{-\mathfrak{c}_\pi}$ and $\epsilon_{\gamma}=\mathfrak{C}_\epsilon T_0^{-\mathfrak{c}_\epsilon}$ for non-negative constants $\mathfrak{C}_\pi$ and $\mathfrak{C}_\epsilon$, and some positive constants $\mathfrak{c}_\pi$ and $\mathfrak{c}_\epsilon$, which are characterized precisely in the supplemental appendix. Theorem \ref{thm: coverage error approximation, plug-in} is also verified using the primitive conditions imposed in Theorem SA-2 in the supplemental appendix, and the associated non-asymptotic constants are characterized in its proof.

As in the previous example, Theorem SA-2 in the supplemental appendix illustrates the kind of primitive conditions needed to quantify in-sample uncertainty using our proposed methods. In this example, we accommodate multiple covariates (equations) in the construction of the SC weights and also allow for AR(1) dependent (stationary) time series data. The only intentionally high-level condition imposed is (SA-2.v), $\P(\P(\|\widehat{\bSigma}-\bSigma\|\leq \epsilon_{\Sigma,1}^{\ttb}|\mathscr{H})\geq 1-\epsilon_{\Sigma,2}^{\ttb})\geq 1-\pi_{\Sigma}^{\ttb}$ for non-negative constants $\epsilon_{\Sigma,1}^{\ttb}$, $\epsilon_{\Sigma,2}^{\ttb}$ and $\pi_{\Sigma}^{\ttb}$, which requires a concentration probability bound for the long-run variance estimator $\widehat{\bSigma}$ used to approximate the quantiles of the conditional (on $\mathscr{H}$) distribution of the bounding random variables $\varsigma_{\mathtt{L}}$ and $\varsigma_{\mathtt{U}}$ via simulations (Theorem \ref{thm: coverage error approximation, plug-in}). This condition is not difficult to verify for specific examples. 

\subsubsection{Cointegration}

Our third and final example illustrates how non-stationary data can also be handled within our framework. See \citet{Tanaka_2017_Book} and references therein for an introduction to non-stationary time series analysis. Suppose that for each $1\leq l\leq M$, $\{a_{t,l}\}_{t=1}^T$, $\{b_{1t,l}\}_{t=1}^T, \cdots, \{b_{Jt,l}\}_{t=1}^T$  are $I(1)$ processes, and $\{c_{1t,l}\}_{t=1}^T, \cdots, \{c_{Kt,l}\}_{t=1}^T$ and $\{u_{t,l}\}_{t=1}^{T}$ are $I(0)$ processes. Therefore, $\bA$ and $\bB$ form a cointegrated system. For simplicity, consider the following example: for each $l=1, \cdots, M$ and $j=1, \cdots, J$,
\begin{align*}
    a_{t,l}  &=\sum_{j=1}^{J}b_{jt,l}w_{0,j}+\sum_{k=1}^{K}c_{kt,l}r_{0,k,l}+u_{t,l},\\
    b_{jt,l} &=b_{j(t-1),l}+v_{jt,l}, 
\end{align*}
where $u_{t,l}$ and $v_{jt,l}$ are stationary unobserved disturbances. In this scenario, $(1, -\bw_0')'$ plays the role of a cointegrating vector such that the linear combination of $\bA$ and $\bB$ is stationary. The normalizing matrix $\bD=\mathrm{diag}\{T_0, \cdots, T_0, \sqrt{T_0}, \cdots, \sqrt{T_0}\}$, where the first $J$ elements are $T_0$ and the remaining ones are $\sqrt{T_0}$. 
Let $\check{\bZ}_t=(\check{\bz}_{t,1}, \cdots, \check{\bz}_{t,M})$ where $\check{\bz}_{t,l}$ is the $((l-1)T_0+t)$th column of $\diag\{T_0^{-1/2}\bI_J, \bI_{KM}\}\bZ'$, for $l=1, \cdots, M$. Recall that $\bu_t=(u_{t,1}, \cdots, u_{t,M})'$. 
Write $\bv_{t,l}=(v_{1t,l}, \cdots, v_{Jt,l})'$, $\bv_t=(\bv_{t,1}', \cdots, \bv_{t, M}')'$, and  $\bc_{t,l}=(c_{1t,l}, \cdots, c_{kt,l})'$. 
We allow some elements in $\bv_t$  to be used in $\{\bc_{t,l}\}_{l=1}^M$. Let $\bq_t$ collect all distinct variables in $\bu_t$, $\bv_t$, $\bc_{t,1}$, $\cdots$, $\bc_{t,M}$.
As in the previous example, a generic variance estimator is
\[\widehat{\bSigma}=\frac{1}{T_0}\sum_{t=1}^{T_0}\check{\bZ}_t\widehat{\V}[\bu_t|\mathscr{H}]\check{\bZ}_t',\]
where $\widehat{\V}[\bu_t|\mathscr{H}]$ is an estimate of $\V[\bu_t|\mathscr{H}]$.

Theorem SA-3 in the supplemental appendix gives more primitive conditions and verifies the high-level conditions of Theorems \ref{thm: coverage error approximation} and \ref{thm: coverage error approximation, plug-in} in the cointegration scenario. More precisely, it provides conditions so that Theorem \ref{thm: coverage error approximation} holds with $\pi_{\gamma}=\mathfrak{C}_{\pi,1}T_0^{-\psi\nu}+\mathfrak{C}_{\pi,2}T_0^{-1}+\pi_{Q,1}+\pi_{Q,2}$ and $\epsilon_{\gamma}=\mathfrak{C}_\epsilon (\log T_0)^{\frac{3}{2}(1+\mathfrak{c}_Q)}T_0^{-1/2}$ for non-negative constants $(\psi,\nu,\pi_{Q,1},\pi_{Q,2},\mathfrak{c}_Q)$ specified in the assumptions of the theorem and non-negative constants $(\mathfrak{C}_{\pi,1},\mathfrak{C}_{\pi,2},\mathfrak{C}_\epsilon)$ characterized in the proof. Similarly, Theorem \ref{thm: coverage error approximation, plug-in} is also verified under more primitive conditions, including a higher-level condition of the form $\P(\P(\|\widehat{\bSigma}-\bSigma\|\leq \epsilon_{\Sigma,1}^{\ttb}|\mathscr{H})\geq 1-\epsilon_{\Sigma,2}^{\ttb})\geq 1-\pi_{\Sigma}^{\ttb}$ for  non-negative constants $\epsilon_{\Sigma,1}^{\ttb}$, $\epsilon_{\Sigma,2}^{\ttb}$ and $\pi_{\Sigma}^{\ttb}$, as in the previous examples.

When $\bC$ is excluded, $\widehat{\bw}$ is a least squares estimator of the cointegrating vector, which is typically biased due to the potential correlation between $\bv_t$ and $\bu_t$. In Theorem SA-3 in the supplemental appendix, we include $\bC$ and allow it to include contemporary $\bv_t$ to correct this bias. More generally, one may augment the regression with $\bv_t$ and its leads and lags, which is termed dynamic OLS in the time series literature. The results for this general case may be established using a similar strategy.

\section{Out-of-Sample Uncertainty} \label{sec:bound e.T}

The unobserved random variable $e_T$ in \eqref{eq: SC Y1T(0)} is a single error term in period $T$, which we interpret as the error from out-of-sample prediction, conditional on $\mathscr{H}= \sigma(\bB, \bC, \bx_T, \bg_T)$. Naturally, in order to set appropriate $M_{2,\mathtt{L}}$ and $M_{2,\mathtt{U}}$ in Lemma \ref{lem: PI}, it is necessary to determine certain features of the conditional distribution $\P[e_T \leq \cdot | \mathscr{H}]$. In turn, determining those features would require strong distributional assumptions between pre-treatment and post-treatment periods, or perhaps across units. In this section we propose principled but agnostic approaches to quantify the uncertainty introduced by the post-treatment unobserved shock $e_T$. Since formalizing the validity of our methods requires strong assumptions, in this paper we recommend a generic sensitivity analysis to incorporate out-of-sample uncertainty to the prediction intervals. In particular, we propose employing three distinct methods for quantifying the uncertainty introduced by $e_T$ as a starting point, and then assessing more generally whether the additional uncertainty would render the prediction intervals large enough to eliminate any statistically significant treatment effect.

Our starting point is a non-asymptotic probability bound on $e_T$ via concentration inequalities. Such textbook results can be found in, for example, \cite{Vershynin_2018_Book} and \cite{Wainwright_2019_Book}. We rely on the following lemma, which provides the desired bounds for $e_T$ under different moment-like conditions.
 
\begin{lem}[Non-Asymptotic Probability Concentration for $e_T$]\label{lem: bound noise}\leavevmode
	\begin{enumerate}[noitemsep]
		\item[\normalfont(G)] If there exists some $\sigma_{\mathscr{H}}>0$ such that
		$\E[\exp(\lambda(e_T-\E[e_T|\mathscr{H}]))|\mathscr{H}]\leq \exp(\sigma_{\mathscr{H}}^2\lambda^2/2)$ a.s. for all $\lambda\in\mathbb{R}$,
		then for any $\varepsilon>0$,
		$\P(|e_T-\E[e_T|\mathscr{H}]|\geq \varepsilon|\mathscr{H})\leq 2\exp(-\varepsilon^2/(2\sigma_{\mathscr{H}}^2))$.
		
		\item[\normalfont(P)] If $\E[|e_T|^m|\mathscr{H}]<\infty$ a.s. for some $m\geq2$, then for any $\varepsilon>0$,
		$\P(|e_T-\E[e_T|\mathscr{H}]|\geq \varepsilon|\mathscr{H})\leq \varepsilon^{-m}\E[|e_T-\E[e_T|\mathscr{H}]|^m|\mathscr{H}]$.
	\end{enumerate}
\end{lem}

This lemma gives (possibly crude) bounds on the necessary features of the conditional distribution of $e_T$ given $\mathscr{H}$. Lemma \ref{lem: bound noise}(G) corresponds to a sub-Gaussian tail assumption, while Lemma \ref{lem: bound noise}(P) exploits only a polynomial bound on moments of $e_t|\mathscr{H}$. In both cases, the only unknowns are the ``center'' and ``scale'' of the distribution: $\E[e_T|\mathscr{H}]$ and $\sigma_{\mathscr{H}}^2$ (or higher-moments), respectively. These unknown features can be estimated or tabulated based on (i) model assumptions and (ii) observed pre-treatment data, at least as an initial step towards a sensitivity analysis.

For practical purposes, we first outline three alternative strategies to assess the uncertainty coming from $e_T$, starting with Lemma \ref{lem: bound noise} and progressively adding more restrictions. After introducing these approaches, we turn to discussing how they can be used as an initial step towards a principled sensitivity analysis for uncertainty quantification of the synthetic control estimator. Section \ref{sec:numerical} illustrates this idea using simulated data and empirical applications.

\begin{itemize}[leftmargin=*]
    \item \textbf{Approach 1: Non-Asymptotic Bounds}. In view of Lemma \ref{lem: bound noise}, we only need to extract some features of $e_T|\mathscr{H}$, namely some conditional moments of the form $\E[|e_T|^m|\mathscr{H}]$ (or $\E[e_T^m|\mathscr{H}]$) for appropriate choice(s) of $m\geq1$. In practice, for example, pre-treatment residuals $\{\widehat{u}_t\}_{t=1}^{T_0}$ could be used to estimate those quantities (e.g., under stationarity and other regularity conditions). Alternatively, the necessary conditional moments could be set using external information, or tabulated across different values to assess the sensitivity of the resulting prediction intervals. Importantly, once $\E[e_T|\mathscr{H}]$ and $\sigma_{\mathscr{H}}^2$ (or higher-moments) are set, then computing $M_{2,\mathtt{L}}$ and $M_{2,\mathtt{U}}$ in Lemma \ref{lem: PI} is straightforward via Lemma \ref{lem: bound noise}.

	\item \textbf{Approach 2: Location-scale Model}. Suppose that $e_T=\E[e_T|\mathscr{H}]+(\V[e_T|\mathscr{H}])^{1/2}\varepsilon_T$ with $\varepsilon_T$ statistically independent of $\mathscr{H}$. This setting imposes restrictions on the distribution of $e_T|\mathscr{H}$, but allows for a much simpler tabulation strategy. Specifically, the bounds in Lemma \ref{lem: PI} can now be set as $M_{2, \mathtt{L}}=\E[e_T|\mathscr{H}]+(\V[e_T|\mathscr{H}])^{1/2}\mathfrak{c}_\varepsilon(\alpha_2/2)$ and $M_{2, \mathtt{U}}=\E[e_T|\mathscr{H}]+(\V[e_T|\mathscr{H}])^{1/2}\mathfrak{c}_\varepsilon(1-\alpha_2/2)$ where $\mathfrak{c}_\varepsilon(\alpha_2/2)$ and $\mathfrak{c}_\varepsilon(1-\alpha_2/2)$ are $\alpha_2/2$  and $(1-\alpha_2/2)$ quantiles of $\varepsilon_t$, respectively, and $\alpha_2$ is the desired pre-specified level. In practice, $\E[e_T|\mathscr{H}]$ and $\V[e_T|\mathscr{H}]$ can be parametrized and estimated using the pre-intervention residuals $\{\widehat{u}_t\}_{t=1}^{T_0}$, or perhaps tabulated using auxiliary information. Once such estimates are available, the appropriate quantiles can be easily obtained using the standardized (estimated) residuals. This approach is likely to deliver more precise prediction intervals when compared to Approach 1, but at the expense of potential misspecification due to the location-scale model used.
	 
	\item \textbf{Approach 3: Quantile Regression}. In view of Lemma \ref{lem: PI}, we only need to determine the $\alpha_2/2$ and $(1-\alpha_2/2)$ conditional quantiles of $e_T|\mathscr{H}$. Consequently, another possibility is to employ quantile regression methods to estimate those quantities using pre-treament data.
\end{itemize}

While the three approaches above are simple and intuitive, they are potentially unsatisfactory because their validity would require arguably strong assumptions on the underlying data generating process linking the pre-treatment and post-treatment data. Such assumptions, however, are difficult to avoid because the ultimate goal is to learn about uncertainty introduced by an unobserved random variable after the treatment began (i.e., $e_T|\mathscr{H}$ for $T>T_0$). Without additional data availability or specific modelling assumptions allowing for transferring information from the pre-treatment period into the post-treatment period, it is difficult to formally set $M_{2,\mathtt{L}}$ and $M_{2,\mathtt{U}}$ in Lemma \ref{lem: PI}.

Nevertheless, it is possible to approach out-of-sample uncertainty quantification as a principled sensitivity analysis, using the methods above as a starting point. Given the formal and detailed in-sample uncertainty quantification developed in the previous section, it is natural to progressively enlarge the final prediction intervals by adding additional out-of-sample uncertainty to then ask the question: how large does the additional out-of-sample uncertainty contribution coming from $e_T|\mathscr{H}$ need to be in order to render the treatment effect $\tau_t$ in \eqref{eq: parameter tau} statistically insignificant? Using the approaches above, or similar ones, it is possible to construct natural initial benchmarks. For instance, the variability displayed by the pre-treatment outcomes or synthetic control residuals can help guide the level of ``reasonable'' out-of-sample uncertainty. Alternatively, in specific applications, natural levels of uncertainty for the outcomes of interests could be available, and hence used to tabulate the additional out-of-sample uncertainty. In Section \ref{sec:numerical} we further discuss and illustrate this idea numerically.

\subsection{Examples}

We revisit the three examples considered in Section \ref{sec:examples} and illustrate how the implementation of the three approaches outlined earlier may accommodate different assumptions on the data generating process. We discuss the outcomes-only case in more detail, which suffices to showcase our basic strategy of out-of-sample uncertainty quantification. For the other two examples, we briefly explain some important conceptual and implementational issues. As mentioned above, these methods rely on strong assumptions and should be viewed as a starting point of a general sensitivity analysis.

\subsubsection{Outcomes-only}
Recall that the data is assumed to be i.i.d. over $1\leq t\leq T$ in this case. The conditional distribution of $e_T$ given $\mathscr{H}$ then reduces to that given the contemporary covariates $\bb_T$ only. Also, we set $\bx_T=\bb_T$ and $\bg_T=(1, \cdots, 1)'$. Then, the out-of-sample error $e_T$ is equivalent to the pseudo-true residual $u_T$. By stationarity of the data, the information about the conditional distribution of $u_T$ can be learned using the pre-intervention residuals. These substantial simplifications facilitate the implementation of the proposed methods for quantifying the out-of-sample uncertainty.

\begin{itemize}[leftmargin=*]
    \item \textbf{Approach 1: Non-Asymptotic Bounds}. In general, we only need to estimate several conditional moments of $u_T$ given $\bb_T$. For example, assume that a (conditional) Gaussian bound holds for $u_T$. If $\E[u_T|\bb_T]=0$, i.e., the SC prediction correctly characterizes the conditional expectation of $a_T$ given $\bb_T$, then an estimate of the conditional variance of $u_T$ suffices to construct a prediction interval for $u_T$. Otherwise, an estimate of $\E[u_T|\bb_T]$ is also required to adjust the location of the prediction interval. These quantities can be estimated using the pre-treatment data. Though the pseudo-true residuals $\{u_t\}_{t=1}^{T_0}$ are not observed, good proxies $\{\widehat{u}_t\}_{t=1}^{T_0}$ are available from the SC fitting. In practice, flexible parametric or nonparametric approaches can be used to estimate these conditional moments. For instance, we can implement a simple linear regression of $\widehat{u}_t$ on $\bb_t$ to estimate $\E[u_t|\bb_t]$. Denote the predicted values by $\widehat{\E}[u_t|\bb_t]$. For the conditional variance, specify a model $\V[u_t|\bb_t]=\exp(\bb_t'\bm\theta_b+\theta_0)$ and implement a regression of $\log((\widehat{u}_t-\widehat{\E}[u_t|\bb_t])^2)$ on $\bb_t$. The predicted conditional variance is guaranteed to be positive. A prediction interval for the out-of-sample error can then be constructed based on Lemma \ref{lem: bound noise}(G).
    
    \item \textbf{Approach 2: Location-scale Model}. Similarly, to implement Approach 2, we only need estimates of the conditional mean and variance of $u_T$ given $\bb_T$, denoted by $\widehat{\E}[u_t|\bb_t]$ and $\widehat{\V}[u_t|\bb_t]$ respectively. They can be obtained using the methods outlined previously. Once they are available, set $M_{2, \mathtt{L}}=\widehat{\E}[u_T|\bb_T]+(\widehat{\V}[u_T|\bb_T])^{1/2}\widehat{\mathfrak{c}}_\varepsilon(\alpha_2/2)$ and $M_{2, \mathtt{U}}=\widehat{\E}[u_T|\bb_T]+(\widehat{\V}[u_T|\bb_T])^{1/2}\widehat{\mathfrak{c}}_\varepsilon(1-\alpha_2/2)$ where $\widehat{\mathfrak{c}}_\varepsilon(\alpha_2/2)$ and $\widehat{\mathfrak{c}}_\varepsilon(1-\alpha_2/2)$ are $\alpha_2/2$  and $(1-\alpha_2/2)$ quantiles of $\{\widehat{\varepsilon}_t\}_{t=1}^{T_0}$ where  $\widehat{\varepsilon}_t=(\widehat{u}_t-\widehat{\E}[u_t|\bb_t])/(\widehat{\V}[u_t|\bb_t])^{1/2}$, respectively.
    
    \item \textbf{Approach 3: Quantile Regression}. We can estimate the $\alpha_2/2$ and $(1-\alpha_2/2)$ conditional quantiles of $u_t$ given $\bb_t$ parametrically or nonparametrically. For instance, assume the $\ell$th quantile of $u_t$ admits a linear form: $Q(\ell|\bb_t)=\bb_t'\theta(\ell)$. Then, we can implement a quantile regression of the pre-treatment residuals $\widehat{u}_t$ on $\bb_t$ for $\ell=\alpha_2/2$ and $(1-\alpha_2/2)$, which suffices to construct a bound on $e_T$. See \citet{Koenker-et-al_2017_Handbook}, and references therein, for a comprehensive discussion of quantile regression methods.
\end{itemize}

\subsubsection{Multi-equation with Weakly Dependent Data}

When data is weakly dependent and multiple features are used in the construction of SC weights, the implementation of the three approaches is similar to that in the outcomes-only case, but two outstanding issues need to be addressed. First, as described in Section \ref{sec: cov. dep. case}, the SC weights are obtained by matching on two pre-intervention features $\{a_{t,1}\}_{t=1}^{T_0}$ and $\{a_{t,2}\}_{t=1}^{T_0}$, while in most SC applications, the final counterfactual prediction is constructed by setting $\bx_T=\bb_T$ (and $\bg_T=\bm{0}$ in this case). Conceptually, the out-of-sample error $e_T=Y_{1T}(0)-\bb_T'\bw_0$ may or may not correspond to the pseudo-true residual $\bu_t$ prior to the treatment. For example, if the pre-treatment outcomes are used in the first equation, then by construction, $e_t$ in this scenario is equivalent to $u_{t,1}$ for $t=1, \cdots, T$. In the pre-intervention period, the residuals $\{\widehat{u}_{t,1}\}_{t=1}^{T_0}$ from the SC fitting play the role of proxies for $\{u_{t,1}\}_{t=1}^{T_0}$. In contrast, if pre-treatment outcomes are not used in any of the two equations, then $e_t$ is generally not the same as $u_{t,1}$ or $u_{t,2}$. Nevertheless, we can manually construct $\widehat{e}_t=Y_{1t}(0)-\bb_t'\widehat{\bw}$ as a proxy for $e_t$ in the pre-intervention period.

Second, the dependence of $e_T$ on $\mathscr{H}$ should be appropriately characterized. Consider a simple scenario where pre-treatment outcomes are used in the construction of SC weights so that $e_t=u_{t,1}$. By our assumptions on $\bm\zeta_{t,u}$ and $\bm\zeta_{t,b}$, $\{u_{t,1}\}_{t=1}^{T}$ is independent of $\{\bb_t\}_{t=1}^T$. If we further assume the two components of $\bm\zeta_{t,u}$ are independent of each other, the conditional distribution of $u_{T,1}$ given $\mathscr{H}$ reduces to its unconditional distribution. Therefore, to implement the three approaches, one only needs to estimate the unconditional mean, variance or quantiles using the residuals $\{\widehat{u}_{t,1}\}_{t=1}^{T_0}$. In practice, however, the independence between $u_{T,1}$ and $\mathscr{H}$ may be unrealistic. Assuming an appropriate weak dependence structure, we can still characterize or approximate the conditional mean, variance or quantiles of $u_{t,1}$ by functions of $\bb_t$ and lags thereof, which could be estimated by parametric or nonparametric regressions using the pre-intervention data.

\subsubsection{Cointegration}

As in the second example, we first determine the pre-intervention analogue to the out-of-sample error $e_T$. For instance, we let $a_{t,1}=Y_{1t}(0)$ and  $b_{jt,1}=Y_{(j+1)t}(0)$ for $j=1, \cdots, N$. In practice, the final counterfactual prediction is often constructed by setting $\bx_T=(Y_{2t}(0), \cdots, Y_{(N+1)t}(0))'$ and $\bg_T=\bm{0}$, i.e., no additional control variables are used in the out-of-sample prediction. Then, we have $e_t=u_{t,1}+\sum_{k=1}^Kc_{kt,1}r_{0,k,1}$ for $t=1, \cdots, T$. In view of the assumptions imposed in Theorem SA-3 in the supplemental appendix, the conditional distribution of $e_t$ given $\mathscr{H}$ reduces to that given the contemporary variables $\{\bv_t,\bc_{t,1}, \cdots, \bc_{t,M}\}$. As in the previous examples, we can estimate its conditional mean, variance or quantiles using various parametric or nonparametric methods.

The assumption that $\{\bq_t\}_{t=1}^T$ is i.i.d. in Theorem SA-3 may be too strong. In practice, as mentioned previously, we may want to augment the regression of the residual $\widehat{e}_t$ by lags (and leads) of $\bv_t$ and $\{\bc_{t,l}\}_{l=1}^M$ and transformations thereof to take into account potential time series dependence.

\section{Numerical Results}\label{sec:numerical}

We illustrate the performance of the proposed prediction intervals with a Monte Carlo experiment and two empirical examples. To implement the methods described in Section \ref{sec:bound w.hat} and \ref{sec:bound e.T}, we take a simple ``plug-in'' estimator $\widehat{\bSigma}$ of the long-run variance $\bSigma$ and employ parametric polynomial regressions to estimate the conditional mean, variance and quantiles of $e_T$ given $\mathscr{H}$ whenever needed. In addition, the choice of the tuning parameter $\varrho$ can be based on a bound implied by optimization. Specifically, since $\widehat{\bdelta}$ must satisfy the basic inequality specified in the definition of $\mathcal{M}_\xi$ (see Lemma \ref{lem: Opt Bounds}), we can construct a threshold $\varrho$ for $\widehat{\bbeta}$ based on some estimates of the variance of $\widehat{\bgamma}$ and the eigenvalues of $\widehat{\bQ}$. More details are discussed below, and we also provide complete replication codes. Last but not least, in view of the small sample size in many SC applications, these practical choices play the role of a reasonable starting point for a principled sensitivity analysis, as illustrated in this section.

\subsection{Simulations}\label{sec:simuls}

We conduct a Monte Carlo investigation of the finite sample performance of our proposed methods. We consider the outcomes-only case where $M=1$, $J=N$, $K=0$, and only the outcome variable is used. Then, $\bA_1=(Y_{11}, Y_{12}, \cdots, Y_{1T_0})'$, $\bb_{j,1}=(Y_{(j+1)1}, Y_{(j+1)2}, \cdots, Y_{(j+1)T_0})'$. We set $T_0=100$, $T_1=1$, and $N=10$. We consider three data generating processes for $b_{jt}:=b_{jt,1}$: $b_{jt}=\rho b_{j(t-1)}+v_{jt}$ with $\rho\in\{0,0.5,1\}$, and where $\bv_t=(v_{1t}, \cdots, v_{Nt})'\thicksim i.i.d. \;\mathsf{N}(0, \bI_N)$. 

To examine the conditional coverage, we first generate a sample of $\{b_{jt}:1\leq t\leq T_0+T_1, 1\leq j\leq N\}$ using one of the three models. We set $5$ evaluation points in the post-treatment period: $\tilde{b}_{1(T_0+1)}:=b_{1(T_0+1)}+\mathtt{c}\cdot \mathsf{sd}(b_{1t})$ where $\mathtt{c}\in\{-1, -0.5, 0, 0.5, 1\}$ and $\mathsf{sd}(b_{1t})$ is the sample standard deviation of $\{b_{1t}\}_{t=1}^{T_0}$. In other words, we construct $5$ designs by varying the value of the first conditioning variable in the last period. Taking each of them as given, we generate the treated unit $a_t:=Y_{1t}=\bb_t'\bw+u_t$ by randomly drawing the error $u_t\thicksim i.i.d.\;\mathsf{N}(0,0.5)$ independent of $\{\bb_t\}_{t=1}^T$. We set $\bw=(0.3, 0.4, 0.3, 0, \cdots, 0)'$. By construction, $\bw$ is exactly equivalent to the pseudo-true weight $\bw_0$ defined in \eqref{eq: pseudo true value} and satisfies the positivity and sum-to-one constraints in the standard SC. We consider $5,000$ simulated datasets. Note that the design $\{\bb_t\}_{t=1}^{T_0}$ is fixed throughout the simulation study, and we only draw a new realization of the error term $u_t$ at each repetition.

For comparison we also investigate the unconditional coverage of the prediction intervals. The models for $a_t$ and $\bb_t$ are the same as described above, but in this case the design $\{\bb_t\}_{t=1}^T$ is not fixed across the simulation study. Instead, at each repetition we randomly draw $\{(a_t, \bb_t)\}_{t=1}^{T_0+1}$. Therefore, the coverage probability obtained in this alternative exercise is \textit{unconditional}, i.e., not conditional on a fixed design. 

In the above construction, the error term $u_t$ is independent of the design and has mean zero. To check the performance of the proposed method in models with misspecification error ($\E[u_t|\mathscr{H}]\neq 0$), we also consider several other data generating processes: for models with $\rho=0$ and $\rho=0.5$, we generate $u_t=0.2b_{1t}+\zeta_{t}$, while for the model with $\rho=1$, $u_t=0.9(b_{1t}-b_{1(t-1)})+\zeta_t$, where $\zeta_t\thicksim i.i.d. \;\mathsf{N}(0, 0.5)$. Notably, in these misspecified models, the weight $\bw$ defined previously is no longer equivalent to the pseudo-true weight $\bw_0$.

We focus on several versions of our proposed prediction intervals for the counterfactual outcome $Y_{1T}(0)$ of the treated unit with $90\%$ nominal (conditional) coverage probability: ``M1'' denotes the prediction interval based on assuming a Gaussian bound and applying the concentration inequality in Lemma \ref{lem: bound noise}(G) (``approach 1''); ``M1-S'' is the same as ``M1'' except that we increase the (estimated) conditional standard deviation $\widehat{\sigma}_{\mathscr{H}}$ by a factor of $2$; ``M2'' denotes the prediction interval based on the location-scale model (``approach 2''); and ``M3'' denotes the prediction interval based on linear quantile regression (``approach 3''). For comparison, we also include the prediction interval ``CONF'', which is based on the conformal method developed in \cite{Chernozhukov-Wuthrich-Zhu_2021_JASA}. In the supplemental appendix, we also report the performance of the prediction interval based on the cross-sectional permutation method proposed in \cite{Abadie-Diamond-Hainmueller_2010_JASA}.

As mentioned previously, we choose the tuning parameter $\varrho$ based on the basic optimization inequality $\varrho=\widehat{\sigma}_u(\log T_0)^{\mathtt{c}}/(\min_{1\leq j\leq J}\widehat{\sigma}_{b_j}T_0^{1/2})$, where $\widehat{\sigma}_u$ is the estimated (unconditional) standard deviation of $u_t$, $\widehat{\sigma}_{b_j}$ is the estimated (unconditional) second moment of $b_{jt}$, and $\mathtt{c}=1$ if $\rho=1$ and $\mathtt{c}=0.5$ if $\rho=0$ or $0.5$. This strategy accommodates the simple data generating processes considered in simulations, and it can be tailored to better suit the assumptions in more complex statistcal models as well. In addition, we use polynomial regression to estimate various features of the conditional distribution of $u_t$ given $\mathscr{H}$. To avoid overfitting, we only use a subset of the $J$ control outcomes which have non-zero weights in $\widehat{\bw}^{\ttb}$. Regarding the long-run variance $\bSigma$, we take a simple plug-in estimator $\widehat{\bSigma}=\bD^{-1}\bZ'\diag\{\tilde{u}_1^2, \cdots, \tilde{u}^2_{T_0}\}\bZ\bD^{-1}$, where $\tilde{u}_t=\widehat{u}_t-\widehat{\E}[u_t|\mathscr{H}]$ and $\widehat{\E}[u_t|\mathscr{H}]$ is the estimate of the conditional mean of $u_t$ given $\mathscr{H}$. 

Panels A and B of Table \ref{table:simul, p1} summarize the results for models with and without misspecification error, respectively, where we use linear regression methods to estimate the conditional mean, variance or quantiles whenever needed. The proposed prediction intervals exhibit good coverage properties throughout different data generating processes and evaluation points, though they are conservative in several cases. 
In contrast, the actual coverage probability of conformal prediction intervals developed in \cite{Chernozhukov-Wuthrich-Zhu_2021_JASA} is lower than the target nominal level in general. See Section SA-3 of the supplement for additional simulation evidence.

\subsection{Empirical Illustration}\label{sec:empapp}

We showcase our methods by reanalyzing two empirical examples from the synthetic control literature. The first example concerns the effect of California's tobacco control program, known as Proposition 99, on per capita cigarette sales \citep[see][for more details]{Abadie-Diamond-Hainmueller_2010_JASA}. The second example corresponds to the economic impact of 1990 German reunification on West Germany \citep[see][for more details]{Abadie_2021_JEL}. To conserve space, the results for the second example are reported in Section SA-4 of the supplement.

The outcome variable of interest is per capita cigarette sales in California, which is arguably non-stationary. We consider both the raw data and the first-differenced data, corresponding to the analysis of levels and growth rates of per capita sales, respectively. In each scenario, we construct (1) the synthetic control prediction $\bx_T'\widehat{\bw}$ as commonly done in the literature; (2) the prediction interval for the (conditionally non-random) ``synthetic control component'' $\bx_T'\bw_0$ only; and (3) three distinct prediction intervals for the counterfactual $Y_{1T}(0)$. More specifically, the prediction intervals for $Y_{1T}(0)$ are implemented using the three methods outlined in Section \ref{sec:bound e.T}: (i) conditional subgaussian bound using Lemma \ref{lem: bound noise}(G), labeled as \textit{approach 1}; (ii) conditional bound based on location-scale model, labeled as \textit{approach 2}; (iii) conditional bound using conditional quantile regression of residuals, labeled as \textit{approach 3}. 

The three methods used to quantify the out-of-sample uncertainty can be viewed as particular instances of a more general sensitivity analysis. In other words, varying the additional uncertainty contribution of $e_T$ in a principled way, researchers can better understand its impact on the construction of the prediction intervals. We also illustrate this approach (``sensitivity analysis''): focusing on approach 1 for concreteness, we rely on Gaussian bounds in Lemma \ref{lem: bound noise}(G) to assess how the prediction intervals behave as the variance of $e_T$ varies. 

Accordingly, we present six plots: the SC prediction $\bx_T'\widehat{\bw}$, the prediction interval (PI) only for the synthetic unit $\bx_T'\bw_0$ with at least $95\%$ nominal coverage probability, the three different constructions of PIs for the counterfactual $Y_{1T}(0)$ with at least $90\%$ nominal coverage probability, and a sensitivity analysis for one chosen post-treatment period. Because the size of the donor pool is larger than the number of available pre-treatment periods, our procedure may rely on loose bounds for Gaussian approximation errors in high-dimensional settings. 

We first consider the raw data of per capita cigarette sales. Figure \ref{fig:California, level}(a) shows the trajectory of per capita sales of the synthetic California (dashed blue) and the actual California (solid black). After 1988, the synthetic California series is above the observed one, suggesting a negative shock of Proposition 99 on cigarette sales in California. Figure \ref{fig:California, level}(b) adds a $95\%$ conservative prediction interval for the synthetic control component of California that takes into account the in-sample uncertainty due to the estimated SC weights. We add the uncertainty associated with $e_T$ in Figures \ref{fig:California, level}(c)-(e). The observed sequence is generally below the prediction intervals for the counterfactual outcome of California, suggesting statistically significant effects of Proposition 99. Figure \ref{fig:California, level}(f) shows the sensitivity analysis of the effect in 1989. The result is robust: the corresponding PIs are well separated from the observed outcome of California if we vary the estimated (conditional) standard deviation of $e_T$ in a relatively wide range.

We also analyze the (log) growth rate of per capita cigarette sales. The result is reported in Figure \ref{fig:California, growth rate}. We can see that the observed growth rate during the post-treatment period is generally lower than the SC prediction, but throughout the three constructions of PIs for the counterfactual outcome, the observed series is within the PIs for most post-treatment periods except in 1989. These empirical findings suggest a statistical significant effect of the tobacco control program on the growth rate of per capita cigarette sales only in year 1989. The sensitivity analysis in Figure \ref{fig:California, growth rate}(f) shows that the significance of the effect in 1989 is robust.

\section{Conclusion}\label{sec:conclusion}

The synthetic control method is part of the standard program evaluation toolkit. Despite its popularity, many important methodological and theoretical developments remain outstanding. We focus on quantifying the uncertainty of the SC method in predicting the main quantity of interest, $\tau_T=Y_{1T}(1)-Y_{1T}(0)$, in the standard SC framework. This quantity is the difference between the observed outcome of the treated unit in a post-treatment period $T$, and the outcome that the treated unit would have had in the same period in the absence of treatment. Because we view $\tau_T$ as a random variable and there is a single treated unit, we propose conditional prediction intervals that offer finite-sample probability guarantees regarding the realization of the counterfactual treated outcome. Our approach takes the SC constrained least squares optimization approach as the starting point. We model the counterfactual of the treated unit in period $T$ as the weighted sum of the untreated units' features at $T$ (with weights estimated with pre-treatment data), and an error term. This decomposition highlights two sources of uncertainty, one from the in-sample estimation of the SC weights in the pre-treatment period, and the other from the post-treatment error that arises due to the unavoidable out-of-sample prediction involved in the SC method, which may include potential misspecification errors from the SC weights. Using finite-sample concentration bounds, we derive prediction intervals that incorporate both sources of uncertainty. Because the uncertainty stemming from the out-of-sample post-treatment error term is hard to handle (specially under general misspecification), we recommend combining the prediction interval for the SC outcome with a principled sensitivity analysis for the post-treatment error. Our empirical illustrations show that our methods perform well using both simulated and real data. A general-purpose software package is underway \citep*{Cattaneo-Feng-Palomba-Titiunik_2021_scpi}.

\begin{appendix}
\numberwithin{thm}{section}
\renewcommand{\thethm}{\Alph{thm}}

\section{Extension to Weakly Dependent Data}\label{sec:extension weakly dependent data}

We generalize Theorem \ref{thm: coverage error approximation} to allow for $\beta$-mixing data. For $\bu_t=(u_{t,1}, \cdots, u_{t,M})'$, define the (conditional on $\mathscr{H}$) mixing coefficient $\mathfrak{b}(\cdot;\mathscr{H})$ by
\[\begin{split}
\mathfrak{b}(k;\mathscr{H})=\max_{1\leq l\leq n-k}\frac{1}{2}\sup\Big\{
&\sum_i\sum_j\Big|\P(\mathcal{E}_i\cap\mathcal{E}_j'|\mathscr{H})-
\P(\mathcal{E}_i|\mathscr{H})\P(\mathcal{E}_j'|\mathscr{H})\Big|:\\
&\{\mathcal{E}_i\} \text{ is a finite partition of }\sigma(\bu_1, \cdots, \bu_l),\\
&\{\mathcal{E}_j'\} \text{ is a finite partition of }\sigma(\bu_{l+k}, \cdots, \bu_{T_0})\Big\}.
\end{split}
\]
See \citet{Pham-Tran_1985_SPA} and \citet{Doukhan_2012_Book} for properties and examples of mixing conditions.

Theorem \ref{thm: coverage error approximation, mixing} below combines a coupling result for dependent data \citep{Berbee_1987_PTRF}, a Berry-Esseen bound for convex sets \citep{Raivc_2019_Bernoulli}, and results on anti-concentration of the Gaussian measure for convex sets \citep{Chernozhukov-Chetverikov-Kato_2015_PTRF,Chernozhukov-Chetverikov-Kato_2017_AoP}, together with the standard ``small-block and large-block'' technique. Decompose the sequence $\{1, \cdots, T_0\}$ into ``large'' and ``small'' blocks: $\mathcal{J}_1=\{1, \cdots, q\}$, $\mathcal{J}_1'=\{q+1, \cdots, q+v\}$, $\cdots$, $\mathcal{J}_m=\{(q+v)(m-1)+1, \cdots, (q+v)(m-1)+q\}$, $\mathcal{J}_m'=\{(q+v)(m-1)+q+1, \cdots, (q+v)m\}$, $\mathcal{J}_{m+1}'=\{(q+v)m+1, \cdots, T_0\}$ where $m=\lfloor T_0/(q+v)\rfloor$, $q>v$ and $q+v\leq T_0/2$. The parameters $q$ and $v$, which depend on $T_0$, control the sizes of the large and small blocks, respectively, and will satisfy certain conditions in the theorem below. To simplify notation, we let $\bs_t=(s_{1t},\cdots, s_{dt})'$ be the summand in $\widehat{\bgamma}-\bgamma$ corresponding to time $t$, and define $\bS_{k,\Box}=\sum_{t\in\mathcal{J}_k}\bs_t$ and $\bS_{k,\diamond}=\sum_{t\in\mathcal{J}_k'}\bs_t$. Accordingly, let $S_{jk,\Box}$ and $S_{jk,\diamond}$ be the $j$th elements of $\bS_{k,\Box}$ and $\bS_{k,\diamond}$, respectively. Let $\bSigma_{\Box}=\sum_{k=1}^{m}\V[\bS_{k,\Box}|\mathscr{H}]$ and introduce
\[\bar{\sigma}^2(q):=\max_{1\leq j\leq d}\frac{1}{m}\sum_{k=1}^{m} \V\Big[q^{-1/2}\sum_{t\in\mathcal{J}_k}s_{jt}\Big|\mathscr{H}\Big],\quad
  \bar{\sigma}^2(v):=\max_{1\leq j\leq d}\frac{1}{m}\sum_{k=1}^m \V\Big[v^{-1/2}\sum_{t\in\mathcal{J}_k'}s_{jt}\Big|\mathscr{H}\Big].
\]

\begin{thm}[Distributional Approximation, Dependent Case]
	\label{thm: coverage error approximation, mixing}
	Assume $\mathcal{W}$ and $\mathcal{R}$ are convex, and $\widehat{\bbeta}$ in \eqref{eq: estimated weight} and $\bbeta_0$ in \eqref{eq: pseudo true value} exist. Let $\psi\geq 3$.
	In addition, for non-negative finite constants $\eta_1$, $\bar{\sigma}$, $\pi_{\gamma,1}$, $\eta_2$, $\pi_{\gamma,2}$, $\eta_3$, $\pi_{\gamma,3}$, $\eta_4$, $\pi_{\gamma,4}$, $\eta_5$, $\pi_{\gamma,5}$ and $\eta_6$, the following conditions hold:
	\begin{enumerate}[label=\normalfont(T\thethm.\roman*),noitemsep]
		\item $\bu_t$ is $\beta$-mixing conditional on $\mathscr{H}$ with mixing coefficient $\mathfrak{b}(\cdot;\mathscr{H})$;
		\item $\P\big(\E[\sum_{j=1}^{d}	\sum_{k=1}^{m}|S_{jk,\diamond}|^\psi|\mathscr{H}]\leq \eta_1,\; \bar{\sigma}^2(v)\leq \bar{\sigma}^2 \big)\geq 1-\pi_{\gamma,1}$;
		\item $\P\big(\max_{1\leq j\leq d}\E[|S_{j(m+1),\diamond}|^{\psi}|\mathscr{H}]\leq \eta_2\big)\geq 1-\pi_{\gamma,2}$;
		\item $\P\big(\sum_{k=1}^{m}\E[\|\bSigma_{\Box}^{-1/2}\bS_{k,\Box}\|^3|\mathscr{H}]\leq	\eta_3(42(d^{1/4}+16))^{-1}\big)\geq 1-\pi_{\gamma,3}$;
		\item $\P\big(\|\bSigma_{\Box}^{-1}\|_{\mathtt{F}}\leq d\eta_4\big)\geq 1-\pi_{\gamma,4}$;
		\item $\P\big(\|\bSigma_{\Box}^{-1/2}\bSigma\bSigma_{\Box}^{-1/2}-\bI_d\|_{\mathtt{F}} \leq 2\eta_5\big)\geq 1-\pi_{\gamma,5}$;
		\item $\max\big\{\eta_3, \eta_5, d\eta_4 [\eta_6^{-1}(\sqrt{mv\bar{\sigma}^2\log d}+\eta_1^{1/\psi}\log d) +(d\eta_2)^{1/\psi}\eta_6^{-1/\psi}]\big\}\leq \eta_6$ and $m\mathfrak{b}(v;\mathscr{H})\leq\eta_6$ a.s. on $\mathscr{H}$.
	\end{enumerate}

    Then, for $\eta_5\in[0,1/4]$,
	\[\P\Big[\P\big(\bp_T'\bD^{-1}\widehat{\bdelta}\leq \mathfrak{c}^\dagger(1-\alpha)\big|\mathscr{H}\big)\geq 1-\alpha-\epsilon_\gamma\Big]\geq 1-\pi_\gamma,\]
	where $\epsilon_\gamma=\mathfrak{C}\eta_6$ for finite positive constant $\mathfrak{C}$, which is characterized in the proof, $\pi_{\gamma}=\sum_{l=1}^{5}\pi_{\gamma,l}$, and  $\mathfrak{c}^\dagger(1-\alpha)$ is the $(1-\alpha)$-quantile of $\varsigma_\mathtt{U}^\dagger=\sup\big\{\bp_T'\bD^{-1}\bdelta:\bdelta\in\mathcal{M}_{\bG}\big\}$ conditional on $\mathscr{H}$.
\end{thm}

Conditions (TA.i) and (TA.iv) are comparable to (T1.i) and (T2.i) in Theorem \ref{thm: coverage error approximation}, respectively. The (conditional) independence assumption is relaxed to (conditional) $\beta$-mixing, and a bound on the conditional third moment of large blocks is imposed. The other conditions in Theorem \ref{thm: coverage error approximation, mixing} are new, and they ensure the small blocks and the last block can be neglected in a proper probability concentration sense.

\end{appendix}

\onehalfspacing
\bibliography{CFT_2021_JASA--bib}
\bibliographystyle{jasa}

\clearpage

\begin{table}[ht]
	\small
	\begin{center}
	\caption{Simulation Evidence, Linear Regression Methods}
	\label{table:simul, p1}
		\setlength{\tabcolsep}{8pt}
		\renewcommand{\arraystretch}{1.07}
		\resizebox{\textwidth}{.9\height}{\input{input/simulation/Table_Simul.txt}}
	\end{center}
		
	\footnotesize\textit{Notes}.
	Conditional mean, variance and quantiles of $u_t$ are estimated based on linear regression methods.
	CP =  coverage probability, AL = average length. 
	``M1": prediction interval for $Y_{1T}(0)$ based on the Gaussian concentration inequality with $90\%$ nominal coverage probability; ``M1-S": the same as ``M1'', but the estimated standard deviation is doubled in the construction; 
	``M2": prediction interval for $Y_{1T}(0)$ based on the location-scale model with $90\%$ nominal coverage probability; ``M3": prediction interval for $Y_{1T}(0)$ based on quantile regression with $90\%$ nominal coverage probability; 
	``CONF'' prediction interval for $Y_{1T}(0)$ based on the conformal method developed in \cite{Chernozhukov-Wuthrich-Zhu_2021_JASA} with $90\%$ nominal coverage probability.
	
\end{table}

\clearpage
\begin{figure}
	\begin{center}\caption{California Tobacco Control: Cigarette Sales Per Capita.\label{fig:California, level}}
		\vspace{-.1in}
		\begin{subfigure}{0.48\textwidth}
			\includegraphics[width=\textwidth]{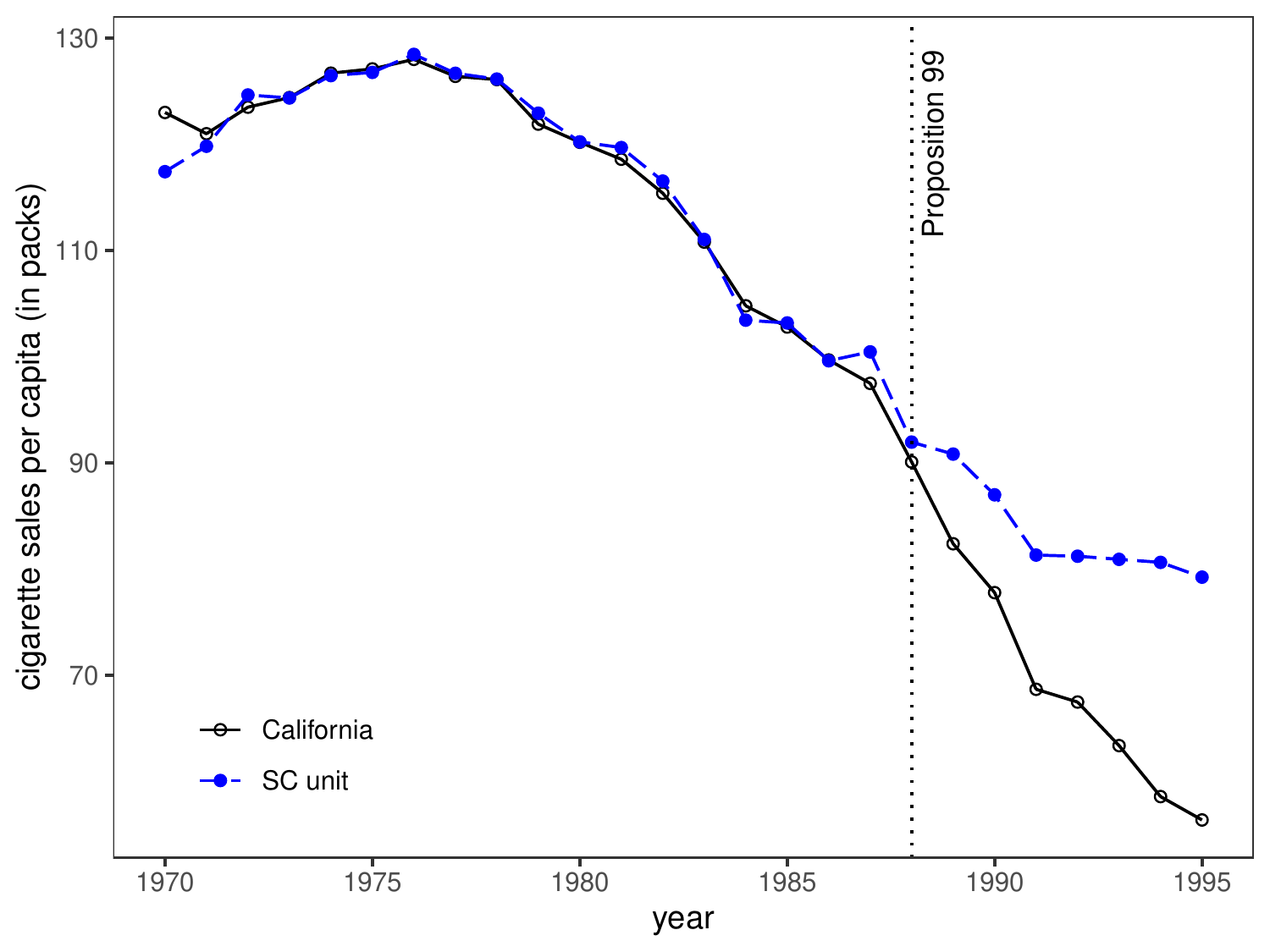}
			\caption{Synthetic California}
		\end{subfigure}
		\begin{subfigure}{0.48\textwidth}
			\includegraphics[width=\textwidth]{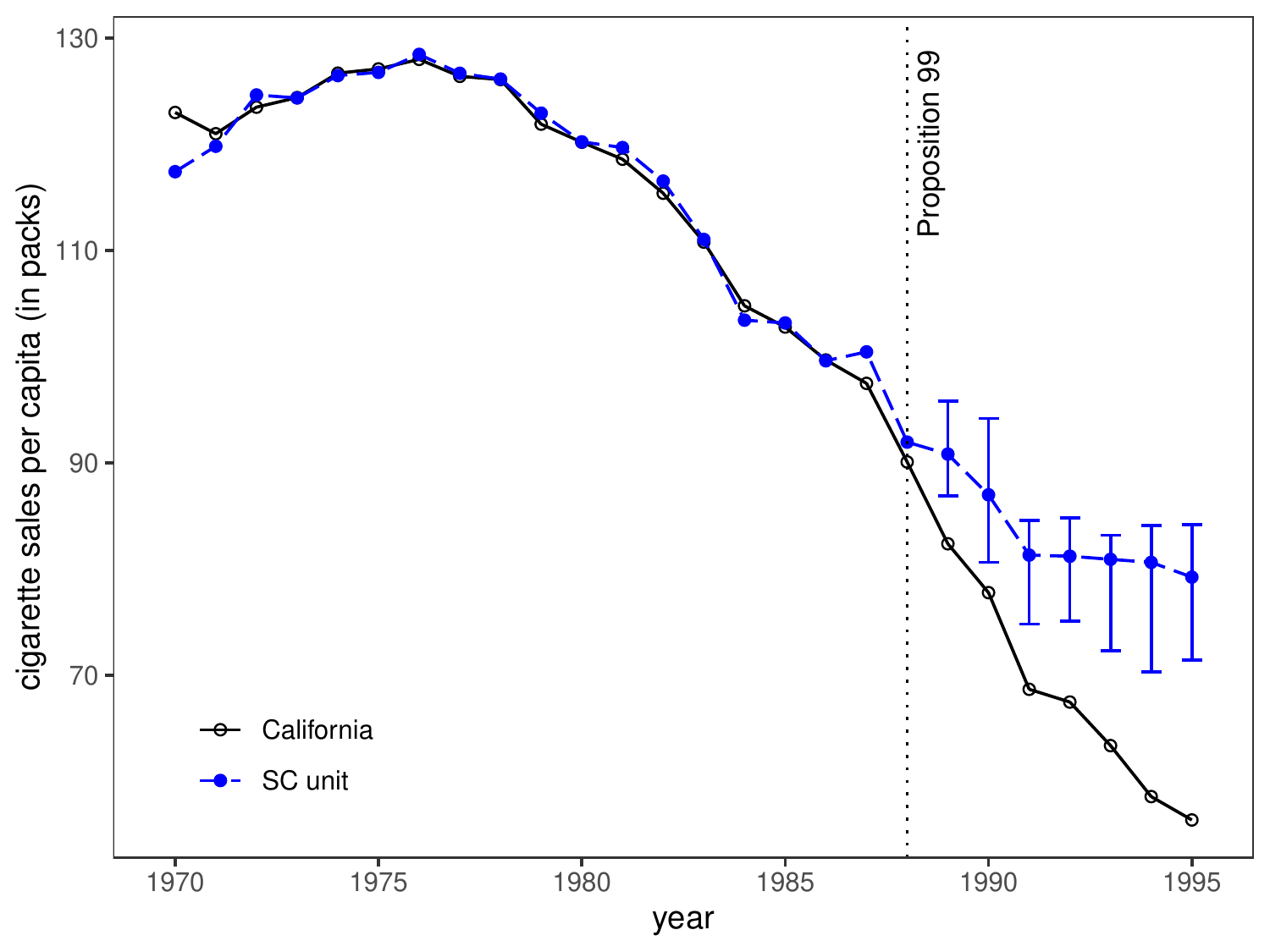}
			\caption{Prediction Interval for $\bx_T'\bw_0$}
		\end{subfigure}\\
		\begin{subfigure}{0.48\textwidth}
			\includegraphics[width=\textwidth]{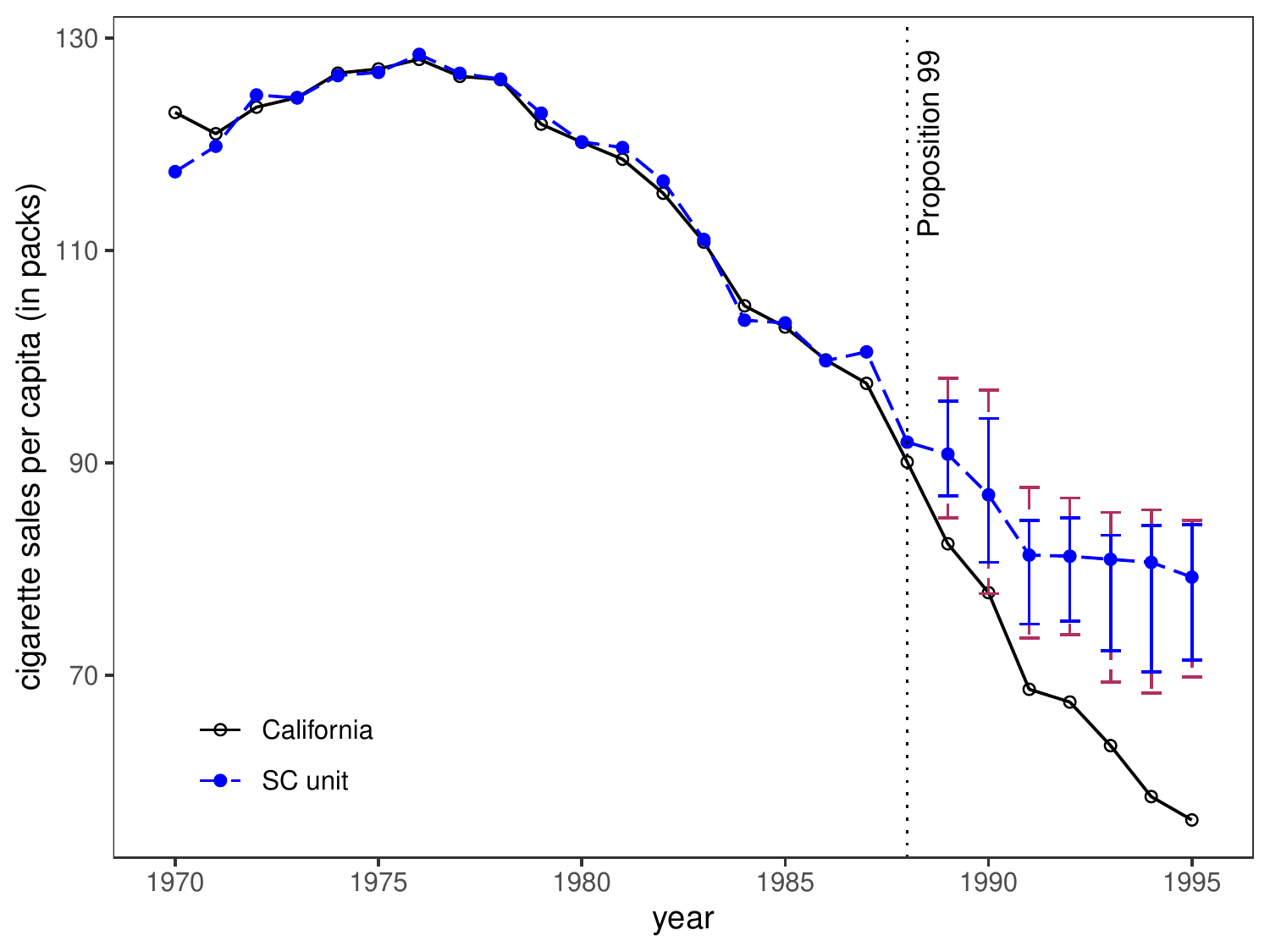}
			\caption{Prediction Interval for $Y_{1T}(0)$, approach 1}
		\end{subfigure}
		\begin{subfigure}{0.48\textwidth}
			\includegraphics[width=\textwidth]{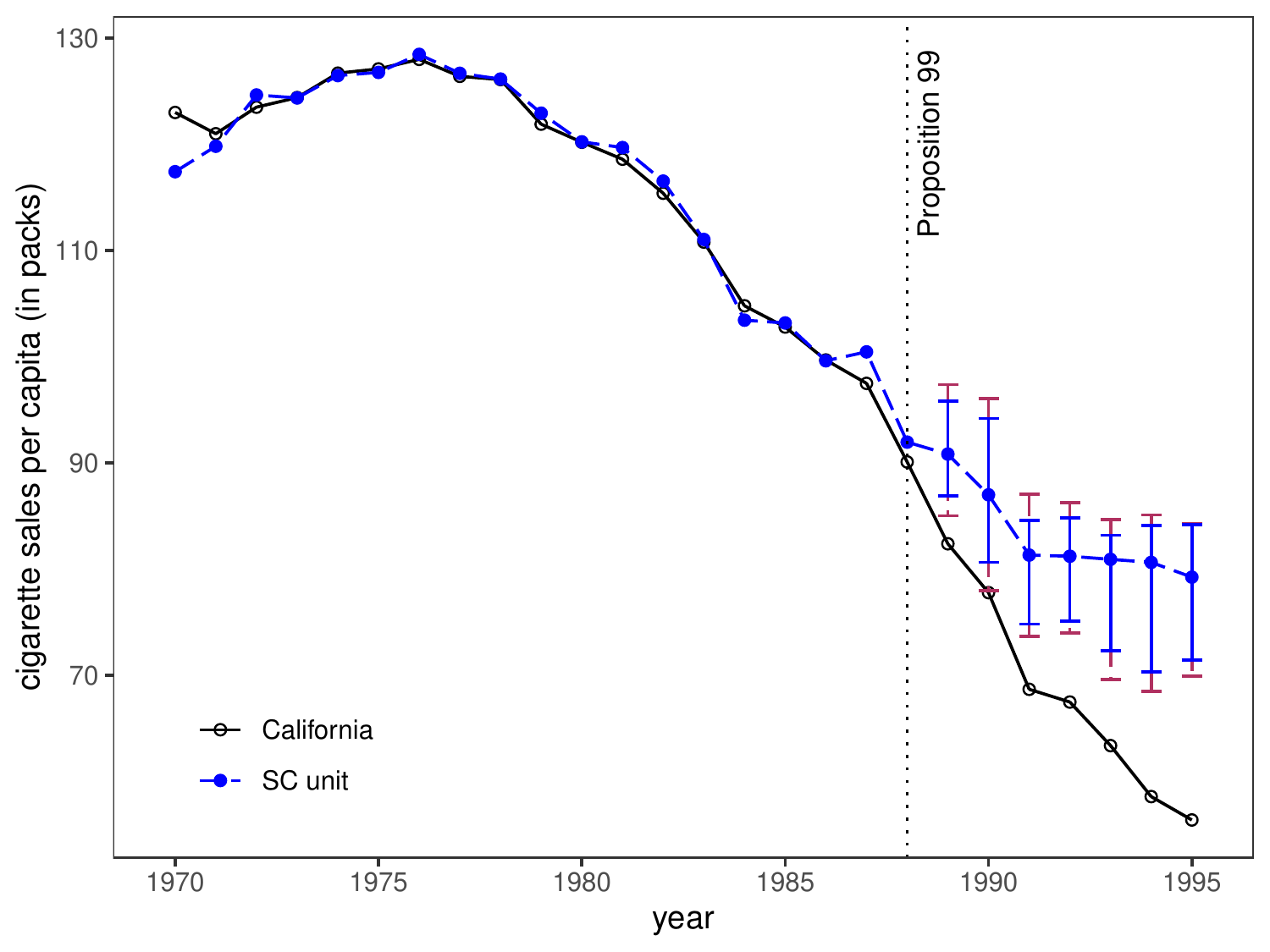}
			\caption{Prediction Interval for $Y_{1T}(0)$, approach 2}
		\end{subfigure}\\
		\begin{subfigure}{0.48\textwidth}
			\includegraphics[width=\textwidth]{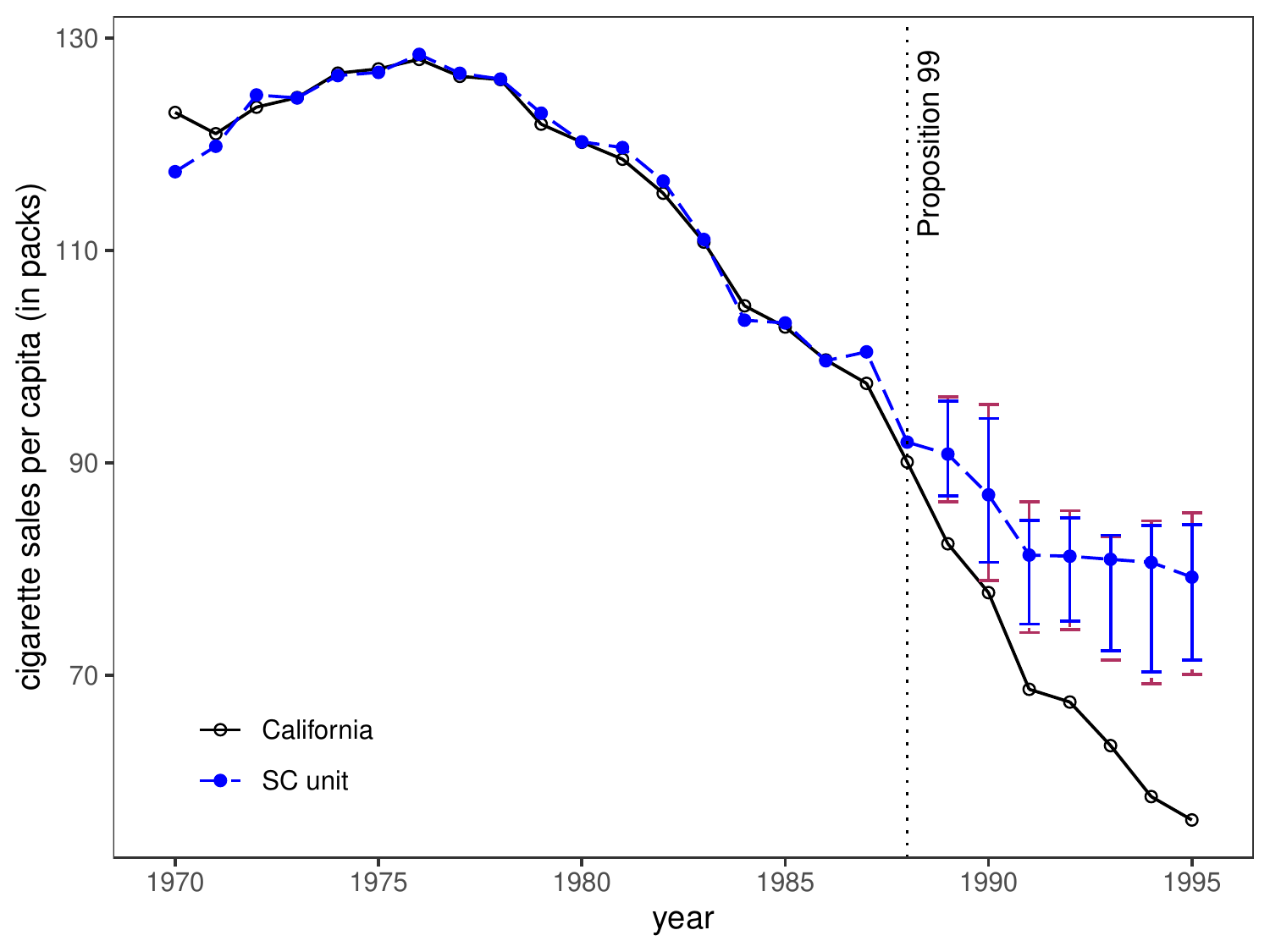}
			\caption{Prediction Interval for $Y_{1T}(0)$, approach 3}
		\end{subfigure}
		\begin{subfigure}{0.48\textwidth}
			\includegraphics[width=\textwidth]{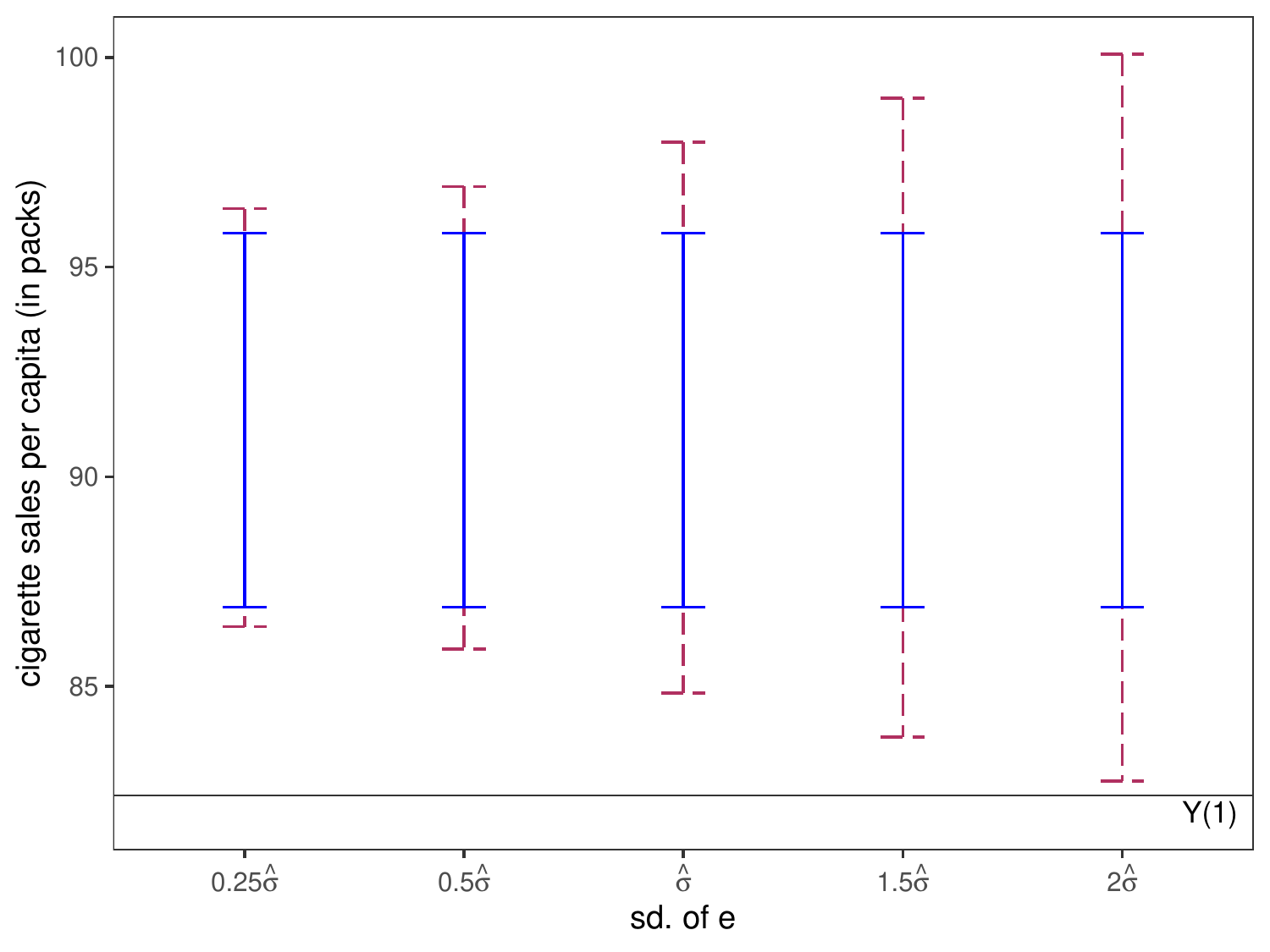}
			\caption{Sensitivity Analysis: PIs in 1989}
		\end{subfigure}
	\end{center}
	\scriptsize\textit{Notes}. Panel (a): Per capita cigarette sales in California and synthetic California. Panel (b): Prediction interval for synthetic California with at least 95\% coverage probability. Panels (c)-(e): Prediction interval for the counterfactual of California for with at least 90\% coverage probability based on three methods described in Section \ref{sec:bound e.T}, respectively. Panel (f): Prediction intervals for the counterfactual California based on approach 1, corresponding to $c\times\sigma_{\mathscr{H}}$, where $c=0.25,0.5,1,1.5,2$. The horizontal solid line represents the observed outcome for the treated.
\end{figure}

\clearpage
\begin{figure}
	\begin{center}\caption{California Tobacco Control: Growth Rate of Cigarette Sales Per Capita.\label{fig:California, growth rate}}
		\vspace{-.1in}
		\begin{subfigure}{0.48\textwidth}
			\includegraphics[width=\textwidth]{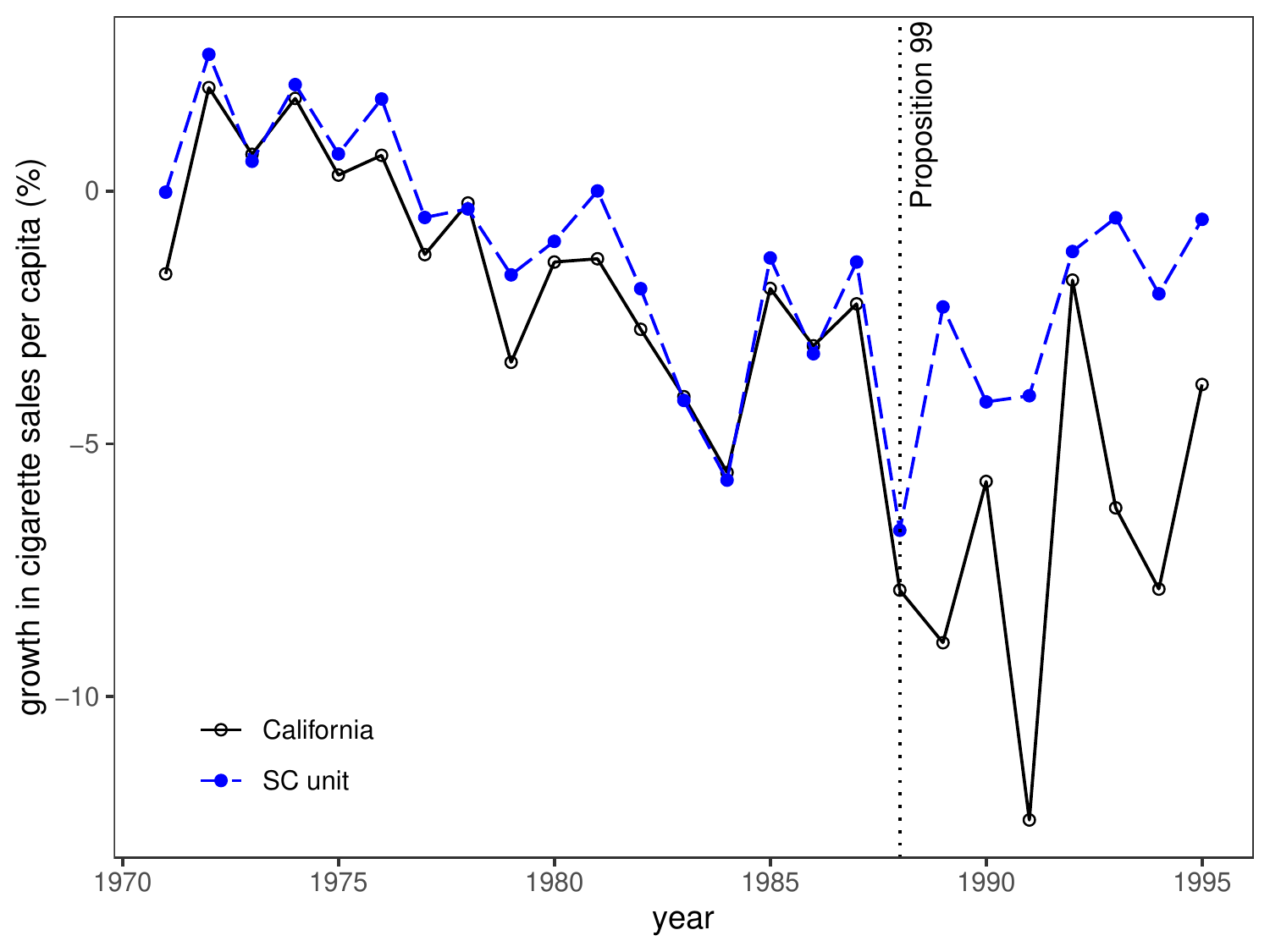}
			\caption{Synthetic California}
		\end{subfigure}
		\begin{subfigure}{0.48\textwidth}
			\includegraphics[width=\textwidth]{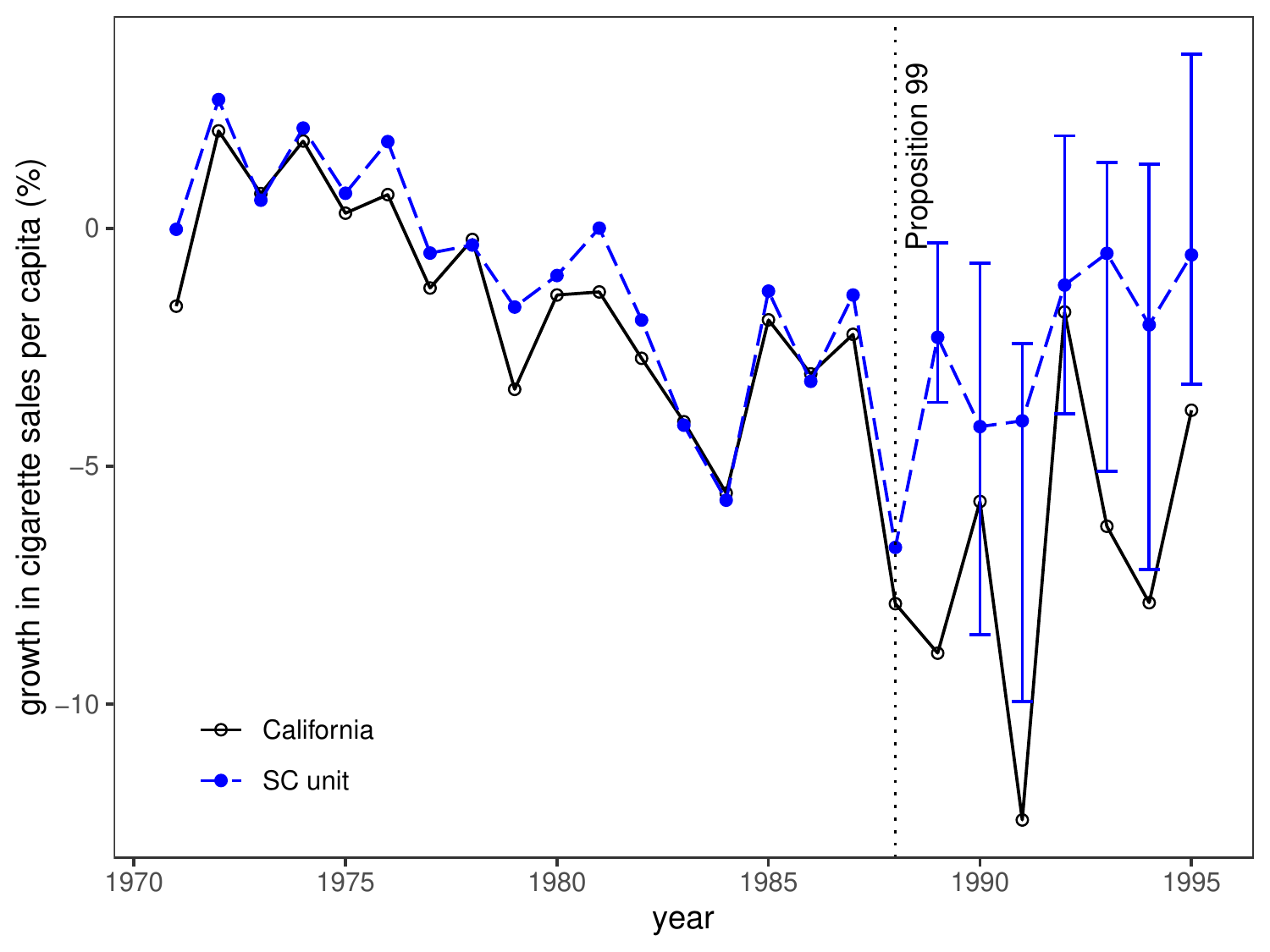}
			\caption{Prediction Interval for $\bx_T'\bw_0$}
		\end{subfigure}\\
		\begin{subfigure}{0.48\textwidth}
			\includegraphics[width=\textwidth]{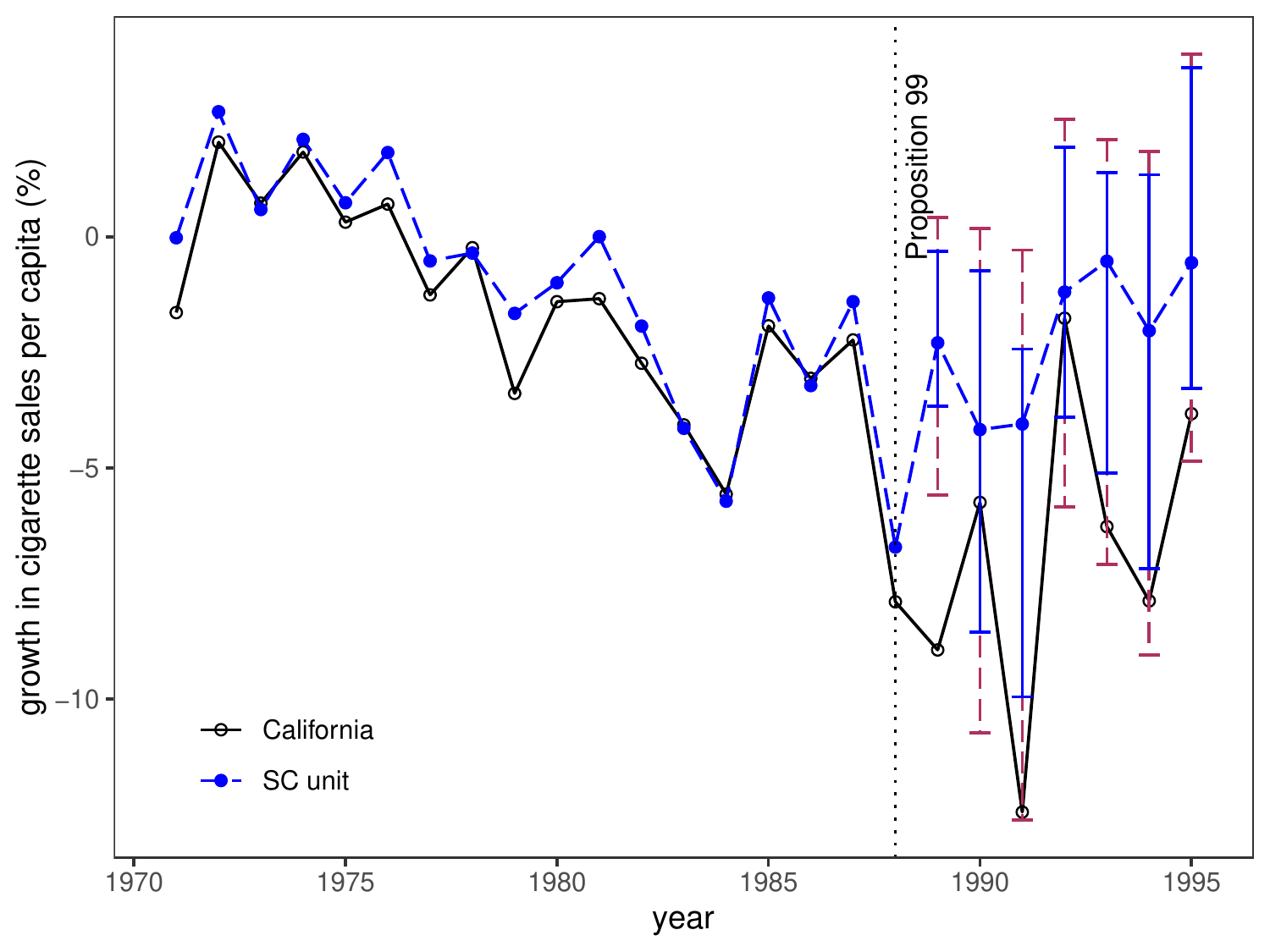}
			\caption{Prediction Interval for $Y_{1T}(0)$, approach 1}
		\end{subfigure}
		\begin{subfigure}{0.48\textwidth}
			\includegraphics[width=\textwidth]{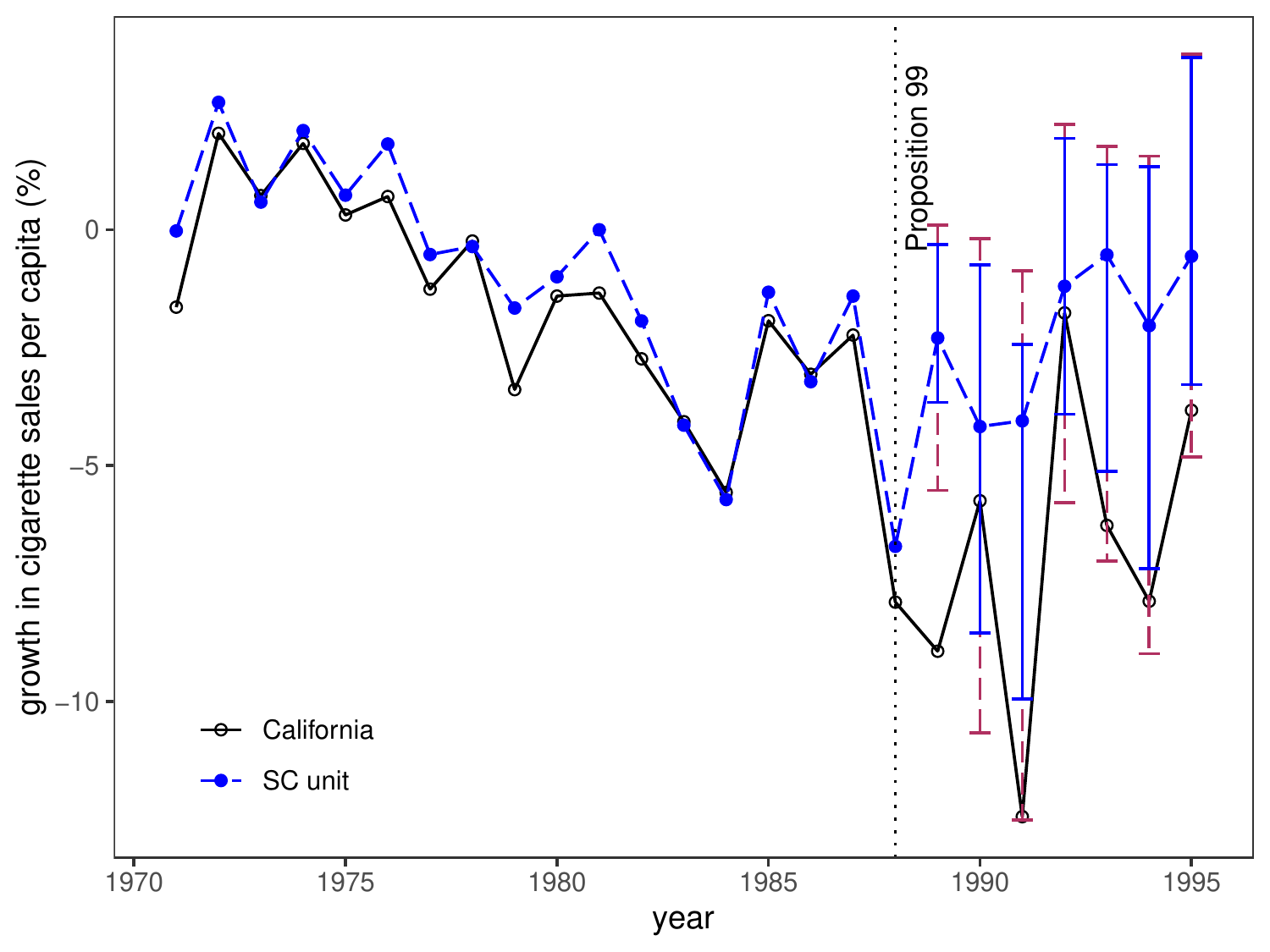}
			\caption{Prediction Interval for $Y_{1T}(0)$, approach 2}
		\end{subfigure}\\
		\begin{subfigure}{0.48\textwidth}
			\includegraphics[width=\textwidth]{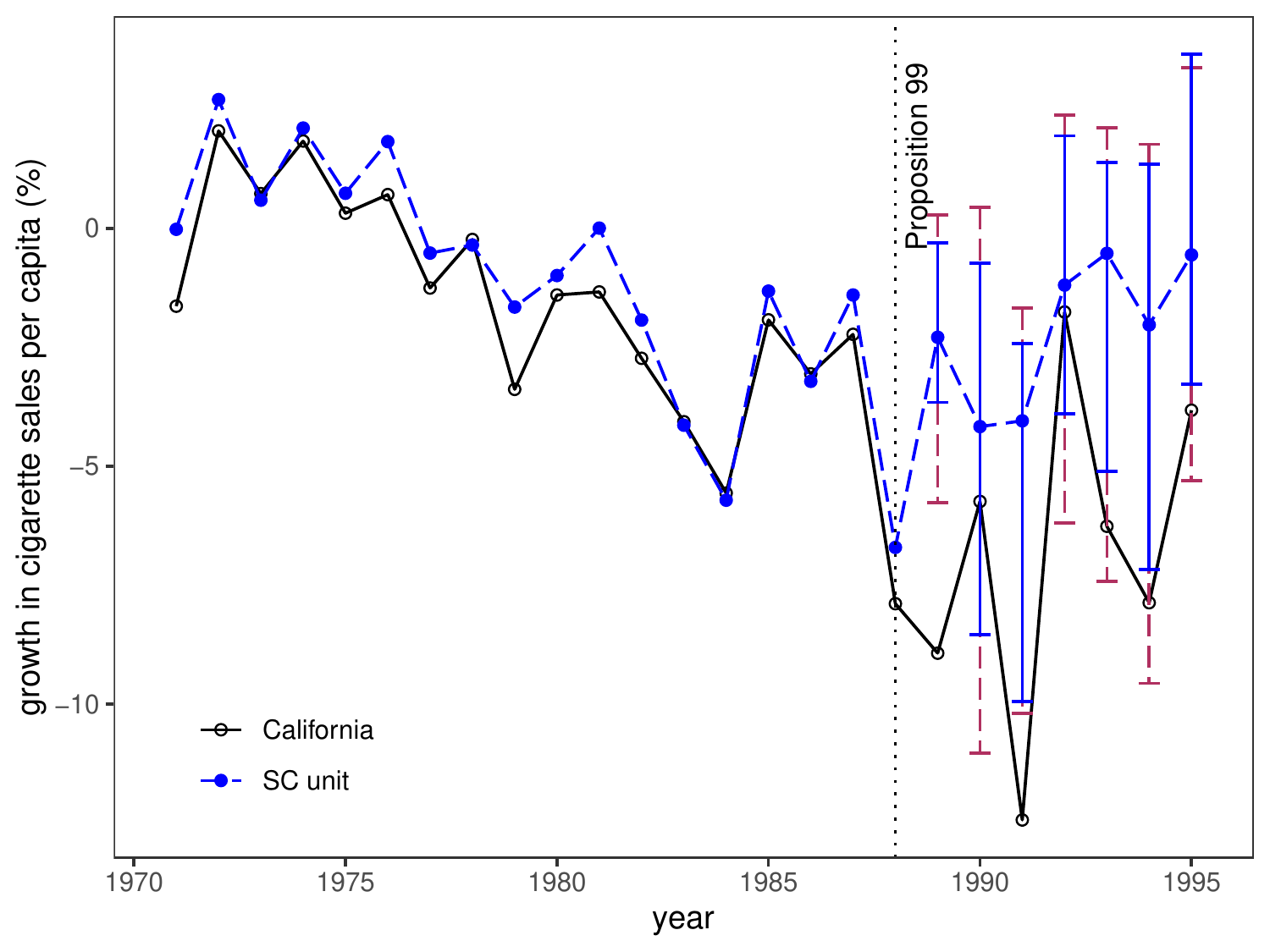}
			\caption{Prediction Interval for $Y_{1T}(0)$, approach 3}
		\end{subfigure}
		\begin{subfigure}{0.48\textwidth}
			\includegraphics[width=\textwidth]{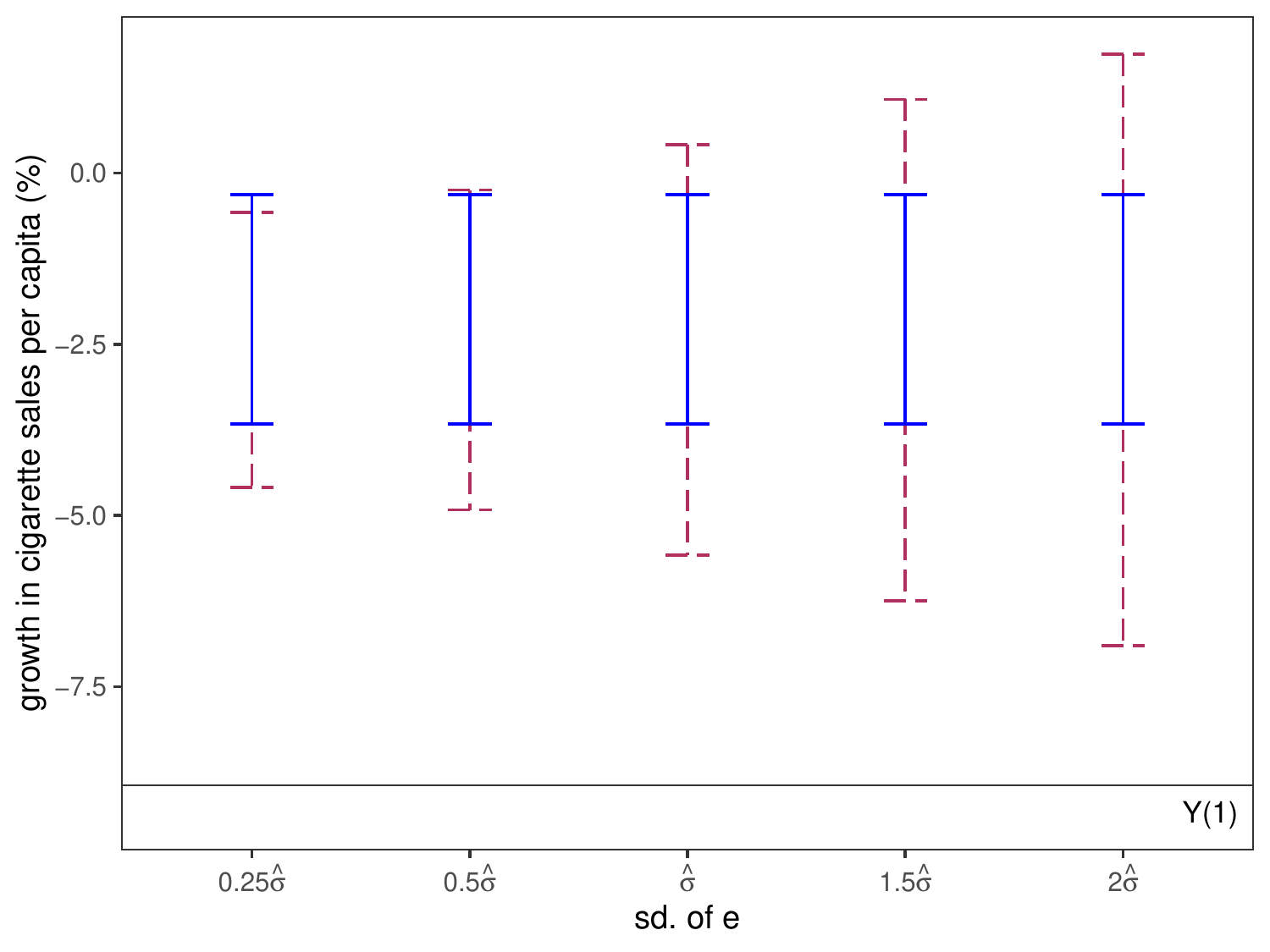}
			\caption{Sensitivity Analysis: PIs in 1989}
		\end{subfigure}
	\end{center}
	\scriptsize\textit{Notes}. Panel (a): Growth rate of per capita cigarette sales in California and synthetic California. Panel (b): Prediction interval for synthetic California with at least 95\% coverage probability. Panels (c)-(e): Prediction interval for the counterfactual of California for with at least 90\% coverage probability based on three methods described in Section \ref{sec:bound e.T}, respectively. Panel (f): Prediction intervals for the counterfactual California based on approach 1, corresponding to $c\times\sigma_{\mathscr{H}}$, where $c=0.25,0.5,1,1.5,2$. The horizontal solid line represents the observed outcome for the treated.
\end{figure}

\end{document}


\title{\vspace{-0.0in}Prediction Intervals for Synthetic Control Methods\thanks{Cattaneo and Titiunik gratefully acknowledges financial support from the National Science Foundation (SES-2019432), and Cattaneo gratefully acknowledges financial support from the National Institutes of Health (R01 GM072611-16).}\\
	Supplemental Appendix
\bigskip }
\author{Matias D. Cattaneo\thanks{Department of Operations Research and Financial Engineering, Princeton University.} \and
	    Yingjie Feng\thanks{School of  Economics and Management, Tsinghua University.}  \and
	    Rocio Titiunik\thanks{Department of Politics, Princeton University.}}
\maketitle

\begin{abstract}
	This supplement contains theoretical proofs of the results discussed in the main paper and additional numerical evidence. Section \ref{sec:examples} provides further details of the three examples discussed in the paper. Section \ref{SA sec: proofs} presents all the proofs. Section \ref{SA sec: simulation} gives additional simulation evidence for the finite-sample performance of the proposed prediction intervals. Section \ref{SA sec: empirical example} provides another empirical application studying the economic impact of 1990 German reunification on West Germany. Finally, Section \ref{SA sec: pseudo true values} provides interpretation in specific settings for the pseudo-true values defined in the main paper.
\end{abstract}

\newpage
\addtocontents{toc}{\protect\thispagestyle{empty}}
\tableofcontents

\thispagestyle{empty}
\clearpage

\doublespacing
\setcounter{page}{1}
\pagestyle{plain}


\section{Details on Examples} \label{sec:examples}

This section provides more details of the three examples discussed in the main paper. Recall that we consider the standard synthetic control constraint: $\mathcal{W}=\{\bw\in\mathbb{R}^N_+: \|\bw\|_1=1\}$ and $\mathcal{R}=\mathbb{R}^{KM}$. For simulation-based inference, we define explicitly a relaxed constraint set based on the original estimated coefficients $\widehat{\bbeta}$: $\Delta^{\star}= \{\bD(\bbeta-\widehat{\bbeta}^{\ttb}):\bbeta=(\bw', \br')', \bw\in\mathbb{R}^N_+,\; \|\bw\|_1=\|\widehat{\bw}^{\ttb}\|_1\}$, where $\widehat{\bbeta}^{\ttb}=(\widehat{\bw}^{\ttb'}, \widehat{\br}')'$, $\widehat{\bw}^{\ttb}=(\widehat{\omega}_2^{\ttb}, \cdots, \widehat{\omega}_{N+1}^{\ttb})'$, $\widehat{\omega}_j^{\ttb}=\widehat{\omega}_j\I(|\widehat{\omega}_j|>\varrho)$, and $\varrho$ is a tuning parameter that ensures the constraint set in the simulation world preserves the local geometry of $\Delta$. Moreover, we set $\bx_T=(Y_{2T}(0), \cdots, Y_{(N+1)T}(0))'$ as it is common in the SC literature. Finally, we let $\mathfrak{C}$, $\mathfrak{C}^{\ttb}$ and $\mathfrak{c}$, with various sub-indexes, denote non-negative finite constants not depending on $T_0$. In simple cases, we give the exact expression of these constants, while in other cases they can be characterized from the proofs of the results.

\subsection{Outcomes-only}

In this example the SC weights are constructed based on past outcomes only, and the model allows for an intercept. Thus, the working model simplifies to 
\[a_{t}=\bb_{t}'\bw_{0} + r_0 + u_t, \qquad t=1, \cdots, T_0,\]
where $a_{t}:=Y_{1t}(0)$, $\bb_{t} :=(Y_{2t}(0), Y_{3t}(0), \ldots, Y_{(N+1)t}(0))'$, and with $M=1$, $K=1$, and $d=N+1$. Let $\bz_t=(\bb_t', 1)'$, $\bbeta_0=(\bw_0', r_0)'$. We further assume independent sampling across time, and thus set $\bD=T_0^{1/2}\bI_d$. A natural variance estimator in this setting is
\[\widehat{\bSigma}=\frac{1}{T_0}\sum_{t=1}^{T_0}\bz_t\bz_t' (\widehat{u}_{t}-\widehat{\E}[u_{t}|\bb_t])^2,\]
where $\widehat{u}_t=a_t-\bz_t'\widehat{\bbeta}$, and $\widehat{\E}[u_{t}|\bb_t]$ denotes some estimate of the conditional mean of the pseudo-residuals.

The following theorem gives precise primitive conditions and verifies the high-level conditions of Theorems 1 and 2 in the main paper. Recall that $\bw_0=(w_{0,1},w_{0,2},\dots,w_{0,J})'$ is defined in Section 2 of the paper, and $\lambda_{\min}(\bM)$ and $\lambda_{\max}(\bM)$ are the minimum and the maximum eigenvalues of a generic square matrix $\bM$.

\begin{thm}[Outcomes-only]\label{thm: exmp independent case}
	Let $\{\bz_t, u_t\}_{t=1}^T$ be i.i.d over $t=1, \cdots, T_0$.
	
	Assume that, for finite non-negative constants $\bar{\eta}_1$, $\bar{\eta}_2$, $\underline{\eta}_1$ and $\underline{\eta}_2$, the following conditions hold:
	\begin{enumerate}[label=\normalfont(\thethm.\roman*),noitemsep]
		\item $\max_{1\leq t\leq T_0}\E[|u_t|^3|\bB]\leq\bar{\eta}_1$ a.s. on $\sigma(\bB)$ and $\E[\|\bz_t\|^6]\leq \bar{\eta}_2$;
		\item $\min_{1\leq t\leq T_0}\V[u_{t}|\bB]\geq\underline{\eta}_1$ a.s. on $\sigma(\bB)$ and $\lambda_{\min}(\E[\bz_t\bz_t'])\geq \underline{\eta}_2$.
	\end{enumerate} 
    Then, the conditions of Theorem 1 hold with
    $\pi_{\gamma}=\mathfrak{C}_{\pi}T_0^{-1}$ and $\epsilon_{\gamma}=\mathfrak{C}_{\epsilon}T_0^{-1/2}$, where $\mathfrak{C}_{\pi}=\frac{d}{\bar{\eta}_2}+\frac{4d^4\bar{\eta}_2}{\underline{\eta}_2^2}$ and 
    $\mathfrak{C}_{\epsilon}=42(d^{1/4}+16)\times
    \frac{2^{5/2}d^{3/2}\bar{\eta}_1\bar{\eta}_2}{(\underline{\eta}_1\underline{\eta}_2)^{3/2}}$. Therefore, condition {\normalfont(T2.i)} of Theorem 2 also holds, while condition {\normalfont(T2.ii)} holds with $\pi_{\delta}^{\ttb}=\mathfrak{C}_{\pi}^{\ttb}T_0^{-1}$, 
    $\epsilon_{\delta}^{\ttb}=\mathfrak{C}_{\epsilon}^{\ttb}T_0^{-1}$, and 
    $\varpi_{\delta}^{\ttb}=\mathfrak{C}_{\varpi}^{\ttb}\sqrt{\log T_0}$,
    where $\mathfrak{C}_{\pi}^{\ttb}=\mathfrak{C}_{\pi}$,
    $\mathfrak{C}_{\epsilon}^{\ttb}=2d$, and
    $\mathfrak{C}_{\varpi}^{\ttb}=8\sqrt{d\bar{\eta}_1\bar{\eta}_2}/\underline{\eta}_2$.
    
    Assume in addition that, for finite non-negative constant $\pi_{w}^{\ttb}$, the following condition holds:
	\begin{enumerate}[label=\normalfont(\thethm.\roman*),noitemsep,resume]
    	\item $\varrho=\varpi_\delta^{\ttb}/\sqrt{T_0}$ and $\P(\min\{|w_{0,j}|:w_{0,j}\neq 0\}\geq\varrho)\geq 1-\pi_{w}^{\ttb}$.
    \end{enumerate}
    Then, condition {\normalfont(T2.iii)} of Theorem 2 holds with $\pi_{\Delta}^{\ttb}=\pi_{\delta}^{\ttb}+\pi_{w}^{\ttb}+\pi_{\gamma}$ and $\epsilon_{\Delta}^{\ttb}=\epsilon_{\delta}^{\ttb}+\epsilon_{\gamma}$. 
    
    Finally, also assume that, for finite non-negative constants $\varpi_u^{\ttb}$, $\epsilon_u^{\ttb}$ and $\pi_u^{\ttb}$, the following conditions hold:
	\begin{enumerate}[label=\normalfont(\thethm.\roman*),noitemsep,resume]
    	\item $\max_{1\leq t\leq T_0}\E[|u_t|^4|\bB]\leq\bar{\eta}_1$ a.s. on $\sigma(\bB)$ and $\E[\|\bz_t\|^{12}]\leq\bar{\eta}_2$;
    	\item $\P[\P(\max_{1\leq t\leq T_0}|\widehat{\E}[u_t|\bb_t]-\E[u_t|\bb_t]|\leq \varpi_u^{\ttb}|\mathscr{H})\geq 1-\epsilon_u^{\ttb}]\geq 1-\pi_u^{\ttb}$.
    \end{enumerate}
    Then, condition {\normalfont(T2.iv)} of Theorem 2 holds with  $\pi_{\gamma}^{\ttb}=\mathfrak{C}_{\pi,1}^{\ttb}T_0^{-1}+\pi_{\delta}^{\ttb}+\pi_u^{\ttb}+\pi_{\gamma}$, 
    $\epsilon_{\gamma,1}^{\ttb}= \mathfrak{C}_{\epsilon,1}^{\ttb}(T_0^{\upsilon-1/2}+\varpi_\delta^{\ttb} T_0^{-1/2}+\varpi_u^{\ttb})$, 
    $\epsilon_{\gamma,2}^{\ttb}= \mathfrak{C}_{\epsilon,2}^{\ttb}T_0^{-2\upsilon}+\mathfrak{C}_{\epsilon,3}^{\ttb}T_0^{-1}+\epsilon_{\delta}^{\ttb}+\epsilon_u^{\ttb}+\epsilon_{\gamma}$ 
    for any $\upsilon\in(0,1/2)$, and non-negative constants $\mathfrak{C}_{\pi,1}^{\ttb}, \mathfrak{C}_{\epsilon,1}^{\ttb}, \mathfrak{C}_{\epsilon,2}^{\ttb}$ and $\mathfrak{C}_{\epsilon,3}^{\ttb}$, which are characterized in the proof.
\end{thm}


\subsection{Multi-equation with Weakly Dependent Data}\label{sec: cov. dep. case}

In our second example, we incorporate pre-intervention covariates in the construction of the SC weights and allow for stationary weakly dependent time series data. We let $M=2$ (two features) and $K=0$ (no additional controls) only for simplicity, which gives the working model
\begin{align*}
   a_{t,1} &=\sum_{j=1}^J b_{jt,1}w_{0,j} + u_{t,1}, \\
   a_{t,2} &=\sum_{j=1}^J b_{jt,2}w_{0,j} + u_{t,2},
\end{align*}
$t=1, \cdots, T_0$. The first equation could naturally correspond to pre-intervention outcomes as in the previous example, i.e., $a_{t,1}:=Y_{1t}(0)$ and $\bb_{t,1} :=(Y_{2t}(0), Y_{3t}(0), \ldots, Y_{(N+1)t}(0))'$, while the second equation could correspond to some other covariate also used to construct $\widehat{\bw}$ as described in Section 2 of the paper. Let $\bb_{t,l}=(b_{1t,l}, \cdots, b_{Jt,l})'$, for $l=1,2$. To provide interpretable primitive conditions, we also assume $\bu_t=(u_{t,1},u_{t,2})'$ and $\bb_t=(\bb_{t,1}',\bb_{t,2}')'$ follow independent first-order stationary autoregressive (AR) processes:
\begin{alignat*}{2}
    \bu_t &= \bH_u\bu_{t-1}+\bm{\zeta}_{t,u}, \qquad &&\bH_u=\diag(\rho_{1,u}, \rho_{2,u}),\\
    \bb_t &= \bH_b\bb_{t-1}+\bm{\zeta}_{t,b}, \qquad &&\bH_b=\diag(\rho_{1,b}, \rho_{2,b}, \cdots, \rho_{J,b}),
\end{alignat*}
where $\bm{\zeta}_{t,u}$ and $\bm{\zeta}_{t,b}$ are i.i.d. over $t$, independent of each other, and $\diag(\cdot)$ denotes a diagonal matrix with the function arguments as the corresponding diagonal elements. Let $\bD=T_0^{1/2}\bI_d$ and recall that $\bU=(u_{1,1}, \cdots, u_{T_0,1}, u_{1,2}, \cdots, u_{T_0, 2})'$ in this case. A natural, generic variance estimator is
\[\widehat{\bSigma}=\frac{1}{T_0}\bZ'\widehat{\V}[\bU|\mathscr{H}]\bZ,\]
where $\widehat{\V}[\bU|\mathscr{H}]$ is an estimate of $\V[\bU|\mathscr{H}]$. In this example, $\bSigma$ corresponds to the (conditional) long-run variance, and naturally $\widehat{\bSigma}$ can be chosen to be any standard estimator thereof.

Because of the time dependence in this example, the following theorem gives primitive conditions that verify the high-level conditions of Theorem A in the appendix (instead of Theorem 1) in the paper, as well as the high-level conditions of Theorem 2 for implementation.

\begin{thm}[Multi-equation with Weakly Dependent Data]\label{thm: exmp dependent case}
	Let $\{\bm{\zeta}_{t,u}\}_{t=1}^T$ and $\{\bm{\zeta}_{t,b}\}_{t=1}^T$ be i.i.d over $t=1, \cdots, T_0$ with mean zero, finite variance, and independent of each other.
	
	Assume that, for finite positive constants $\varphi$ and $\varphi'$, and finite non-negative constants $\bar{\eta}_0, \bar{\eta}_1$, $\bar{\eta}_2$, $\underline{\eta}_1$ and $\underline{\eta}_2$, the following conditions hold:
	\begin{enumerate}[label=\normalfont(\thethm.\roman*),noitemsep]
		\item $\|\bH_b\|<1$, $\|\bH_u\|<1$,  and $\bm{\zeta}_{t,u}$ and $\bm{\zeta}_{t,b}$ have densities $f_u(\cdot)$ and $f_b(\cdot)$ satisfying $\int\|\bx\|^\varphi f_u(\bx)d\bx<\infty$, $\int\|\bx\|^\varphi f_b(\bx)d\bx<\infty$, $\int|f_u(\bx)-f_u(\bx-\bm\theta)|d\bx\leq\bar{\eta}_0\|\bm\theta\|^{\varphi'}$ for all $\bm\theta\in\mathbb{R}^2$,  and $\int|f_b(\bx)-f_b(\bx-\bm\theta)|d\bx\leq\bar{\eta}_0\|\bm\theta\|^{\varphi'}$ for all $\bm\theta\in\mathbb{R}^J$;
		\item  $\max_{1\leq t\leq T_0}\E[\|\bu_t\|^4|\bB]\leq\bar{\eta}_1$ a.s. on $\sigma(\bB)$ and $\E[\|\bb_{t}\|^6]\leq \bar{\eta}_2$;
		\item $\min_{1\leq t\leq T_0}\lambda_{\min}(\V[\bU|\bB])\geq\underline{\eta}_1$ a.s. on $\sigma(\bB)$ and $\min_{l=1,2}\lambda_{\min}(\E[\bb_{t,l}\bb_{t,l}'])\geq \underline{\eta}_2$.
	\end{enumerate} 
	Then, the conditions of Theorem A hold with $\pi_{\gamma}=\mathfrak{C}_\pi T_0^{-\mathfrak{c}_\pi}$ and 
	$\epsilon_{\gamma}=\mathfrak{C}_\epsilon T_0^{-\mathfrak{c}_\epsilon}$ for non-negative constants $\mathfrak{C}_\pi$ and $\mathfrak{C}_\epsilon$, and some positive constants $\mathfrak{c}_\pi$ and $\mathfrak{c}_\epsilon$, which are characterized in the proof. Therefore, condition {\normalfont(T2.i)} of Theorem 2 also holds, while condition {\normalfont(T2.ii)} holds with $\pi_{\delta}^{\ttb}=\mathfrak{C}_{\pi}^{\ttb}T_0^{-\mathfrak{c}_\pi^{\ttb}}$,
	$\epsilon_{\delta}^{\ttb}=\mathfrak{C}_\epsilon^{\ttb}T_0^{-1}$, $\varpi_{\delta}^{\ttb}=\mathfrak{C}_{\varpi}^{\ttb}\sqrt{\log T_0}$ for non-negative constants $\mathfrak{C}_\pi^{\ttb}$,  $\mathfrak{C}_\epsilon^{\ttb}$ and $\mathfrak{C}_{\varpi}^{\ttb}$, and positive constant $\mathfrak{c}_\pi^{\ttb}$, which are also characterized in the proof.
	
    Assume in addition that, for finite non-negative constant $\pi_{w}^{\ttb}$, the following condition holds:
	\begin{enumerate}[label=\normalfont(\thethm.\roman*),noitemsep,resume]
		\item $\varrho=\varpi_\delta^{\ttb}/\sqrt{T_0}$ and $\P(\min\{|w_{0,j}|: w_{0,j}\neq 0\}\geq \varrho)\geq 1-\pi_w^{\ttb}$.
	\end{enumerate}
    Then, condition {\normalfont(T2.iii)} of Theorem 2 holds with $\pi_{\Delta}^{\ttb}=\pi_{\delta}^{\ttb}+\pi_w^{\ttb}+\pi_{\gamma}$ and  $\epsilon_{\Delta}^{\ttb}=\epsilon_{\delta}^{\ttb}+\epsilon_{\gamma}$. 
 
    Finally, also assume that, for finite non-negative constants $\epsilon_{\Sigma,1}^{\ttb}$, $\epsilon_{\Sigma,2}^{\ttb}$ and $\pi_{\Sigma}^{\ttb}$, the following condition holds:
    \begin{enumerate}[label=\normalfont(\thethm.\roman*),noitemsep,resume]
    	\item $\P(\P(\|\widehat{\bSigma}-\bSigma\|\leq \epsilon_{\Sigma,1}^{\ttb}|\mathscr{H})\geq 1-\epsilon_{\Sigma,2}^{\ttb})\geq 1-\pi_{\Sigma}^{\ttb}$.
    \end{enumerate}
    Then, condition {\normalfont(T2.iv)} of Theorem 2 holds with  $\epsilon_{\gamma,1}^{\ttb}=\sqrt{d}\epsilon_{\Sigma,1}^{\ttb}/(2\underline{\eta}_1\underline{\eta}_2)$, $\epsilon_{\gamma,2}^{\ttb}=\epsilon_{\Sigma,2}^{\ttb}$ and $\pi_{\gamma}^{\ttb}=\pi_{\Sigma}^{\ttb}+\pi_{\delta}^{\ttb}$.
\end{thm}


\subsection{Cointegration}

Our third and final example illustrates how non-stationary data can also be handled within our framework. Suppose that for each $1\leq l\leq M$, $\{a_{t,l}\}_{t=1}^T$, $\{b_{1t,l}\}_{t=1}^T, \cdots, \{b_{Jt,l}\}_{t=1}^T$  are $I(1)$ processes, and $\{c_{1t,l}\}_{t=1}^T, \cdots, \{c_{Kt,l}\}_{t=1}^T$ and $\{u_{t,l}\}_{t=1}^{T}$ are $I(0)$ processes. Therefore, $\bA$ and $\bB$ form a cointegrated system. For simplicity, consider the following example: for each $l=1, \cdots, M$ and $j=1, \cdots, J$,
\begin{align*}
    a_{t,l}  &=\sum_{j=1}^{J}b_{jt,l}w_{0,j}+\sum_{k=1}^{K}c_{kt,l}r_{0,k,l}+u_{t,l},\\
    b_{jt,l} &=b_{j(t-1),l}+v_{jt,l}, 
\end{align*}
where $u_{t,l}$ and $v_{jt,l}$ are stationary unobserved disturbances. In this scenario, $(1, -\bw_0')'$ plays the role of a cointegrating vector such that the linear combination of $\bA$ and $\bB$ is stationary. The normalizing matrix $\bD=\mathrm{diag}\{T_0, \cdots, T_0, \sqrt{T_0}, \cdots, \sqrt{T_0}\}$, where the first $J$ elements are $T_0$ and the remaining ones are $\sqrt{T_0}$. Let $\check{\bZ}_t=(\check{\bz}_{t,1}, \cdots, \check{\bz}_{t,M})$ where $\check{\bz}_{t,l}=(\check{\bz}_{t,\triangleleft,l}', \check{\bz}_{t,\triangleright,l}')'$ be the $((l-1)T_0+t)$th column of $\diag\{T_0^{-1/2}\bI_J, \bI_{KM}\}\bZ'$. We use  $\check{\bz}_{t,\triangleleft,l}$ to denote the vector of the first $J$ elements of $\check{\bz}_{t}$, which corresponds to the nonstationary components. The remaining vector is denoted by $\check{\bz}_{t,\triangleright,l}$, which corresponds to the stationary components. Recall $\bu_t=(u_{t,1}, \cdots, u_{t,M})'$. 
Write $\bv_{t,l}=(v_{1t,l}, \cdots, v_{Jt,l})'$, $\bv_t=(\bv_{t,1}', \cdots, \bv_{t, M}')'$, $\bc_{t,l}=(c_{1t,l}, \cdots, c_{kt,l})'$, $\bC_t=(\bc_{t,1}, \cdots, \bc_{t,M})$. We allow some elements in $\bv_t$  to be used in $\bC_t$. Let $\bq_t$ collect all distinct variables in $\bu_t$, $\bv_t$, $\bc_{t,1}$, $\cdots$, $\bc_{t,M}$. Moreover, define
\[
\bQ_{\triangleleft}=\frac{1}{T_0}\sum_{l=1}^{M}\sum_{t=1}^{T_0}\bG_l(t/T_0)\bG_l(t/T_0)',
\]
where $\bG=(\bG_1', \cdots, \bG_M')'$ is a mean-zero Brownian motion on $[0,1]$ with variance $\E[\bv_t\bv_t']$.

As in the precious example, we consider the generic variance estimator
\[\widehat{\bSigma}=\frac{1}{T_0}\sum_{t=1}^{T_0}\check{\bZ}_t\widehat{\V}[\bu_t|\mathscr{H}]\check{\bZ}_t',\]
where $\widehat{\V}[\bu_t|\mathscr{H}]$ is an estimate of $\V[\bu_t|\mathscr{H}]$.

The following theorem gives more primitive conditions and verifies the high-level conditions of Theorems 1 and 2 in the paper for the cointegration scenario.

\begin{thm}[Cointegration]\label{thm: exmp cointegration}
	Assume that $\{\bq_t\}_{t=1}^T$ is i.i.d over $t=1, \cdots, T_0$ with mean zero and finite variance.
	
	Assume that, for $\psi\geq 3$ and finite non-negative constants $\bar{\eta}_1$, $\bar{\eta}_2$, $\underline{\eta}_1$, $\underline{\eta}_2$ and $\mathfrak{c}_Q$, $\pi_{Q,1}$, finite positive constants $\mathfrak{C}_{Q}$ and $\pi_{Q,2}$, and constant $\nu_Q\in(0,1/2)$, the following conditions hold:
	\begin{enumerate}[label=\normalfont(\thethm.\roman*),noitemsep]
		\item $\max_{1\leq t\leq T_0}\E[\|\bu_t\|^\psi|\bB,\bC]\leq\bar{\eta}_1$ a.s. on $\sigma(\bB,\bC)$ and $\E[\|\bq_t\|^\psi]\leq\bar{\eta}_2$;
		\item $\min_{1\leq t\leq T_0}\lambda_{\min}(\V[\bu_t|\bB, \bC])\geq \underline{\eta}_1$ a.s. on $\sigma(\bB, \bC)$ and $\lambda_{\min}(\E[\bq_t\bq_t'])\geq \underline{\eta}_2$;
		\item $\P\big[(\log T_0)^{-\mathfrak{c}_Q}\leq\lambda_{\min}(\bQ_{\triangleleft})\leq\lambda_{\max}(\bQ_{\triangleleft})\leq(\log T_0)^{\mathfrak{c}_Q}\big]\geq 1-\pi_{Q,1}$;
		\item $\P\big[\|\frac{1}{T_0}\sum_{l=1}^{M}\sum_{t=1}^{T_0} \check{\bz}_{t,\triangleleft,l}\check{\bz}_{t,\triangleright,l}'\| \leq \mathfrak{C}_{Q}T_0^{-1/2+\nu_Q}\big]\geq 1-\pi_{Q,2}$.
	\end{enumerate}
	Then, conditions {\normalfont(T1.i)} and {\normalfont(T1.ii)} of Theorem 1 hold with $\pi_{\gamma}=\mathfrak{C}_{\pi,1}T_0^{-\psi\nu}+\mathfrak{C}_{\pi,2}T_0^{-1}+\pi_{Q,1}+\pi_{Q,2}$ and $\epsilon_{\gamma}=\mathfrak{C}_\epsilon (\log T_0)^{\frac{3}{2}(1+\mathfrak{c}_Q)}T_0^{-1/2}$ for finite non-negative constants $\mathfrak{C}_{\pi,1}$, $\mathfrak{C}_{\pi,2}$ and $\mathfrak{C}_\epsilon$, which are characterized in the proof, and for any $\nu\in(0,1/2-1/\psi)$. Therefore, condition {\normalfont(T2.i)} of Theorem 2 also holds, while condition {\normalfont(T2.ii)} holds with $\varpi_{\delta}^{\ttb}=\mathfrak{C}_\varpi^{\ttb}(\log T_0)^{2\mathfrak{c}_Q+1/2}$, $\epsilon_{\delta}^{\ttb}=\mathfrak{C}_\epsilon^{\ttb}T_0^{-1}$, and $\pi_{\delta}^{\ttb}=\pi_{\gamma}$, for finite non-negative constants $\mathfrak{C}_{\varpi}^{\ttb}$ and $\mathfrak{C}_\epsilon^{\ttb}$, which are also characterized in the proof.
	 
    Assume in addition that, for finite non-negative constant $\pi_{w}^{\ttb}$, the following condition holds:
	\begin{enumerate}[label=\normalfont(\thethm.\roman*),noitemsep,resume]
		\item $\varrho= \varpi_{\delta}^{\ttb}/T_0$ and $\P(\min\{|w_{0,j}|: w_{0,j}\neq 0\}\geq \varrho)\geq 1-\pi_w^{\ttb}$.
	\end{enumerate}
    Then, condition {\normalfont(T2.iii)} of Theorem 2 holds with $\pi_{\Delta}^{\ttb}=\pi_{\delta}^{\ttb}+\pi_w^{\ttb}+\pi_{\gamma}$ and $\epsilon_{\Delta}^{\ttb}=\epsilon_{\delta}^{\ttb}+\epsilon_{\gamma}$.
    
    Finally, also assume that, for finite non-negative constants $\epsilon_{\Sigma,1}^{\ttb}$, $\epsilon_{\Sigma,2}^{\ttb}$ and $\pi_{\Sigma}^{\ttb}$, the following condition holds:
    \begin{enumerate}[label=\normalfont(\thethm.\roman*),noitemsep,resume]
  	    \item $\P(\P(\|\widehat{\bSigma}-\bSigma\|\leq \epsilon_{\Sigma,1}^{\ttb}|\mathscr{H})\geq 1-\epsilon_{\Sigma,2}^{\ttb})\geq 1-\pi_{\Sigma}^{\ttb}$. 
    \end{enumerate}
    Then, condition {\normalfont(T2.iv)} of Theorem 2 holds with 
	$\epsilon_{\gamma,1}^{\ttb}=\mathfrak{C}_{\epsilon,1}^{\ttb}(\log T_0)^{\mathfrak{c}_Q}\epsilon_{\Sigma,1}^{\ttb}$, $\epsilon_{\gamma,2}^{\ttb}=\epsilon_{\Sigma,2}^{\ttb}$, and $\pi_{\gamma}^{\ttb}=\pi_{\Sigma}^{\ttb}+\pi_{\delta}^{\ttb}$, for finite non-negative constant $\mathfrak{C}_{\epsilon,1}^{\ttb}$, which is characterized in the proof.
\end{thm}


\section{Proofs}\label{SA sec: proofs}

\subsection{Proof of Lemma 1}
Let $\mathcal{E}$ be the event on which 
\begin{alignat*}{2}
	&\P\Big[M_{1,\mathtt{L}}\leq\bp_T'(\bbeta_0-\widehat{\bbeta})\leq M_{1,\mathtt{U}} \; &&\Big| \; \mathscr{H} \Big]\geq 1-\alpha_1, \qquad \text{and}\\
	&\P\Big[M_{2,\mathtt{L}}\leq e_T\leq M_{2,\mathtt{U}} \; &&\Big| \; \mathscr{H} \Big] \geq 1-\alpha_2.
\end{alignat*}  
By assumption, $\P(\mathcal{E})\geq 1-\pi_1-\pi_2$.
On $\mathcal{E}$, we have that
\[
\begin{split}
	&\P\Big[M_{1,\mathtt{L}}+M_{2,\mathtt{L}}\leq
	\bp_T'(\bbeta_0-\widehat{\bbeta})+e_T\leq M_{1,\mathtt{U}}+M_{2,\mathtt{U}}\Big|\mathscr{H}\Big]\\
	=\,&1-\P\Big[
	\Big\{\bp_T'(\bbeta_0-\widehat{\bbeta})+e_T>
	M_{1,\mathtt{U}}+M_{2,\mathtt{U}}\Big\}\cup\\ &\hspace{2.8em}\Big\{\bp_T'(\bbeta_0-\widehat{\bbeta})+e_T<M_{1,\mathtt{L}}+M_{2,\mathtt{L}}\Big\}\Big|\mathscr{H}\Big]\\
	\geq\,&1-\P\Big[
	\Big\{\bp_T'(\bbeta_0-\widehat{\bbeta})>M_{1,\mathtt{U}}\Big\}
	\cup \Big\{e_T> M_{2,\mathtt{U}}\Big\}\cup\\
	&\hspace{2.8em}\Big\{\bp_T'(\bbeta_0-\widehat{\bbeta})<M_{1,\mathtt{L}}\Big\}
	\cup \Big\{e_T< M_{2,\mathtt{L}}\Big\}\Big|\mathscr{H}\Big]\\
	\geq \,&1-\P\Big[\Big\{\bp_T'(\bbeta_0-\widehat{\bbeta})> M_{1,\mathtt{U}}\Big\}\cup
	\Big\{\bp_T'(\bbeta_0-\widehat{\bbeta})<M_{1,\mathtt{L}}\Big\}\Big|\mathscr{H}\Big]-\\
	&\hspace{1.7em}\P\Big[\Big\{e_T>M_{2,\mathtt{U}}\Big\}\cup
	\Big\{e_T<M_{2,\mathtt{L}}\Big\}\Big|\mathscr{H}\Big].
\end{split}
\]
Then, the result directly follows.\qed

\subsection{Proof of Lemma 2}
By definition of $\widehat{\bbeta}$, $(\bA-\bZ\widehat{\bbeta})'(\bA-\bZ\widehat{\bbeta})\leq (\bA-\bZ\bbeta_0)'(\bA-\bZ\bbeta_0)$, implying that $\widehat{\bdelta}'\widehat{\bQ}\widehat{\bdelta} \leq 2\widehat{\bgamma}'\widehat{\bdelta}=2\bgamma'\widehat{\bdelta}+2(\widehat{\bgamma}-\bgamma)'\widehat{\bdelta}$. On the other hand, for any $\bbeta_\alpha=\bbeta_0+\alpha(\widehat{\bbeta}-\bbeta_0)$, $\alpha\in(0,1]$, since $\mathcal{W}\times \mathcal{R}$ is convex, $\bbeta_\alpha\in\mathcal{W}\times\mathcal{R}$. By definition of $\bw_0$, $\E_{\widehat{\bbeta}}[(\bA-\bZ\bbeta_0)'(\bA-\bZ\bbeta_0)|\mathscr{H}]\leq \E_{\widehat{\bbeta}}[(\bA-\bZ\bbeta_\alpha)'(\bA-\bZ\bbeta_\alpha)|\mathscr{H}]$, where $\E_{\widehat{\bbeta}}[\cdot|\mathscr{H}]$ denotes the expectation against the conditional distribution of $(\bA, \bZ)$ given $\mathscr{H}$ with $\widehat{\bbeta}$ treated as fixed. Then, we have $\alpha\widehat{\bdelta}'\widehat{\bQ}\widehat{\bdelta}\geq 2\bgamma'\widehat{\bdelta}$. Since it holds for any $\alpha\in(0,1]$, $\bgamma'\widehat{\bdelta}\leq 0$. Consequently, we have $\widehat{\bdelta}'\widehat{\bQ}\widehat{\bdelta}\leq 2(\widehat{\bgamma}-\bgamma)'\widehat{\bdelta}$.

For any $\bxi_1, \bxi_2\in\mathbb{R}^d$ such that $\sup_{\bdelta\in\mathcal{M}_{\bxi_1}}\bp_T'\bD^{-1}\bdelta\leq \kappa$ and
$\sup_{\bdelta\in\mathcal{M}_{\bxi_2}}\bp_T'\bD^{-1}\bdelta\leq \kappa$. Let $\tilde{\bxi}=\alpha\bxi_1+(1-\alpha)\bxi_2$ for $\alpha\in[0,1]$. Consider $\mathcal{M}_{\tilde{\bxi}}=\{\bdelta\in\Delta:\bdelta'\widehat{\bQ}\bdelta-2\tilde{\bxi}'\bdelta\leq 0\}$. For any $\bdelta\in\mathcal{M}_{\tilde{\bxi}}$, 
$$
\Big(\alpha\bdelta'\widehat{\bQ}\bdelta-2\alpha\bxi_1'\bdelta\Big)+
\Big((1-\alpha)\bdelta'\widehat{\bQ}\bdelta-2(1-\alpha)\bxi_2'\bdelta\Big)
\leq 0.
$$
It immediately follows that either $\alpha\bdelta'\widehat{\bQ}\bdelta-2\alpha\bxi_1'\bdelta\leq 0$ or $((1-\alpha)\bdelta'\widehat{\bQ}\bdelta-2(1-\alpha)\bxi_2'\bdelta)
\leq 0$, which implies that $\bdelta\in\mathcal{M}_{\bxi_1}$ or $\bdelta\in\mathcal{M}_{\bxi_2}$. In either case, $\bp_T'\bD^{-1}\bdelta\leq \kappa$. Therefore, 
$\{\bxi: \sup_{\bdelta\in\mathcal{M}_{\bxi}}\bp_T'\bD^{-1}\bdelta\leq\kappa\}$ is convex. The proof for $\{\bxi:\inf_{\bdelta\in\mathcal{M}_{\bxi}}\bp_T'\bD^{-1}\bdelta\geq \kappa\}$ is similar. 
\qed

\subsection{Proof of Theorem 1}
	We write $\tilde{u}_{t,l}=u_{t,l}-\E[u_{t,l}|\mathscr{H}]$.
	Fix $\widehat{\bQ}$ and $\bp_T$,
	define $\mathcal{A}_\kappa=\{\bxi\in\mathbb{R}^d:\sup_{\bdelta\in\mathcal{M}_{\bxi}}\bp_T'\bD^{-1}\bdelta\leq \kappa\}$ for $\kappa\in\mathbb{R}$.	$\{\widehat{\bgamma}-\bgamma\in\mathcal{A}_\kappa\}
	\subseteq\{\bp_T'\bD^{-1}\widehat{\bdelta}\leq \kappa\}$ for any $\kappa$. By Berry-Esseen Theorem for convext sets \citep*{Raivc_2019_Bernoulli},
	\[
	\Big|\P(\widehat{\bgamma}-\bgamma\in\mathcal{A}_\kappa|\mathscr{H})-
	\P(\bG\in\mathcal{A}_\kappa|\mathscr{H})\Big|
	\leq 42(d^{1/4}+16)\sum_{t=1}^{T_0}\E\Big[\Big\|\sum_{l=1}^{M}\tilde{\bz}_{t,l}\tilde{u}_{t,l}\Big\|^3|\mathscr{H}\Big].
	\]
	Recall that $\tilde{\bz}_{t,l}$ is the $(t+(l-1)T_0)$th column of $\bSigma^{-1/2}\bD^{-1}\bZ'$, $\bG|\mathscr{H}\sim\mathsf{N}(0, \bSigma)$, and $\bSigma=\V[\widehat{\bgamma}|\mathscr{H}]$. 
	
	Then, by condition (T1.ii), with probability at least $1-\pi_{\gamma}$ over $\mathscr{H}$,
	\[
	\Big|\P(\widehat{\bgamma}-\bgamma\in\mathcal{A}_\kappa|\mathscr{H})-
	\P(\bG\in\mathcal{A}_\kappa|\mathscr{H})\Big|\leq 
	\epsilon_{\gamma}.
	\]
	It yields the desired result: for all $\kappa$,
	\[
	\P(\bp_T'(\widehat{\bbeta}-\bbeta_0)\leq \kappa|\mathscr{H})\geq
	\P(\widehat{\bgamma}-\bgamma\in\mathcal{A}_{\kappa}|\mathscr{H})
	\geq\P(\bG\in\mathcal{A}_{\kappa}|\mathscr{H})-\epsilon_{\gamma}
	= \P(\varsigma_{\mathtt{U}}^{\dagger}\leq \kappa|\mathscr{H})-\epsilon_\gamma.
	\]
\qed

\begin{remark}[Lower Bound for Theorem 1]
The lower bound for $\bp_T'(\widehat{\bbeta}-\bbeta_0)$ follows similarly. Specifically, under the same conditions of Theorem 1, we have 
\[
\P\Big[\P(\bp_T'\bD^{-1}\widehat{\bdelta}\geq\mathfrak{c}^\dagger(\alpha)|\mathscr{H})\geq 1-(\alpha+\epsilon_\gamma)\Big]\geq 1-\pi_\gamma,
\]
where $\mathfrak{c}^\dagger(\alpha)$ denotes the $\alpha$-quantile of $\varsigma_{\mathtt{L}}^\dagger=\inf\{\bp_T'\bD^{-1}\bdelta:\bdelta\in\mathcal{M}_{\bG}\}$ conditional on $\mathscr{H}$, with $\mathcal{M}_{\bG}=\{\bdelta\in\Delta:\ell^\dagger(\bdelta)\leq 0\}$, $\ell^\dagger(\bdelta)=\bdelta'\widehat{\bQ}\bdelta-2\bG'\bdelta$, and $\bG|\mathscr{H}\sim\mathsf{N}(\bm{0}, \bSigma)$.
\end{remark}

\subsection{Proof of Theorem 2}

	We first introduce some notation. Define $\mathcal{A}_{\kappa}=\{\bxi\in\mathbb{R}^d:
	\sup_{\bdelta\in\tilde{\mathcal{M}}_{\bxi}^{\star}}\bp_T'\bD^{-1}\bdelta\leq \kappa\}$ where 
	$$
	\tilde{\mathcal{M}}_{\bxi}^\star=\Big\{\bdelta\in\Delta^{\ttb}:\|\bdelta\|\leq\varpi_{\delta}^{\ttb}, \;
	\bdelta'\widehat{\bQ}\bdelta-2\bxi'\bdelta\leq 0\Big\}.
	$$
	Accordingly, define
	\[
	\begin{split}
		&\tilde{\varsigma}_\mathtt{U}^{\ttb}=\sup\Big\{\bp_T'\bD^{-1}\bdelta: \bdelta\in\Delta^{\ttb},\,\|\bdelta\|\leq \varpi_{\delta}^{\ttb},\,
		\ell^{\ttb}(\bdelta)\leq 0 \Big\}\;\text{and}\\
		&\tilde{\varsigma}_\mathtt{U}^{\dagger}=\sup\Big\{\bp_T'\bD^{-1}\bdelta: \bdelta\in\Delta,\,\|\bdelta\|\leq \varpi_{\delta}^{\ttb},\,
		\ell^{\dagger}(\bdelta)\leq 0 \Big\}.
	\end{split}
	\]
	Let $\tilde{\mathfrak{c}}^{\ttb}(1-\alpha)$ be the $(1-\alpha)$-quantile of $\tilde{\varsigma}_{\mathtt{U}}^{\ttb}$ conditional on the data. Similarly, $\tilde{\mathfrak{c}}^{\dagger}(1-\alpha)$ is the $(1-\alpha)$-quantile of $\tilde{\varsigma}_{\mathtt{U}}^{\dagger}$ conditional on $\mathscr{H}$.
	
	Let $\P_1=\mathsf{N}(0, \widehat{\bSigma})$ and $\P_2=\mathsf{N}(0, \bSigma)$. Then, by Pinsker's inequality, for any $\kappa$,
	\[
	\Big|\P(\bG^{\ttb}\in\mathcal{A}_{\kappa}|\mathscr{H}, \bA)-\P(\bG\in\mathcal{A}_{\kappa}|\mathscr{H})\Big|
	\leq\sqrt{\mathbb{KL}(\P_1, \P_2)/2},
	\]
	where $\mathbb{KL}(\cdot, \cdot)$ is the Kullback-Leibler divergence.
	Define $\bvGam=\bSigma^{-1/2}\widehat{\bSigma}\bSigma^{-1/2}$.
	Note that 
	\[
	\mathbb{KL}(\P_1, \P_2)=-0.5\log\{\det(\bvGam)\}+0.5\tr(\bvGam-\bI_d)
	=0.5\sum_{j=1}^{d}(\lambda_j-\log(\lambda_j+1)),
	\]
	where $\lambda_j$ is the $j$th largest eigenvalue of $\bvGam-\bI_d$.
	By condition (T2.iv), with probability over $\mathscr{H}$ at least $1-\pi_{\gamma}^{\ttb}$, it holds that with $\P(\cdot|\mathscr{H})$-probability at least $1-\epsilon_{\gamma,2}^{\ttb}$, $|\lambda_1|\leq 0.5$ and 
	$$
	\mathbb{KL}(\P_1, \P_2)\leq 0.5\sum_{j=1}^{d}\lambda_j^2\leq (2\epsilon_{\gamma,1}^{\ttb})^2/2.
	$$
	In view of condition (T2.iii), for any $\kappa$,
	\[
	\P\Big\{\P\Big(|\P^{\ttb}(\tilde{\varsigma}_{\mathtt{U}}^{\ttb}\leq \kappa)-\P^{\ttb}(\tilde{\varsigma}_{\mathtt{U}}^{\dagger}\leq\kappa)|
	\leq \epsilon_{\gamma,1}^{\ttb}\Big|\mathscr{H}\Big)\geq 1-\epsilon_{\gamma,2}^{\ttb}-\epsilon_{\Delta}^{\ttb}\Big\}\geq 1-\pi_{\gamma}^{\ttb}-\pi_{\Delta}^{\ttb}.
	\]
	Note that by construction, $\mathfrak{c}^{\ttb}(1-\alpha)\geq\tilde{\mathfrak{c}}^{\ttb}(1-\alpha)$. Then, we have 
	\begin{equation} \label{eq: proof simul. fact 1}
		\P\Big\{\P\Big(\mathfrak{c}^{\ttb}(1-\alpha)\geq
		\tilde{\mathfrak{c}}^{\dagger}(1-\alpha-\epsilon_{\gamma,1}^{\ttb})
		\Big|\mathscr{H}\Big)\geq 1-\epsilon_{\gamma,2}^{\ttb}-\epsilon_{\Delta}^{\ttb}\Big\}\geq 1-\pi_{\gamma}^{\ttb}-\pi_{\Delta}^{\ttb}.
	\end{equation}
	By condition (T2.ii), with probability over $\mathscr{H}$ at least $1-\pi_{\delta}^{\ttb}$,
	\[
	\P\Big(\mathcal{M}_\bG=\{\bdelta\in\Delta: \|\bdelta\|\leq \varpi_{\delta}^{\ttb}, \ell^{\dagger}(\bdelta)\leq 0\}\Big|\mathscr{H}\Big)\geq 1-\epsilon_{\delta}^{\ttb},
	\]
	which implies that for any $\kappa$,
	$\P(\tilde{\varsigma}_{\mathtt{U}}^{\dagger}\leq \kappa|\mathscr{H})\leq
	\P(\varsigma_{\mathtt{U}}^{\dagger}\leq \kappa|\mathscr{H})
	+\epsilon_{\delta}^{\ttb}$. Thus,
	\begin{equation}\label{eq: proof simul. fact 2}
		\P\Big(\tilde{\mathfrak{c}}^{\dagger}(1-\alpha-\epsilon_{\gamma,1}^\star)\geq\mathfrak{c}^\dagger(1-\alpha-\epsilon_{\gamma,1}^{\ttb}-\epsilon_{\delta}^{\ttb})\Big)\geq 1-\pi_{\delta}^\star.
	\end{equation}
	
	Therefore, on an event $\mathcal{A}\in\mathscr{H}$ with $\P(\mathcal{A})\geq 1-\pi_{\gamma}-\pi_{\gamma}^{\ttb}-\pi_{\Delta}^{\ttb}-\pi_{\delta}^{\ttb}$,
	\[
	\begin{split}
		\P\Big(\bp_T'(\widehat{\bbeta}-\bbeta_0)
		\leq \mathfrak{c}^{\ttb}(1-\alpha)\Big|\mathscr{H}\Big)
		&\geq \P\Big(\bp_T'\bD^{-1}\widehat{\bdelta}\leq \tilde{\mathfrak{c}}^{\dagger}(1-\alpha-\epsilon_{\gamma,1}^{\ttb})\Big|\mathscr{H}\Big)-\epsilon_{\gamma,2}^{\ttb}-\epsilon_{\Delta}^{\ttb}\\
		&\geq \P\Big(\bp_T'\bD^{-1}\widehat{\bdelta}\leq \mathfrak{c}^{\dagger}(1-\alpha-\epsilon_{\gamma,1}^{\ttb}-\epsilon_{\delta}^{\ttb})\Big|\mathscr{H}\Big)-\epsilon_{\gamma,2}^{\ttb}-\epsilon_{\Delta}^{\ttb}\\
		&\geq 1-\alpha-\epsilon_{\gamma,1}^{\ttb}-\epsilon_{\delta}^{\ttb}-\epsilon_{\gamma}-\epsilon_{\gamma,2}^{\ttb}-\epsilon_{\Delta}^{\ttb},
	\end{split}
	\]
	where the first line follows by inequality \eqref{eq: proof simul. fact 1}, the second by inequality \eqref{eq: proof simul. fact 2}, and the third by condition (T2.i). Then the proof is complete.
\qed

\begin{remark}[Lower Bound for Theorem 2]
The lower bound for $\bp_T'(\widehat{\bbeta}-\bbeta_0)$ follows similarly. Specifically, we replace  condition \normalfont{(T2.i)} in Theorem 2 with $\P[\P(\bp_T'\bD^{-1}\widehat{\bdelta}\geq \mathfrak{c}^\dagger(\alpha)|\mathscr{H})\geq 1-\alpha-\epsilon_\gamma]\geq 1-\pi_\gamma$, and the other conditions in Theorem 2 remain the same. Under these conditions, we have for $\epsilon_\gamma\in[0,0.25]$,
\[
\P\Big[\P(\bp_T'\bD^{-1}\widehat{\bdelta}\geq \mathfrak{c}^{\ttb}(\alpha)|\mathscr{H})\geq 1-\alpha-\epsilon\Big]\geq 1-\pi,
\]
where $\mathfrak{c}^{\ttb}(\alpha)$ denotes the $\alpha$-quantile of $\varsigma_{\mathtt{L}}^{\ttb}=\inf\{\bp_T'\bD^{-1}\bdelta:\bdelta\in\Delta^{\ttb},\ell^{\ttb}(\bdelta)\leq 0\}$ conditional on the data. $\pi$ and $\epsilon$ are defined as in Theorem 2.
\end{remark}

\subsection{Proof of Theorem A}

	Throughout this proof, we write $\mathfrak{b}(\cdot):=\mathfrak{b}(\cdot;\mathscr{H})$ and assume $\bu_t$ is correctly centered, i.e., $\E[\bu_t|\mathscr{H}]=0$, only for ease of notation. Let $\|\bv\|_\infty$ denote the sup-norm of a generic vector $\bv$. 
	As in the proof of Theorem 1, fix $\widehat{\bQ}$ and $\bp_T$,
	define $\mathcal{A}_\kappa=\{\bxi\in\mathbb{R}^d:\sup_{\bdelta\in\mathcal{M}_{\bxi}}\bp_T'\bD^{-1}\bdelta\leq \kappa\}$ for each $\kappa\in\mathbb{R}$.	Since $\{\widehat{\bgamma}-\bgamma\in\mathcal{A}_\kappa\}\subseteq
	\{\bp_T'(\widehat{\bbeta}-\bbeta_0)\leq \kappa\}$ for any $\kappa$, 
	\[
	\P(\bp_T'(\widehat{\bbeta}-\bbeta_0)\leq \kappa|\mathscr{H})\geq
	\P(\widehat{\bgamma}-\bgamma\in\mathcal{A}_{\kappa}|\mathscr{H}).
	\]
	It suffices to approximate $\P(\widehat{\bgamma}-\bgamma\in\mathcal{A}_{\kappa}|\mathscr{H})$.
	
	According to the blocking design, let $\bh_k=(\bu_t)_{t\in\mathcal{J}_k}$ and $\bhs_k=(\bu_t)_{t\in\mathcal{J}_k'}$ be vectors that contain all $\bu_t$'s with $t$ in $\mathcal{J}_k$ and $\mathcal{J}_k'$ respectively. By Berbee's coupling lemma \citep*{Berbee_1987_PTRF}, on an enlarged probability space, there exists  $\{\bh_k^*\}_{k=1}^m$ that is independent over $k$ conditional on $\mathscr{H}$ such that for each $k$, $\bh_k$ and $\bh_k^*$  are identically distributed (conditional on $\mathscr{H}$), and $\P(\bh_k\neq \bh_k^*|\mathscr{H})\leq \mathfrak{b}(v)$. Similarly, on an enlarged probability space, there exists $\{\bhs_k^*\}_{k=1}^{m}$ that is independent over $k$ conditional on $\mathscr{H}$ such that for each $k$, $\bhs_k$ and $\bhs_k^*$ are identically distributed (conditional on $\mathscr{H}$) and $\P(\bhs_k\neq \bhs_k^*|\mathscr{H})\leq \mathfrak{b}(q)$. Let $\bu_t^*$ be the vector in $\bh_k^*$ and $\bhs_k^*$ that corresponds to $\bu_t$. 
	Accordingly, $\bs_t^*$ is the same as $\bs_t$ except that $\bu_t$ is replaced by $\bu_t^*$. Also, $\bS_{k,\Box}^*=\sum_{t\in\mathcal{J}_k}\bs_t^*$ and $\bS_{k,\diamond}^*=\sum_{j\in\mathcal{J}_k'}\bs_t^*$. 
	
	By the coupling argument, we have
	\begin{align*}\allowdisplaybreaks
		\P\Big(\widehat{\bgamma}-\bgamma\in\mathcal{A}_\kappa|\mathscr{H}\Big)
		\leq\; &
		\P\Big(\Big\|\sum_{k=1}^{m} \bS_{k,\diamond}\Big\|_\infty\geq \xi_1\Big|\mathscr{H}\Big)+
		\P\Big(\Big\|\bS_{m+1,1}\Big\|_\infty\geq \xi_2\Big|\mathscr{H}\Big)+\\
		&\P\Big(\sum_{k=1}^{m} \bS_k\in\mathcal{A}_\kappa^{\xi_1+\xi_2}\Big|\mathscr{H}\Big)\\
		\leq\; &
		\P\Big(\Big\|\sum_{k=1}^{m} \bS_{k,\diamond}^*\Big\|_\infty\geq \xi_1\Big|\mathscr{H}\Big)+
		\P\Big(\Big\|\bS_{m+1,1}\Big\|_\infty\geq \xi_2\Big|\mathscr{H}\Big)\\
		&\P\Big(\sum_{k=1}^{m} \bS_{k,\Box}^*\in\mathcal{A}_\kappa^{\xi_1+\xi_2}\Big|\mathscr{H}\Big)+
		m\mathfrak{b}(v)+m\mathfrak{b}(q)\\
		=:\,&I+II+III+m\mathfrak{b}(v)+m\mathfrak{b}(q),
	\end{align*}
	where $\mathcal{A}_\kappa^{\xi_1+\xi_2}=\{\bv: \rho(\bv, \mathcal{A}_{\kappa})\leq \xi_1+\xi_2\}$ is the $(\xi_1+\xi_2)$-enlargement of $\mathcal{A}_\kappa$ for $\xi_1, \xi_2\geq0$ and $\rho(\bv, \mathcal{\mathcal{A}_\kappa})=\inf_{\bv'\in\mathcal{A}_\kappa}
	\|\bv-\bv'\|$.
	
	For $I$, by Markov's inequality, condition (TA.ii), and Lemma 8 of \citet*{Chernozhukov-Chetverikov-Kato_2015_PTRF}, for some absolute constant $\mathfrak{C}_1>0$, for any $\xi_1>0$,
	\[
	\P\Big(\Big\|\sum_{k=1}^{m}
	\bS_{k,\diamond}^{*}\Big\|_\infty\geq\xi_1\Big|\mathscr{H}\Big)
	\leq \mathfrak{C}_1\xi_1^{-1}\Big(\sqrt{mv\bar{\sigma}^2\log d}+
	\eta_1^{1/\psi}\log d\Big),
	\]
	where we use the fact that 
	\[
	\begin{split}
		(\E[(\max_{1\leq k\leq m}
		\|\bS_{k,\diamond}^*\|_\infty)^2|\mathscr{H}])^{1/2}
		&\leq(\E[\max_{1\leq j\leq d}\max_{1\leq k\leq m}
		|S^*_{jk,\diamond}|^\psi|\mathscr{H}])^{1/\psi}\\
		&\leq \Big(\E\Big[\sum_{j=1}^{d}\sum_{k=1}^{m}|S_{jk,\diamond}|^\psi\Big|\mathscr{H}\Big]\Big)^{1/\psi}
		\leq \eta_1^{1/\psi},
	\end{split}
	\]
	which holds with probability over $\mathscr{H}$ at least $1-\pi_{\gamma,1}$.
	Taking $\xi_1=\mathfrak{C}_1\eta_6^{-1}\Big(\sqrt{mv\bar{\sigma}^2\log d}+\eta_1^{1/\psi}\log d\Big)$, we have 
	$I\leq \eta_6$.
	
	For $II$, take $\xi_2=(d\eta_2)^{1/\psi}\eta_6^{-1/\psi}$.
	By Markov's inequality and condition (TA.iii), with probability over $\mathscr{H}$ at least $1-\pi_{\gamma,2}$,
	\[
	II\leq d\max_{1\leq j\leq d}
	\P\Big(\Big|\sum_{t\in\mathcal{J}_{m+1}}s_{jt}\Big|
	\geq \xi_2|\mathscr{H}\Big)\leq 
	\max_{1\leq j\leq d}
	\frac{\E[|\sum_{t\in\mathcal{J}_{m+1}}s_{jt}|^{\psi}|\mathscr{H}]\eta_6}
	{\eta_2}
	\leq \eta_6. 
	\]
	
	For $III$, since $\mathcal{A}_\kappa^{\xi_1+\xi_2}$ is also convex and $\{\bS_{k,\Box}^*\}$ is independent over $k$, repeating the argument in the proof of Theorem 1 and using condition (TA.iv), we have that for any $\xi_1> 0$ and $\xi_2> 0$, with probability over $\mathscr{H}$ at least $1-\pi_{\gamma,3}$,
	\[
	\Big|III-
	\P(\bG_{\Box}\in\mathcal{A}_\kappa^{\xi_1+\xi_2}|\mathscr{H})\Big|
	\leq 42(d^{1/4}+16)\sum_{k=1}^{m}\E\Big[
	\Big\|\bSigma_{\Box}^{-1/2}\bS_{k,\Box}
	\Big\|^3\Big|\mathscr{H}\Big]\leq \eta_3
	\]
	for $\bG_{\Box}|\mathscr{H}\sim \mathsf{N}(0, \bSigma_{\Box})$.
	
	Combining the results above, we have with probability over $\mathscr{H}$ at least $1-\pi_{\gamma,1}-\pi_{\gamma,2}-\pi_{\gamma,3}$,
	\[
	\P\Big(\widehat{\bgamma}-\bgamma\in\mathcal{A}_\kappa|\mathscr{H}\Big)
	\leq \P\Big(\bG_{\Box}\in\mathcal{A}_{\kappa}^{\xi_1+\xi_2}\Big|\mathscr{H}\Big)
	+5\eta_6.
	\]
	
	On the other hand, for any $\xi>0$, define
	\[
	\mathcal{A}_{\kappa}^{-\xi}=\Big\{\bh: \mathcal{B}(\bh,\xi)\subseteq\mathcal{A}_\kappa\Big\}.
	\]
	
	Suppose that $\mathcal{A}_\kappa^{-\xi_1-\xi_2}\neq \emptyset$.
	Note that for any $\bh_1,\bh_2\in\mathcal{A}_\kappa^{-\xi}$ and $\omega\in[0,1]$, define $\bh'=\omega\bh_1+(1-\omega)\bh_2$. For any $\bh''\in\mathcal{B}(\bh',\xi)$, $$
	\bh''=\bh'+(\bh''-\bh')=\omega(\bh_1+(\bh''-\bh'))+(1-\omega)(\bh_2+(\bh''-\bh')).
	$$ 
	Since $\|\bh''-\bh'\|\leq \xi$, $\bh_1+(\bh''-\bh')\in\mathcal{B}(\bh_1,\xi)\subseteq\mathcal{A}_\kappa$.
	Similarly, $\bh_2+(\bh''-\bh')\in\mathcal{A}_\kappa$. Therefore, $\bh''\in\mathcal{A}_\kappa$. Since it holds for arbitrary $\bh''\in\mathcal{B}(\bh',\xi)$, $\mathcal{A}_\kappa^{-\xi}$ is convex.
	Then, repeat the coupling argument and apply the Berry-Esseen inequality again. We have the following inequalities hold with probability over $\mathscr{H}$ at least $1-\pi_{\gamma,3}$:
	\[
	\begin{split}
		&\,\P(\bG_{\Box}\in\mathcal{A}_\kappa^{-(\xi_1+\xi_2)}|\mathscr{H})-\eta_3
		\leq\P\Big(\sum_{k=1}^{m} \bS_{k,\Box}^*\in\mathcal{A}_\kappa^{-(\xi_1+\xi_2)}\Big|\mathscr{H}\Big)\\ \leq
		&\,\P\Big(\sum_{k=1}^{m} \bS_{k,\Box}\in\mathcal{A}_\kappa^{-(\xi_1+\xi_2)}\Big|\mathscr{H}\Big)+
		m\mathfrak{b}(v)\\
		\leq &\,\P\Big(\widehat{\bgamma}-\bgamma\in\mathcal{A}_\kappa\Big|\mathscr{H}\Big)+ \P\Big(\Big\|\sum_{k=1}^{m} \bS_{k,\diamond}\Big\|_\infty\geq \xi_1\Big|\mathscr{H}\Big)+
		\P\Big(\Big\|\bS_{m+1,\diamond}\Big\|_\infty\geq \xi_2\Big|\mathscr{H}\Big)+m\mathfrak{b}(v).
	\end{split}
	\]
	Using the bounds obtained previously, the above implies that with probability over $\mathscr{H}$ at least $1-\pi_{\gamma,1}-\pi_{\gamma,2}-\pi_{\gamma,3}$,
	\[
	\P(\widehat{\bgamma}-\bgamma\in\mathcal{A}_\kappa|\mathscr{H})\geq
	\P(\bG_{\Box}\in\mathcal{A}_{\kappa}^{-(\xi_1-\xi_2)}|\mathscr{H})-5\eta_6.
	\]
	
	By condition (TA.v) and Anti-Concentration of the Gaussian measure for convex sets (see, e.g., Lemma A.2 of \citet*{Chernozhukov-Chetverikov-Kato_2017_AoP}), for some absolute constant $\mathfrak{C}_2>0$, with probability over $\mathscr{H}$ at least $1-\pi_{\gamma,4}$,
	\[
	\begin{split}
		&\P(\bG_{\Box}\in\mathcal{A}_\kappa^{\xi_1+\xi_2}|\mathscr{H})\leq \P(\bG_{\Box}\in\mathcal{A}_\kappa|\mathscr{H})+\mathfrak{C}_2d\eta_4(\xi_1+\xi_2),\quad \text{and}\\
		&\P(\bG_{\Box}\in\mathcal{A}_\kappa^{-(\xi_1+\xi_2)}|\mathscr{H})\geq
		\P(\bG_{\Box}\in\mathcal{A}_\kappa)-\mathfrak{C}_2d\eta_4(\xi_1+\xi_2).
	\end{split}
	\]
	The result for $\mathcal{A}_\kappa^{-(\xi_1+\xi_2)}=\emptyset$ trivially follows. Then, we have with probability over $\mathscr{H}$ at least $1-\pi_{\gamma,1}-\pi_{\gamma,2}-\pi_{\gamma,3}-\pi_{\gamma,4}$,
	\[
	\Big|\P(\widehat{\bgamma}-\bgamma\in\mathcal{A}_\kappa|\mathscr{H})-\P(\bG_{\Box}\in\mathcal{A}_\kappa|\mathscr{H})\Big|\leq 5\eta_6+\mathfrak{C}_2d\eta_4(\xi_1+\xi_2)\leq \mathfrak{C}_3\eta_6
	\]
	for some constant $\mathfrak{C}_3>0$.
	
	Finally, as in the proof of Theorem 2, letting $\P_1=\mathsf{N}(0, \bSigma)$ and $\P_2=\mathsf{N}(0, \bSigma_{\Box})$, by Pinsker's inequality, for any $\kappa$,
	\[
	\Big|\P(\bG\in\mathcal{A}_{\kappa}|\mathscr{H})-\P(\bG_{\Box}\in\mathcal{A}_{\kappa}|\mathscr{H})\Big|
	\leq\sqrt{\mathbb{KL}(\P_1, \P_2)/2},
	\]
	where  
	\[
	\mathbb{KL}(\P_1, \P_2)=-0.5\log\{\det(\bvGam)\}+0.5\tr(\bvGam-\bI_d)
	=0.5\sum_{j=1}^{d}(\lambda_j-\log(\lambda_j+1)),
	\]
	where $\bvGam=\bSigma_{\Box}^{-1/2}\bSigma\bSigma_{\Box}^{-1/2}$ and $\lambda_j$ is the $j$th largest eigenvalue of $\bvGam-\bI_d$.
	By condition (TA.vi), with probability over $\mathscr{H}$ at least $1-\pi_{\gamma,5}$, $|\lambda_1|\leq 0.5$ and  
	$$
	\mathbb{KL}(\P_1, \P_2)\leq 0.5\sum_{j=1}^{d}\lambda_j^2\leq (2\eta_5)^2/2.
	$$
	Then the desired result follows.
\qed

\begin{remark}[Lower Bound for Theorem A]
   The lower bound for $\bp_T'\bD^{-1}\widehat{\bdelta}$ follows similarly. Specifically, under the same conditions of Theorem A, we have for $\eta_5\in[0,0.25]$
   \[
     \P\Big[\P(\bp_T'\bD^{-1}\widehat{\bdelta}\geq\mathfrak{c}^\dagger(\alpha)|\mathscr{H})
     \geq 1-\alpha-\epsilon_\gamma\Big]\geq 1-\pi_\gamma,
   \]
   where $\mathfrak{c}^\dagger(\alpha)$ denotes the $\alpha$-quantile of $\varsigma_{\mathtt{L}}^\dagger=\inf\{\bp_T'\bD^{-1}\bdelta:\bdelta\in\mathcal{M}_{\bG}\}$ conditional on $\mathscr{H}$, with $\mathcal{M}_{\bG}=\{\bdelta\in\Delta:\ell^\dagger(\bdelta)\leq 0\}$, $\ell^\dagger(\bdelta)=\bdelta'\widehat{\bQ}\bdelta-2\bG'\bdelta$, and $\bG|\mathscr{H}\sim\mathsf{N}(\bm{0}, \bSigma)$. $\epsilon_\gamma$ and $\pi_\gamma$ are defined as in Theorem A.
\end{remark}

\subsection{Proof of Theorem \ref{thm: exmp independent case}}
	Note that given the assumption, conditional on $\bB$, $\bb_T$ is independent of the pre-treatment data. Therefore, when we verify the conditions of Theorems 1 and 2, the analysis conditional on $\mathscr{H}$ is the same as that conditional on $\bB$ only. 
	
	(1) We first verify the conditions of Theorem 1. Condition (T1.i) trivially holds. To show condition (T1.ii), we note by condition (SA-1.i), with probability (over $\mathscr{H}$) one,
	\[
	\E\Big[\|\tilde{\bz}_{t}\tilde{u}_{t}\|^3\Big|\mathscr{H}\Big]
	\leq \|\tilde{\bz}_{t}\|^{3}\E[|\tilde{u}_{t}|^3|\mathscr{H}]\leq 
	\bar{\eta}_1 \|\tilde{\bz}_{t}\|^3
	\]
	for any $t$. Then, it suffices to bound 
	$$
	\sum_{t=1}^{T_0}\|\tilde{\bz}_{t}\|^3\leq \|\bSigma^{-1/2}\|^{3}
	\bigg(\frac{1}{T_0^{3/2}}\sum_{t=1}^{T_0}\|\bz_{t}\|^3
	\bigg).
	$$ 
	Note that 
	$\frac{1}{T_0}\sum_{t=1}^{T_0}\|\bz_{t}\|^3\leq
	\frac{\sqrt{d}}{T_0}\sum_{t=1}^{T_0}\sum_{j=1}^{d}|z_{jt}|^3$ where $z_{jt}$ is the $j$th element of $\bz_{t}$. By Markov's inequality,
	\[
	\P\Big(\Big|\frac{1}{T_0}\sum_{t=1}^{T_0}|z_{jt}|^3-\E[|z_{jt}|^3]\Big|\geq\bar{\eta}_2\Big)\leq \frac{1}{T_0\bar{\eta}_2}.
	\]
	Thus, with probability at least $1-d(\bar{\eta}_2 T_0)^{-1}$, 
	$\frac{1}{T_0}\sum_{t=1}^{T_0}\sum_{j=1}^{d}|z_{jt}|^3\leq 2d\bar{\eta}_2$, and then
	$\frac{1}{T_0}\sum_{t=1}^{T_0}\|\bz_{t}\|^3\leq
	2d^{3/2}\bar{\eta}_2$.
	
	On the other hand, by the lower bound on the conditional variance of $u_{t}$, $\lambda_{\min}(\bSigma)\geq \underline{\eta}_1\lambda_{\min}(\widehat{\bQ})$. By Markov's inequality and union bounds,
	\[
	\P\Big(\Big\|\frac{1}{T_0}\sum_{t=1}^{T_0}
	\bz_t\bz_t'-\E[\bz_t\bz_t']\Big\|_{\mathtt{F}}\geq d\underline{\eta}_2/(2d)\Big)\leq 
	\frac{4d^4\bar{\eta}_2}{\underline{\eta}_2^2T_0}.
	\]
	It follows that $\lambda_{\min}(\bSigma)\geq \underline{\eta}_1\underline{\eta}_2/2$ with probability over $\mathscr{H}$ at least $1-\frac{4d^4\bar{\eta}_2}{\underline{\eta}_2^2T_0}$. Then, condition (T1.ii) holds.
	
	(2) Now, we verify the conditions of Theorem 2. Condition (T2.i) follows by part (1). For conditions (T2.ii) and (T2.iii), note that $\bdelta\in\mathcal{M}_{\bG}$ implies that
	\begin{equation}\label{eq: basic inequality}
	\lambda_{\min}(\widehat{\bQ})\|\bdelta\|^2\leq
	\bdelta\widehat{\bQ}\bdelta\leq 2\bG'\bdelta\leq  
	2\|\bG\|\|\bdelta\|.
	\end{equation}
	Let $\sigma_{\max}^2$ be the largest diagonal element of $\bSigma$. Note that for Gaussian random variables,
	\[
	\P\Big(\|\bG\|\geq\sqrt{2d\log T_0}\sigma_{\max}\Big|\mathscr{H}\Big)\leq 2d\exp\Big(-\frac{2\sigma_{\max}^2\log T_0}{2\sigma_{\max}^2}\Big)
	\leq 2d/T_0.
	\]
	
	On the other hand, note that by Markov's inequality and union bounds,
	\[
	\P\Big(\max_{1\leq j\leq d}
	\Big|\frac{1}{T_0}\sum_{t=1}^{T_0}z_{jt}^2-\E[z_{jt}^2]\Big|
	\geq\bar{\eta}_2\Big)\leq\frac{d\bar{\eta}_2}{\bar{\eta}_2^2T_0}.
	\]
	Thus, $\sigma_{\max}^2\leq 2\bar{\eta}_2\bar{\eta}_1$ with probability over $\mathscr{H}$ at least $1-\frac{d}{\bar{\eta}_2T_0}$.
	
	Using the results in part (1), we have for $\bdelta\in\mathcal{M}_{\bG}$, with probability over $\mathscr{H}$ at least $1-\frac{4d^4\bar{\eta}_2}{\underline{\eta}_2^2T_0}-\frac{d}{\bar{\eta}_2T_0}$, 
	\[
	\P\Big(\|\bdelta\|\leq \frac{8\sqrt{d\log T_0
			\bar{\eta}_1\bar{\eta}_2}}{\underline{\eta}_2}\Big|\mathscr{H}\Big)
	\geq 1-\frac{2d}{T_0}. 
	\]
	Thus, condition (T2.ii) holds. 
	
	For condition (T2.iii), we note that $\widehat{\bdelta}$ satisfies the same basic inequality \eqref{eq: basic inequality} except $\bG$ needs to be replaced by $\widehat{\bgamma}-\bgamma$. Noting condition (T1.ii) proved in part (1), we can see the same (upper) bound also holds for $\widehat{\bdelta}$ with $\P(\cdot|\mathscr{H})$-probability at least  $1-2d/T_0-\epsilon_\gamma$, with probability over $\mathscr{H}$ at least $1-\frac{4d^4\bar{\eta}_2}{\underline{\eta}_2^2T_0}-\frac{d}{\bar{\eta}_2T_0}-\pi_\gamma$.
	
	Finally, consider the variance estimator $\widehat{\bSigma}$. 
	Since $\|\bSigma^{-1/2}\widehat{\bSigma}\bSigma^{-1/2}-\bI_d\|\leq
	\lambda_{\min}(\bSigma)^{-1}\|\widehat{\bSigma}-\bSigma\|
	$,
	\[
	\tr\Big[(\bSigma^{-1/2}\widehat{\bSigma}\bSigma^{-1/2}-\bI_d)^2\Big]
	\leq d\lambda_{\min}(\bSigma)^{-2}\|\widehat{\bSigma}-\bSigma\|^2.
	\]
	
	$\widehat{\bSigma}-\bSigma$ can be decomposed into two parts. First consider $\frac{1}{T_0}\sum_{t=1}^{T_0}\bz_t\bz_t'\tilde{u}_t^2-\bSigma=\frac{1}{T_0}\sum_{t=1}^{T_0}\bz_t\bz_t'(\tilde{u}_t^2-\sigma^2_t)$ for $\tilde{u}_t=u_t-\E[u_t|\bb_t]$ and $\sigma_t^2=\E[\tilde{u}_t^2|\bb_t]$. 
	Applying Markov's inequality again, we have with probability over $\mathscr{H}$ at least $1-\frac{d^2}{\bar{\eta}_2T_0}$,
	$\max_{1\leq j,j'\leq d}
	\frac{1}{T_0}\sum_{t=1}^{T_0}z_{jt}^2z_{j't}^2\leq 2\bar{\eta}_2$.
	Then, it follows by Markov's inequality and condition (SA-1.iv) that with probability over $\mathscr{H}$ at least  $1-\frac{d^2}{\bar{\eta}_2T_0}$,
	\[
	\P\Big(\Big\|\frac{1}{T_0}\sum_{t=1}^{T_0}\bz_t\bz_t'(\tilde{u}_t^2-\sigma^2_t)\Big\|_{\mathtt{F}}\geq \frac{d}{T_0^{1/2-\upsilon}}\Big|\mathscr{H}\Big)\leq 
	\frac{d^2\bar{\eta}_1T_0^{1-2\upsilon}}{T_0}
	\max_{1\leq j,j'\leq d}\frac{1}{T_0}\sum_{t=1}^{T_0}z_{jt}^2z_{j't}^2
	\leq\frac{2d^2\bar{\eta}_1\bar{\eta}_2}{T_0^{2\upsilon}}.
	\]
	Next, note that
	\[
	\begin{split}
		&(\widehat{u}_t-\widehat{\E}[u_t|\bb_t])^2-\tilde{u}_t^2\\
		=&\Big((\widehat{u}_t-u_t)-(\widehat{\E}[u_t|\bb_t]-\E[u_t|\bb_t])\Big)
		\Big(\widehat{u}_t+u_t-\widehat{\E}[u_t|\bb_t]-\E[u_t|\bb_t]\Big)\\
		=&\Big(\bz_t'(\bbeta_0-\widehat{\bbeta})-(\widehat{\E}[u_t|\bb_t]-\E[u_t|\bb_t])\Big)
		\Big(\bz_t'(\bbeta_0-\widehat{\bbeta})+2u_t-\widehat{\E}[u_t|\bb_t]-\E[u_t|\bb_t]\Big)\\
		=&:q_t(q_t+2\tilde{u}_t).
	\end{split}
	\]
	Write $q_{t}=q_{t,1}+q_{t,2}$, $q_{t,1}=\bz_t'(\bbeta_0-\widehat{\bbeta})$, and $q_{t,2}=\widehat{\E}[u_t|\bb_t]-\E[u_t|\bb_t]$.
	Then, 
	\[
	\begin{split}
		\Big\|\frac{1}{T_0}\sum_{t=1}^{T_0}\bz_t\bz_t'\Big((\widehat{u}_t-\widehat{\E}[u_t|\bb_t])^2-\tilde{u}_t^2\Big)\Big\|&=
		\Big\|\frac{1}{T_0}\sum_{t=1}^{T_0}\bz_t\bz_t'(q_t^2+2q_t\tilde{u}_t)\Big\|\\
		&\leq\Big\|\frac{1}{T_0}\sum_{t=1}^{T_0}\bz_t\bz_t'q_t^2\Big\|+
		2\Big\|\frac{1}{T_0}\sum_{t=1}^{T_0}\bz_t\bz_t'\tilde{u}_tq_t\Big\|.
	\end{split}
	\]
	For the first term, 
	\[
	\begin{split}
		\Big\|\frac{1}{T_0}\sum_{t=1}^{T_0}\bz_t\bz_t'q_t^2\Big\|&\leq 
		2\Big\|\frac{1}{T_0}\sum_{t=1}^{T_0}\bz_t\bz_t'q_{t,1}^2\Big\|+
		2\Big\|\frac{1}{T_0}\sum_{t=1}^{T_0}\bz_t\bz_t'q_{t,2}^2\Big\|\\
		&\leq 2\Big\|\frac{1}{T_0}\sum_{t=1}^{T_0}\bz_t\bz_t'\|\bz_t\|^2\Big\|
		\|\bbeta_0-\widehat{\bbeta}\|^2+
		2\Big\|\frac{1}{T_0}\sum_{t=1}^{T_0}\bz_t\bz_t'\Big\|\max_{1\leq t\leq T_0} q_{t,2}^2.
	\end{split}
	\]
	Using the bound on $\widehat{\bdelta}$ explained before and condition (SA-1.v), we have with probability over $\mathscr{H}$ at least $1-\mathfrak{C}_1'/T_0-\pi_{\delta}^{\ttb}-\pi_\gamma-\pi_{u}^{\ttb}$, $\P(\|\frac{1}{T_0}\sum_{t=1}^{T_0}\bz_t\bz_t'q_t^2\| 
	\leq\mathfrak{C}_1((\varpi_{\delta}^{\ttb})^2/T_0+(\varpi_u^{\ttb})^2)|\mathscr{H})\geq 1-\epsilon_{\delta}^{\ttb}-\epsilon_\gamma-\epsilon_u^{\ttb}$ for some  constants $\mathfrak{C}_1>0$ and $\mathfrak{C}_1'>0$.
	
	For the second term,
	\[
	\begin{split}
		2\Big\|\frac{1}{T_0}\sum_{t=1}^{T_0}\bz_t\bz_t'\tilde{u}_tq_t\Big\|&\leq
		2\Big\|\frac{1}{T_0}\sum_{t=1}^{T_0}\bz_t\bz_t'\tilde{u}_tq_{t,1}\Big\|+
		2\Big\|\frac{1}{T_0}\sum_{t=1}^{T_0}\bz_t\bz_t'\tilde{u}_tq_{t,2}\Big\|\\
		&\leq \Big(\Big\|\frac{1}{T_0}\sum_{t=1}^{T_0}\bz_t\bz_t'\|\bz_t\|\Big\|+
		\Big\|\frac{1}{T_0}\sum_{t=1}^{T_0}\bz_t\bz_t'\|\bz_t\|\tilde{u}_t^2\Big\|\Big)
		\|\widehat{\bbeta}-\bbeta_0\|	
		\\
		&\quad+\max_{1\leq t\leq T_0}|q_{t,2}|
		\Big(\Big\|\frac{1}{T_0}\sum_{t=1}^{T_0}\bz_t\bz_t'\Big\|+
		\Big\|\frac{1}{T_0}\sum_{t=1}^{T_0}\bz_t\bz_t'\tilde{u}_t^2\Big\|\Big).	
	\end{split}
	\]
	Note that we use the fact that for any number $a$, $2|a|\leq 1+a^2$. 
	Using the previous argument again, we have with probability over $\mathscr{H}$ at least $1-\mathfrak{C}_2'T_0^{-1}-\pi_{\delta}^{\ttb}-\pi_\gamma-\pi_u^{\ttb}$, with $\P(\cdot|\mathscr{H})$-probability at least $1-\mathfrak{C}_2''T_0^{-1}-\epsilon_{\delta}^{\ttb}-\epsilon_\gamma-\epsilon_u^{\ttb}$, the above is bounded by $\mathfrak{C}_2(\varpi_{\delta}^{\ttb}/\sqrt{T_0}+\varpi_u^{\ttb})$ for some constants $\mathfrak{C}_2$, $\mathfrak{C}_2'$, and $\mathfrak{C}_2''$. Then, the result follows.
\qed

\subsection{Proof of Theorem \ref{thm: exmp dependent case}}

	Throughout this proof, $\mathfrak{C}_1$, $\mathfrak{c}_1$, $\cdots$ denote some constants independent of $T_0$. By assumption, the analysis conditional on $\mathscr{H}$ can be replaced by that conditional on $\bB$.
	
	(1) We first verify the conditions of Theorem A.  Under condition (T4.i),
	by Theorem 3.1 of \citet*{Pham-Tran_1985_SPA}, $\bu_t$ is $\beta$-mixing at an exponential rate $\mathfrak{b}(k):=\mathfrak{b}(k;\mathscr{H})=\exp(-\mathfrak{c}_1k)$ (even conditional on $\mathscr{H}$). Similarly, 
	$\bb_t$ is also $\beta$-mixing at an exponential rate $\exp(-\mathfrak{c}_2k)$.
	
	For condition (TA.ii) of Theorem A, let  $\rho=\rho_{u,1}\wedge\rho_{u,2}$ and $\sigma^2=\E[\zeta_{t,u,1}^2]\vee\E[\zeta_{t,u,2}^2]$ where $\bm{\zeta}_{t,u}=(\zeta_{t,u,1}, \zeta_{t,u,2})'$. Note that
	\[
	\begin{split}
		\bar{\sigma}^2(v)&=
		\max_{1\leq j\leq d}
		\frac{1}{T_0vm}\sum_{k=1}^{m}\V\Big[\sum_{t\in\mathcal{J}_k'}(b_{jt,1}u_{t,1}+b_{jt,2}u_{t,2})\Big|\bB\Big]\\
		&\leq\max_{1\leq j\leq d}
		\frac{2}{T_0vm}\sum_{l=1}^{2}\sum_{k=1}^{m}\Big\{\sum_{t\in\mathcal{J}_k'}
		\frac{\sigma^2}{1-\rho^2}b_{jt,l}^2+
		\sum_{\ell=1}^{v-1}
		\Big(\sum_{t-t'=\ell \atop t,t'\in\mathcal{J}_k'}b_{jt,l}b_{jt',l}\Big)
		\frac{\rho^{\ell}\sigma^2}{1-\rho^2}\Big\}\\
		&\leq \max_{1\leq j\leq d}
		\frac{2}{T_0vm}\sum_{l=1}^{2}\sum_{k=1}^{m}
		\Big\{\sum_{t\in\mathcal{J}_k'}
		\frac{\sigma^2}{1-\rho^2}b_{jt,l}^2+
		2\sum_{\ell=1}^{v-1}\Big(
		\sum_{t\in\mathcal{J}_k'}b_{jt,l}^2\Big)
		\frac{|\rho|^{\ell}\sigma^2}{1-\rho^2}\Big\}.
	\end{split}
	\] 
	It is bounded by $\mathfrak{C}_1 T_0^{-1}$
	with probability over $\mathscr{H}$ at least $1-\mathfrak{c}_3m^{-1}$,
	since by the coupling argument used in the proof of Theorem A and Markov's inequality,
	\[
	\P\Big(\Big|\frac{1}{vm}\sum_{k=1}^{m}\sum_{t\in\mathcal{J}_k'}(b_{jt,l}^2-\E[b_{jt,l}^2])\Big|\geq \bar\eta_1\Big)\leq 
	\frac{1}{\bar\eta_1^2m^2}\sum_{k=1}^{m}\frac{1}{v^2}\E\Big[\Big(
	\sum_{t\in\mathcal{J}_k'}(b_{jt,l}-\E[b_{jt,l}])\Big)^2\Big]+
	m\exp(-\mathfrak{c}_2q)
	\]
	which is bounded by $\mathfrak{c}_4m^{-1}$ for some $\mathfrak{c}_4$ large enough.
	
	On the other hand, for $\psi=3$, we have 
	\[
	\begin{split}
		\E\Big[\sum_{j=1}^{d}\sum_{k=1}^{m}|S_{jk,\diamond}|^\psi\Big|\mathscr{H}\Big]&=
		\E\Big[\sum_{j=1}^{d}\sum_{k=1}^{m}\Big|\sum_{t\in\mathcal{J}_k'}s_{jt}\Big|^{3}\Big|\mathscr{H}\Big]\\
		&\leq 4mdT_0^{-3/2}
		\frac{1}{md}\sum_{j=1}^{d}\sum_{k=1}^{m}
		\sum_{l=1}^{2}\E\Big[\Big|\sum_{t\in\mathcal{J}_k'}b_{jt,l}u_{t,l}\Big|^3|\mathscr{H}\Big]\\
		&\leq 4mdT_0^{-3/2}
		\frac{1}{md}\sum_{j=1}^{d}\sum_{k=1}^{m}
		\sum_{l=1}^{2}\max\Big\{\sum_{t\in\mathcal{J}_k'}(\E[|b_{jt,l}u_{t,l}|^{4}|\mathscr{H}])^{\frac{3}{4}},\\
		&\hspace*{15em}
		\Big(\sum_{t\in\mathcal{J}_k'}(\E[|b_{jt,l}u_{t,l}|^{3}|\mathscr{H}])^{\frac{2}{3}}\Big)^{\frac{3}{2}}\Big\}.
	\end{split}
	\]
	The second line uses H\"{o}lder's inequality, and  the third line uses Rosenthal inequality for strong mixing sequences (see, e.g., Theorem 2 in Section 1.4 of \citet*{Doukhan_2012_Book}) and the fact that $\beta$-mixing implies strong mixing. 
	Again, by the coupling argument and Markov's inequality, with probability over $\mathscr{H}$ at least $1-\mathfrak{c}_5m^{-1}$, the last line is bounded by $\eta_1:=\mathfrak{C}_2m(v/T_0)^{3/2}$.
	
	For condition (TA.iii), note that by Rosenthal inequality,
	\[
	\begin{split}
		\max_{1\leq j\leq d}\E\Big[|S_{j(m+1),\diamond}|^{3}\Big|\mathscr{H}\Big]
		&\leq\max_{1\leq j\leq d}
		4T_0^{-3/2}\sum_{l=1}^{2}\E\Big[\Big|\sum_{t\in\mathcal{J}_{m+1}'}
		b_{jt,l}u_{t,l}\Big|^3|\mathscr{H}\Big]\\
		&\leq\max_{1\leq j\leq d}
		4T_0^{-3/2}\sum_{l=1}^{2}
		\max\Big\{\sum_{t\in\mathcal{J}_{m+1}'}(\E[|b_{jt,l}u_{t,l}|^{4}|\mathscr{H}])^{\frac{3}{4}},\\
		&\hspace*{11em}\Big(\sum_{t\in\mathcal{J}_{m+1}'}(\E[|b_{jt,l}u_{t,l}|^{3}|\mathscr{H}])^{\frac{2}{3}}\Big)^{\frac{3}{2}}\Big\}.
	\end{split}
	\]
	Applying Markov's inequality to $\sum_{t\in\mathcal{J}_{m+1}'}|b_{jt,l}|^3$ and $\sum_{t\in\mathcal{J}_{m+1}'}|b_{jt,l}|^2$, by the moment conditions imposed, we have $\max_{1\leq j\leq d}\E[|S_{j(m+1),\diamond}|^{3}|\mathscr{H}]$ is bounded by
	$\eta_2:=\mathfrak{C}_3(q/T_0)^{3/2}T_0^{\mathfrak{c}_6}$ with probability over $\mathscr{H}$ at least $1-\mathfrak{c}_7T_0^{-2\mathfrak{c}_6}$ for any small $\mathfrak{c}_6>0$.
	
	For condition (TA.iv), note that by H\"{o}lder's inequality and Rosenthal inequality,
	\[
	\begin{split}
		\sum_{k=1}^{m}\E[\|\bS_{k,\Box}\|^3|\mathscr{H}]
		&=\frac{1}{\sqrt{T_0}}\frac{m}{T_0}\frac{1}{m}\sum_{k=1}^{m}
		\E\Big[\Big(\sum_{j=1}^{d}\Big(\sum_{l=1}^{2}\sum_{t\in\mathcal{J}_k}b_{jt,l}u_{t,l}\Big)^2\Big)^{3/2}\Big|\mathscr{H}\Big]\\
		&\leq\frac{4d^{3/2}}{\sqrt{T_0}}\frac{m}{T_0}\frac{1}{m}\sum_{k=1}^{m}
		\frac{1}{d}\sum_{j=1}^{d}\sum_{l=1}^{2}\E\Big[\Big|\sum_{t\in\mathcal{J}_k}b_{jt,l}u_{t,l}\Big|^3\Big|\mathscr{H}\Big]\\
		&\leq\frac{4\mathfrak{C}_4d^{3/2}}{\sqrt{T_0}}\frac{m}{T_0}\frac{1}{m}\sum_{k=1}^{m}
		\frac{1}{d}\sum_{j=1}^{d}\sum_{l=1}^{2}\max\Big\{\sum_{t\in\mathcal{J}_k}(\E[|b_{jt,l}u_{t,l}|^{4}|\mathscr{H}])^{\frac{3}{4}},\\
		&\hspace*{16em}
		\Big(\sum_{t\in\mathcal{J}_k}(\E[|b_{jt,l}u_{t,l}|^{3}|\mathscr{H}])^{\frac{2}{3}}\Big)^{\frac{3}{2}}\Big\}.
	\end{split}
	\]
	Again, by the coupling argument and Markov's inequality, the last term is bounded by $\mathfrak{C}_5m^{-1/2}$ with probability over $\mathscr{H}$ at least $1-\mathfrak{c}_8m^{-1}$.
	
	On the other hand, 
	\[
	\lambda_{\min}(\bSigma_{\Box})
	=\lambda_{\min}\Big(
	\sum_{k=1}^{m}\V[\bS_{k,\Box}|\mathscr{H}]\Big)
	\geq \underline{\eta}_1\lambda_{\min}\Big(\frac{1}{T_0}\sum_{k=1}^{m}\sum_{l=1}^{2}\sum_{t\in\mathcal{J}_k}\bb_{t,l}\bb_{t,l}'\Big)\\
	\]
	Note that the minimum eigenvalue of  $\E[\frac{1}{q}\sum_{l=1}^{2}\sum_{t\in\mathcal{J}_k}\bb_{t,l}\bb_{t,l}']=\sum_{l=1}^{2}\E[\bb_{t,l}\bb_{t,l}']$ is bounded from below by $2\underline{\eta}_2$. Then,
	by the coupling argument, with probability over $\mathscr{H}$ at least $1-\mathfrak{c}_9m^{-1}$, the last term is bounded from below by $\underline{\eta}_1\underline{\eta}_2$. Therefore, we can set $\eta_3=\mathfrak{C}_6m^{-1/2}$.
	The above argument also shows that condition (TA.v) holds with $\eta_4=\underline{\eta}_1\underline{\eta}_2$.
	
	For condition (TA.vi), note that
	\begin{align*}\allowdisplaybreaks
		\bSigma=\;&\V\Big[\frac{1}{\sqrt{T_0}}\Big(\sum_{l=1}^{2}\sum_{k=1}^{m}\sum_{t\in\mathcal{J}_k}\bb_{t,l}u_{t,l}+\sum_{l=1}^{2}\sum_{k=1}^{m+1}\sum_{t\in\mathcal{J}_k'}\bb_{t,l}u_{t,l}\Big)\Big|\mathscr{H}\Big]\\
		=\;&\frac{1}{T_0}\V\Big[\sum_{l=1}^{2}\sum_{k=1}^{m}\sum_{t\in\mathcal{J}_k}\bb_{t,l}u_{t,l}\Big|\mathscr{H}\Big]+\frac{1}{T_0}\V\Big[\sum_{l=1}^{2}\sum_{k=1}^{m+1}\sum_{t\in\mathcal{J}_k'}\bb_{t,l}u_{t,l}\Big|\mathscr{H}\Big]+\\
		&\frac{2}{T_0}\Cov\Big[\sum_{l=1}^{2}\sum_{k=1}^{m}\sum_{t\in\mathcal{J}_k}\bb_{t,l}u_{t,l},\; \sum_{l=1}^{2}\sum_{k=1}^{m+1}\sum_{t\in\mathcal{J}_k'}\bb_{t,l}u_{t,l}\Big|\mathscr{H}\Big]\\
		=\;&\bSigma_{\Box}+\frac{2}{T_0}
		\sum_{k,k'\atop k<k'}
		\Cov\Big[\sum_{l=1}^{2}\sum_{t\in\mathcal{J}_k}\bb_{t,l}u_{t,l},\;
		\sum_{l=1}^{2}\sum_{t\in\mathcal{J}_k}\bb_{t,l}u_{t,l}\Big|\mathscr{H}\Big]+\\
		&\frac{1}{T_0}\V\Big[\sum_{l=1}^{2}\sum_{k=1}^{m+1}\sum_{t\in\mathcal{J}_k'}\bb_{t,l}u_{t,l}\Big|\mathscr{H}\Big]+
		\frac{2}{T_0}\Cov\Big[\sum_{l=1}^{2}\sum_{k=1}^{m}\sum_{t\in\mathcal{J}_k}\bb_{t,l}u_{t,l},\; \sum_{l=1}^{2}\sum_{k=1}^{m+1}\sum_{t\in\mathcal{J}_k'}\bb_{t,l}u_{t,l}\Big|\mathscr{H}\Big]\\
		=\;&\bSigma_{\Box}+2\mathrm{I}+\mathrm{II}+2\mathrm{III}.
	\end{align*}
	We consider a generic $(j,j')$th element in $\bSigma$. For $\mathrm{I}$, the $(j,j')$th element can be bounded as follows: 
	\begin{align*}
		|\mathrm{I}[j,j']|:=\;&\Big|\frac{1}{T_0}\sum_{k,k'\atop k<k'}
		\sum_{t\in\mathcal{J}_k,\atop t'\in\mathcal{J}_{k'}}
		\sum_{l=1}^2\sum_{l'=1}^2b_{jt,l}b_{j't',l'}\Cov[u_{t,l},u_{t',l'}|\mathscr{H}]\Big|\\
		\leq \;&\frac{1}{T_0}\sum_{k,k'\atop k<k'}
		\sum_{t\in\mathcal{J}_k\atop t'\in\mathcal{J}_{k'}}
		\sum_{l=1}^{2}(b_{jt,l}^2+b_{j't',l}^2)|\rho|^{|t-t'|}\sigma^2/(1-\rho^2)\\
		= \;&\frac{\sigma^2}{1-\rho^2}\Big\{\frac{1}{T_0}
		\sum_{k=1}^{m-1}\sum_{t\in\mathcal{J}_k}
		\Big(\sum_{l=1}^{2}b_{jt,l}^2\Big)\sum_{k':k'>k}
		\sum_{t'\in\mathcal{J}_{k'}}
		|\rho|^{|t-t'|}+\\
		&\hspace{3.5em}\frac{1}{T_0}\sum_{k'=2}^{m}\sum_{t'\in\mathcal{J}_{k'}}
		\Big(\sum_{l=1}^{2}b_{j't',l}^2\Big)\sum_{k:k<k'}\sum_{t\in\mathcal{J}_k}|\rho|^{|t-t'|}\Big\}.
	\end{align*}
	Note that by the property of geometric series and the moment condition,
	$$\E\Big[\frac{1}{T_0}\sum_{k=1}^{m-1}\sum_{t\in\mathcal{J}_k}
	\Big(\sum_{l=1}^{2}b_{jt,l}^2\Big)\sum_{k':k'>k}
	\sum_{t'\in\mathcal{J}_{k'}}
	|\rho|^{|t-t'|}\Big]\leq\mathfrak{C}_7|\rho|^v/q.$$
	Also note that by coupling argument, the moment condition, and Markov's inequality,
	\[
	\P\Big(\frac{1}{T_0}\sum_{k=1}^{m-1}\sum_{t\in\mathcal{J}_k}
	\Big(\sum_{l=1}^{2}(b_{jt,l}^2-\E[b_{jt,l}^2])|\rho|^{t-(k-1)q+v+1}\Big)\geq |\rho|^v/q\Big)\leq \mathfrak{c}_{10}m^{-1}.
	\]
	The second term in the decomposition of $\mathrm{I}$ can be treated similarly.
	Therefore, with probability over $\mathscr{H}$ at least $1-\mathfrak{c}_{11}m^{-1}$, $\|\mathrm{I}\|\leq \mathfrak{C}_{8}|\rho|^v/q$.
	
	Regarding $\mathrm{II}$, the covariance between blocks can be analyzed  exactly the same way as above, and the analysis of the sum of variance of each blocks is similar to that of $\bar{\sigma}^2(v)$. It follows that with probability over $\mathscr{H}$ at least $1-\mathfrak{c}_{12}m^{-1}$, $\|\mathrm{II}\|\leq \mathfrak{C}_9(v/q+m^{-1}+|\rho|^q/q)$.
	The analysis of $\mathrm{III}$ is similar to that of $\mathrm{I}$, yielding that with probability over $\mathscr{H}$ at least $1-\mathfrak{c}_{13}m^{-1}$,  $\|\mathrm{III}\|\leq \mathfrak{C}_{10}q^{-1}$.
	Using these results, we conclude that $\|\bSigma-\bSigma_{\Box}\|\leq \mathfrak{C}_{11}(v/q+m^{-1})=:\eta_5$ with probability over $\mathscr{H}$ at least $1-\mathfrak{c}_{14}m^{-1}$. Then, condition (TA.vi) holds. Also note that when $q/v$ and $m$ are sufficiently large, $\eta_5\leq 0.25$. As specified below, this is true when $T_0$ is large enough.
	
	Finally, to satisfy condition (TA.vii), we can take $m=\mathfrak{C}_{12} T_0^{2/5}$, $v=\mathfrak{C}_{13}T_0^{\mathfrak{c}_{15}}$, and let $\mathfrak{c}_6$ and  $\mathfrak{c}_{15}$ be sufficiently small. Then, we can take $\eta_6=\mathfrak{C}_{14} T_0^{-3/20-\mathfrak{c}_{16}}$ for sufficiently small $\mathfrak{c}_{16}$. The proof for the first part is complete.
	
	(2) Next, we consider Theorem 2. Condition (T2.i) holds by the argument given in part (1) of this proof. As in the proof of Theorem \ref{thm: exmp independent case}, conditions (T2.ii) and (T2.iii) of Theorem 2 can be verified using the basic inequality
	\[
	\lambda_{\min}(\widehat{\bQ})\|\bdelta\|^2\leq
	\bdelta\widehat{\bQ}\bdelta\leq 2\bG'\bdelta\leq  
	2\|\bG\|\|\bdelta\|.
	\]
	It has been shown in part (1) that the minimum eigenvalue of $\widehat{\bQ}$ is bounded from below by some constant with probability over $\mathscr{H}$ at least $1-\mathfrak{c}_9m^{-1}$. 
	On the other hand, let $\sigma_{\max}^2$ be the largest diagonal element of $\bSigma$. Note that for Gaussian random variables,
	\[
	\P\Big(\|\bG\|\geq\sqrt{2d\log T_0}\sigma_{\max}\Big|\mathscr{H}\Big)\leq 2d\exp\Big(-\frac{2\sigma_{\max}^2\log T_0}{2\sigma_{\max}^2}\Big)
	\leq 2d/T_0.
	\]
	By the argument given in part (1), we have with probability over $\mathscr{H}$ at least $1-\mathfrak{c}_9m^{-1}-\mathfrak{c}_{14}m^{-1}$, $\sigma_{\max}^2$ is bounded by some constant. Then, conditions (T2.ii) and (T2.iii) hold.
	
	For condition (T2.iv), since 
	\[
	\tr\Big[(\bSigma^{-1/2}\widehat{\bSigma}\bSigma^{-1/2}-\bI_d)^2\Big]
	\leq d\lambda_{\min}(\bSigma)^{-2}\|\widehat{\bSigma}-\bSigma\|^2.
	\]
	The result follows by the lower bound on the minimum eigenvalue of $\bSigma$ and the assumption imposed in the theorem.
\qed

\subsection{Proof of Theorem \ref{thm: exmp cointegration}}

	Throughout this proof, $\mathfrak{C}_1$, $\mathfrak{c}_1, \cdots$ denote some constants independent of $T_0$. 
	First note that given the assumptions on $\{u_{t,l}\}$ and $\{v_{jt,l}\}$, conditional on $\bB$, $\bp_T$ is independent of the pre-treatment data. Therefore, when we verify the conditions of Theorems 1 and 2, the analysis conditional on $\mathscr{H}$ is the same as that conditional on $\bB$ and $\bC$ only. 
	
	(1) We consider Theorem 1 first. Condition (T1.i) directly follows from the assumptions on $\bq_t$. For condition (T1.ii), as in the proof of Theorem \ref{thm: exmp independent case}, we only need to bound 
	$\frac{1}{T_0^{3/2}}\sum_{t=1}^{T_0}\sum_{l=1}^{M}
	\|\check{\bz}_{t,l}\|^3$ and $\|\bSigma^{-1}\|$. First consider $\frac{1}{T_0}\sum_{t=1}^{T_0}\sum_{l=1}^{M}
	\|\check{\bz}_{t,l}\|^3$.
	Note that  
	\[
	\frac{1}{T_0}\sum_{t=1}^{T_0}\sum_{l=1}^{M}\|\check{\bz}_{t,l}\|^3\leq\frac{\sqrt{d}}{T_0}\sum_{t=1}^{T_0}\sum_{l=1}^{M}\sum_{j=1}^{d}|\check{z}_{jt,l}|^3,
	\] 
	where $\check{z}_{jt,l}$ is the $j$th element of $\check{\bz}_{t,l}$. 
	
	For the stationary components, i.e., $\{c_{kt,l}\}$, since the third moment is bounded, by Markov's inequality, with probability over $\mathscr{H}$ at least $1-\mathfrak{c}_1T_0^{-1}$, $\frac{1}{T_0}\sum_{l=1}^{M}\sum_{j=J+1}^{d}\sum_{t=1}^{T_0}
	|\tilde{z}_{jt,l}|^3\leq\mathfrak{C}_1$.
	
	On the other hand, note that the non-stationary components $\check{\bz}_{t,\triangleleft,l}$ can be understood as a multivariate partial sum process indexed by $t$. Write $\check{\bZ}_{\triangleleft}(t)=(\check{\bz}_{t,\triangleleft,1}', \cdots, \check{\bz}_{t,\triangleleft,M}')'$.  By strong approximation of partial sum processes, for any $0<\nu<1/2-1/\psi$,
	\[
	\P\Big(\max_{1\leq t\leq T_0}
	\|\check{\bZ}_{\triangleleft}(t)-\bG(t/T_0)\|\geq T_0^{-\frac{1}{2}+\frac{1}{\psi}+\nu} \Big)\leq \mathfrak{C}_2T_0^{-\psi\nu},
	\]
	where $\bG(\cdot)=(\bG_1(\cdot)', \cdots, \bG_M(\cdot)')'$ is a $(JM)$-dimensional Brownian motion on $[0,1]$ with the variance $\E[\bv_t\bv_t']$. On the other hand, it is well known that for each $1\leq j\leq J$, $1\leq l\leq M$, for any $m>0$,
	\[
	\P(\max_{0\leq r\leq 1}|G_{j,l}(r)|>m)\leq 2\P(\max_{0\leq r\leq 1}G_{j,l}(r)>m)=2\P(|G_{j,l}(1)|>m),
	\]
	where $G_{j,l}(\cdot)$ is the $j$th element of $\bG_l(\cdot)$. 
	Using the tail bound for Gaussian distributions, we can set $m=\sqrt{2\log (2JMT_0)\sigma_{\max}^2}$ where $\sigma_{\max}^2$ is the largest variance of $\{v_{jt,l}:1\leq j\leq J, 1\leq l\leq M\}$, which leads to 
	$\max_{0\leq r\leq 1}|G_{j,l}(r)|\leq m$ 
	with probability over $\mathscr{H}$ at least $1-(JMT_0)^{-1}$. Therefore, we conclude that 
	\[
	\frac{1}{T_0}\sum_{t=1}^{T_0}\sum_{l=1}^{M}\sum_{j=1}^{J}|\check{z}_{jt,l}|^3
	\leq \mathfrak{C}_3\Big(\sqrt{2\log (2JMT_0)\sigma^2_{\max}}+\mathfrak{C}_2T_0^{\frac{1}{\psi}
		-\frac{1}{2}+\nu}\Big)^3
	\]
	with probability over $\mathscr{H}$ at least $1-T_0^{-1}-\mathfrak{c}_2T_0^{-\psi\nu}$.

	Finally, we consider $\bSigma$. By assumption in the theorem, $\lambda_{\min}(\bSigma)\geq\underline{\eta}_1\widehat{\bQ}$. Partition $\widehat{\bQ}$ into
	\[
	\begin{bmatrix}
		\widehat{\bQ}_{11}&\widehat{\bQ}_{12}\\
		\widehat{\bQ}_{12}&\widehat{\bQ}_{22}
	\end{bmatrix}.
	\]
	$\widehat{\bQ}_{11}\in\mathbb{R}^{J\times J}$ corresponds to the Gram of the non-stationary component:
	\[
	\widehat{\bQ}_{11}=\sum_{l=1}^{M}\Big(\frac{1}{T}\sum_{t=1}^{T_0}
	\check{\bz}_{t,\triangleleft,l}\check{\bz}_{t,\triangleleft,l}'\Big).
	\]
	Again, by strong approximation used previously, with probability over $\mathscr{H}$ at least $1-\mathfrak{c}_2T_0^{-\psi\nu}-T_0^{-1}$,
	$$
	\bigg\|\frac{1}{T_0}\sum_{t=1}^{T_0}\check{\bz}_{t,\triangleleft,l}\check{\bz}_{t,\triangleleft,l}'-\frac{1}{T_0}\sum_{t=1}^{T_0}\bG_l\Big(\frac{t}{T_0}\Big)\bG_l\Big(\frac{t}{T_0}\Big)'\bigg\|\leq 
	\mathfrak{C}_4T_0^{\frac{1}{\psi}-\frac{1}{2}+\nu}\sqrt{\log T_0}.
	$$ 
	Then, by the condition in the theorem, with probability over $\mathscr{H}$ at least $1-\mathfrak{c}_2T_0^{-\psi\nu}-T_0^{-1}-\pi_{Q,1}$,  $\lambda_{\min}(\widehat{\bQ}_{11})\geq (\log T_0)^{-\mathfrak{c}_Q}/2$ for $T_0$ large enough so that $\mathfrak{C}_4T_0^{\frac{1}{\psi}-\frac{1}{2}+\nu}\sqrt{\log T_0}\leq (\log T_0)^{-\mathfrak{c}_Q}/2$.
	
	Note that $\lambda_{\min}(\sum_{l=1}^{M}\E[\check{\bz}_{t,\triangleright,l}\check{\bz}_{t,\triangleright,l}])$ is bounded away from zero. Then,
	by Markov's inequality and the argument used in the proof of Theorem \ref{thm: exmp independent case},
	$\lambda_{\min}(\widehat{\bQ}_{22})\geq \mathfrak{C}_5$ with probability at least $1-\mathfrak{c}_3T_0^{-1}$. 
	
	For off-diagonal blocks, note that 
	\[
	\widehat{\bQ}_{12}=\sum_{l=1}^{M}\Big\{\Big(\frac{1}{T_0}\sum_{t=1}^{T_0}\check{\bz}_{t,\triangleleft,l}(\E[\check{\bz}_{t,\triangleright,l}])'\Big)+
	\Big(\frac{1}{T_0}\sum_{t=1}^{T_0}\check{\bz}_{t,\triangleleft,l}(\check{\bz}_{t,\triangleright,l}-\E[\check{\bz}_{t,\triangleright,l}])'\Big)\Big\}.
	\]
	The first term, by assumption, is zero. For
	the second term, by the condition in the theorem, $\|\frac{1}{T_0}\sum_{t=1}^{T_0}\check{\bz}_{t,\triangleleft,l}(\check{\bz}_{t,\triangleright,l}-\E[\check{\bz}_{t,\triangleright,l}])'\|\leq
	\mathfrak{C}_6T_0^{-1/2+\nu_Q}$ with probability over $\mathscr{H}$ at least $1-\pi_{Q,2}$. Thus, we have with probability over $\mathscr{H}$ at least $1-\mathfrak{c}_2T_0^{-\psi\nu}-\mathfrak{c}_4T_0^{-1}-\pi_{Q,1}-\pi_{Q,2}$, $\lambda_{\min}(\widehat{\bQ})\geq \mathfrak{C}_7(\log T_0)^{-\mathfrak{c}_Q}/2$.
	Therefore, we can take $\epsilon_{\gamma}=\mathfrak{C}_\epsilon (\log T_0)^{\frac{3}{2}(1+\mathfrak{c}_Q)}T_0^{-1/2}$ for some non-negative finite constant $\mathfrak{C}_\epsilon$.

	(ii) Next, we consider Theorem 2. Condition (T2.i) holds by part (1). As in the proof of Theorem \ref{thm: exmp independent case}, condition (T2.ii) can be established using the results given in the previous step, the Gaussian tail bound, and the basic inequality 
	\[
	\lambda_{\min}(\widehat{\bQ})\|\bdelta\|^2\leq
	\bdelta\widehat{\bQ}\bdelta\leq
	2\bG'\bdelta\leq
	2\|\bG\|\|\bdelta\|.
	\]
	Recall that $\lambda_{\min}(\widehat{\bQ})\geq\mathfrak{C}_7(\log T_0)^{-\mathfrak{c}_Q}$ with probability over $\mathscr{H}$ at least $1-\mathfrak{c}_2T_0^{-\psi\nu}-\mathfrak{c}_4T_0^{-1}-\pi_{Q,1}-\pi_{Q,2}$. On the other hand, by a similar argument, with probability over $\mathscr{H}$ at least $1-\mathfrak{c}_2T_0^{-\psi\nu}-\mathfrak{c}_4T_0^{-1}-\pi_{Q,1}-\pi_{Q,2}$,  $\lambda_{\max}(\widehat{\bQ})\leq (\log T_0)^{\mathfrak{c}_Q}$. Then, $\varpi_{\delta}^{\ttb}=\mathfrak{C}_8(\log T_0)^{2\mathfrak{c}_Q+0.5}$,
	$\pi_\delta^{\ttb}=\mathfrak{c}_2T_0^{-\psi\nu}-\mathfrak{c}_4T_0^{-1}-\pi_{Q,1}-\pi_{Q,2}$, and $\epsilon_{\delta}^{\ttb}=\mathfrak{C}_9T_0^{-1}$. Condition (T2.iii) holds by the construction of the thresholding rule and condition (SA-5.v). Finally, for condition (T2.iv), note that
	\[
	\tr\Big[(\bSigma^{-1/2}\widehat{\bSigma}\bSigma^{-1/2}-\bI_d)^2\Big]
	\leq d\lambda_{\min}(\bSigma)^{-2}\|\widehat{\bSigma}-\bSigma\|^2.
	\]
	Since $\lambda_{\min}(\bSigma)\geq \underline{\eta}_1\lambda_{\min}(\widehat{\bQ})\geq\underline{\eta}_1\mathfrak{C}_7(\log T_0)^{-\mathfrak{c}_Q}$ with probability over $\mathscr{H}$ at least $1-\pi_{\delta}^{\ttb}$,  
	the result follows. 
\qed

\section{Additional Simulation Evidence}\label{SA sec: simulation}

The data generating processes have been described in Section 6.1 of the main paper. We consider models with misspecification error and make use of different methods to estimate the conditional mean, variance and quantiles of $u_t$ given $\mathscr{H}$ whenever needed. Specifically, Tables \ref{table:simul, mis, p0} and \ref{table:simul, mis, p2} are based on zero-order (``constant'') and second-order (``quadratic'') polynomial regression methods. The proposed prediction intervals generally perform well with high coverage probability, though they are very conservative in several cases. Note that in models with $\rho=1$, conditional coverage is more difficult to achieve, since by construction of the evaluation points, we introduce potentially large ``shocks'' on the control outcomes and thus on the out-of-sample error $e_T$ on purpose. As expected, the constant regression fails to correct the misspecification error, thus leading to very poor coverage. Also, compared with the results based on linear regression reported in the main paper, the prediction intervals based on quadratic regression do not perform well, probably due to the overfitting issue. 

For comparison, we also include two other prediction intervals: ``PERM" denotes the prediction interval based on the cross-sectional permutation
method proposed in \cite{Abadie-Diamond-Hainmueller_2010_JASA}, and ``CONF" denotes the prediction interval based on the conformal method developed in 
\cite{Chernozhukov-Wuthrich-Zhu_2021_JASA}. As expected, the prediction interval ``PERM" provides much lower actual coverage probability than the nominal level, since it relies on a cross-sectional permutation
and does not apply to the causal inference framework considered in this paper. In contrast, the conformal prediction interval developed in \cite{Chernozhukov-Wuthrich-Zhu_2021_JASA} performs better, though its
actual coverage probability is still lower than the target nominal level in general.

\section{Empirical Example: 1990 German Reunification}\label{SA sec: empirical example}

In this example, we analyze the economic impact on West Germany of 1990 German reunification \citep[see][for more details]{Abadie_2021_JEL}. The key variable of interest is real per capita GDP of West Germany. We first consider the raw data of per capita GDP. Figure \ref{fig:Germany, level}(a) shows the per capita GDP of the synthetic West Germany (dashed blue) and the actual West Germany (solid black). The synthetic control prediction is able to closely approximate the observed trajectory of per capita GDP of West Germany prior to the treatment, which can be expected since the data are non-stationary and may contain (deterministic or stochastic) common trends. After 1990, the synthetic West Germany is above the actual one, suggesting a negative shock on West Germany after reunification. Figure \ref{fig:Germany, level}(b) adds a $95\%$ conservative prediction interval that takes into account the in-sample uncertainty due to the estimated SC weights, and  we add the uncertainty associated with the out-of-sample error $e_T$ in Figures \ref{fig:Germany, level}(c)-(e).  The constructed PIs for the counterfactual outcome of West Germany are not always separated from the observed sequence. To further assess the robustness of the result, we present the sensitivity analysis of the effect in 1993 in Figure \ref{fig:Germany, level}(f). The constructed PIs cover the observed per capita GDP of West Germany, unless we shrink the estimated (conditional) standard deviation of $e_T$ by a factor of $0.25$.

Then, we apply our methods to the (log) GDP growth rate time series. The raw data are transformed by taking the (log) difference operator. Figure \ref{fig:Germany, growth rate}(a) shows that the growth rate of per capita GDP of the synthetic West Germany is above that of the actual West Germany after 1990, suggesting a negative economic shock on West Germany after reunification. The PIs in Figures \ref{fig:Germany, growth rate}(c)-(e) cover the observed sequence for most post-treatment periods, which do not support statistically significant (negative) effects of reunification on West Germany. Figure \ref{fig:Germany, growth rate}(f) shows the sensitivity analysis of the effect in 1993. We can see that, though the PI constructed based on approach 1 in Figure \ref{fig:Germany, growth rate}(c) appears to suggest a significant effect, this result is not very robust. The corresponding PI covers the observed GDP growth rate of West Germany if we inflate the estimated (conditional) standard deviation of $e_T$ by a factor of $1.5$.

\section{Pseudo-True Values of SC Weights}\label{SA sec: pseudo true values}

Recall that in the main paper we define
\[
\bbeta_0 := (\bw_0', \br_0')' = \underset{\bw\in\mathcal{W},\, \br\in\mathcal{R}}{\arg\min}\; \E[(\bA-\bB\bw-\bC\br)'(\bA-\bB\bw-\bC\br)|\mathscr{H}],
\]
which is the estimand of the SC estimator $\widehat{\bbeta}$. This definition emphasizes the fact that the prediction is conditional on $\mathscr{H}$ and the randomness related to $\mathscr{H}$ is retained in $\bbeta_0$. In this section we consider two examples to give a better understanding of $\bbeta_0$.

\subsection{Linear Factor Model}

Consider a linear factor model with one common factor $\mu_t$:
\[
y_{it}=\lambda_t\mu_i + \varepsilon_{it}, \qquad 1\leq i\leq N+1,\; 1\leq t\leq T.
\]
For simplicity, assume that $\lambda_t$ is i.i.d. over $1\leq t\leq T$, $\varepsilon_{it}$ is i.i.d. across $1\leq i\leq N+1$ and over $1\leq t\leq T$, and $\{\lambda_t:1\leq t\leq T\}$ and $\{\varepsilon_{it}:1\leq i\leq N+1, 1\leq t\leq T\}$ are independent of each other. In addition, we assume that $\E[\varepsilon_{it}]=0$ and the factor loadings $\{\mu_i\}_{i=1}^{N+1}$ are fixed, only the pre-intervention outcomes are used in the SC estimation, and no additional covariates are specified in $\bC$. Then, $\mathscr{H}=\{(Y_{2t}(0), \cdots, Y_{(N+1)t}(0)): 1\leq t\leq T\}$. Accordingly, $\bw_0$ is given by the following expression:
\[
\bw_0=\underset{\bw\in\mathcal{W}}{\argmin}\; \frac{1}{T_0}\sum_{t=1}^{T_0}\E\Big[\Big((\mu_1-\bmu_c'\bw)\lambda_t+(\varepsilon_{1t}-\bm{\varepsilon}_{t,c}'\bw)\Big)^2\Big|\mathscr{H}\Big],
\]
where $\bmu_c=(\mu_2, \cdots, \mu_{N+1})'$ and $\bm{\varepsilon}_{t,c}=(\varepsilon_{2t}, \cdots, \varepsilon_{(N+1)t})'$, assuming the expectation (exists and) is finite.
Define $\mathscr{H}_t=\{Y_{2t}(0), \cdots, Y_{(N+1)t}(0)\}$. Then, $\bw_0$ can be further written as
\begin{align*}
\bw_0=\underset{\bw\in\mathcal{W}}{\argmin}\; \Big\{
&\frac{1}{T_0}\sum_{t=1}^{T_0}\E[\lambda_t^2|\mathscr{H}_t](\mu_1-\bmu_c'\bw)^2 + \E[\varepsilon_{11}^2]+\bw'\Big(\frac{1}{T_0}\sum_{t=1}^{T_0}\E[\bm{\varepsilon_{t,c}}\bm{\varepsilon_{t,c}}'|\mathscr{H}_t]\Big)\bw\\
&\qquad - 2(\mu_1-\bmu_c'\bw)\Big(\frac{1}{T_0}\sum_{t=1}^{T_0}\E[
\lambda_t\bm{\varepsilon}_{t,c}'|\mathscr{H}_t]\Big)\bw
\Big\},
\end{align*}
assuming these expectations are finite. Given our assumptions on $\lambda_t$ and $\varepsilon_{it}$, we expect that 
$$\frac{1}{T_0}\sum_{t=1}^{T_0}\E[\lambda_t^2|\mathscr{H}_t] \approx \E[\lambda_t^2], \quad
\frac{1}{T_0}\sum_{t=1}^{T_0}\E[\bm{\varepsilon}_{t,c}\bm{\varepsilon}_{t,c}'|\mathscr{H}_t] \approx \E[\varepsilon_{1t}^2]\bI_N, \quad\text{and}\quad
\frac{1}{T_0}\sum_{t=1}^{T_0}\E[\lambda_t\bm{\varepsilon}_{t,c}'|\mathscr{H}_t]\approx \bm{0}
$$ 
with high probability when $T_0$ is large, which can be shown under additional moment conditions.
As a consequence, our expression for $\bw_0$ is similar to that in \cite{Ferman-Pinto_2021_wp}.

\subsection{AR(1) Model}

Consider a simple autoregressive model: for $\rho\in(0,1)$,
\[
y_{it}=\rho y_{i(t-1)}+\varepsilon_{it}, \qquad 1\leq i\leq N+1, 1\leq t\leq T,
\]
where $\varepsilon_{it}$ is i.i.d. over $1\leq t\leq T$.
Only the pre-intervention outcomes are used in the SC estimation, and no additional covariates are specified in $\bC$. Then, $\mathscr{H}=\{(Y_{2t}(0), \cdots, Y_{(N+1)t}(0)): 1\leq t\leq T\}$. Given this setup, $\bw_0$ can be written as
\[
\bw_0=\underset{\bw\in\mathcal{W}}{\argmin}\; \frac{1}{T_0}\sum_{t=1}^{T_0}\E\Big[(y_{it}-\bY_{t,c}'\bw)^2\Big|\mathscr{H}_1, \ldots,\mathscr{H}_{T_0}\Big],
\]
where $\bY_{t,c}=(Y_{2t}(0), \cdots, Y_{(N+1)t}(0))'$ and $\mathscr{H}_{t}=\{Y_{2t}(0), \cdots, Y_{(N+1)t}(0)\}$.
In other words, the information in $\mathscr{H}$ that is related to periods later than $T_0$ does not affect the definition of $\bw_0$ and thus can be dropped from $\mathscr{H}$.

\bibliography{CFT_2021_JASA--bib}
\bibliographystyle{jasa}


\clearpage
\begin{table}[h]
	\small
	\begin{center}
	\caption{Simulation, Misspecification Error, Constant Regression Methods}
	\label{table:simul, mis, p0}
		\setlength{\tabcolsep}{8pt}
		\renewcommand{\arraystretch}{1.07}
		\resizebox{\textwidth}{.8\height}{\input{input/simulation/Table_Main3.txt}}
	\end{center}
		\smallskip
		
	\footnotesize\textit{Notes}.
	Conditional mean, variance and quantiles of $u_t$ are estimated based on constant regression methods.
	CP =  coverage probability, AL = average length. 
	``M1": prediction interval for $Y_{1T}(0)$ based on the Gaussian concentration inequality with $90\%$ nominal coverage probability; ``M1-S": the same as ``M1'', but the estimated standard deviation is doubled in the construction; 
	``M2": prediction interval for $Y_{1T}(0)$ based on the location-scale model with $90\%$ nominal coverage probability; ``M3": prediction interval for $Y_{1T}(0)$ based on quantile regression with $90\%$ nominal coverage probability; 
	``PERM": prediction interval for $Y_{1T}(0)$ based on the permutation method proposed in \citet*{Abadie-Diamond-Hainmueller_2010_JASA} with $90\%$ nominal coverage probability;
	``CONF'' prediction interval for $Y_{1T}(0)$ based on the conformal method developed in \citet*{Chernozhukov-Wuthrich-Zhu_2021_JASA} with $90\%$ nominal coverage probability.
\end{table}

\clearpage
\begin{table}[h]
	\small
	\begin{center}
	\caption{Simulation, Misspecification Error, Quadratic Regression Methods}
	\label{table:simul, mis, p2}
		\setlength{\tabcolsep}{8pt}
		\renewcommand{\arraystretch}{1.07}
		\resizebox{\textwidth}{.8\height}{\input{input/simulation/Table_Main4.txt}}
	\end{center}
		\smallskip
		
	\footnotesize\textit{Notes}.
	Conditional mean, variance and quantiles of $u_t$ are estimated based on quadratic regression methods.
	CP =  coverage probability, AL = average length. 
	``M1": prediction interval for $Y_{1T}(0)$ based on the Gaussian concentration inequality with $90\%$ nominal coverage probability; ``M1-S": the same as ``M1'', but the estimated standard deviation is doubled in the construction; 
	``M2": prediction interval for $Y_{1T}(0)$ based on the location-scale model with $90\%$ nominal coverage probability; ``M3": prediction interval for $Y_{1T}(0)$ based on quantile regression with $90\%$ nominal coverage probability; 
	``PERM": prediction interval for $Y_{1T}(0)$ based on the permutation method proposed in \citet*{Abadie-Diamond-Hainmueller_2010_JASA} with $90\%$ nominal coverage probability;
	``CONF'' prediction interval for $Y_{1T}(0)$ based on the conformal method developed in \citet*{Chernozhukov-Wuthrich-Zhu_2021_JASA} with $90\%$ nominal coverage probability.
\end{table}

\clearpage
\begin{figure}
	\begin{center}\caption{1990 German Reunification: GDP Per Capita.\label{fig:Germany, level}}
		\vspace{-.1in}
		\begin{subfigure}{0.48\textwidth}
			\includegraphics[width=\textwidth]{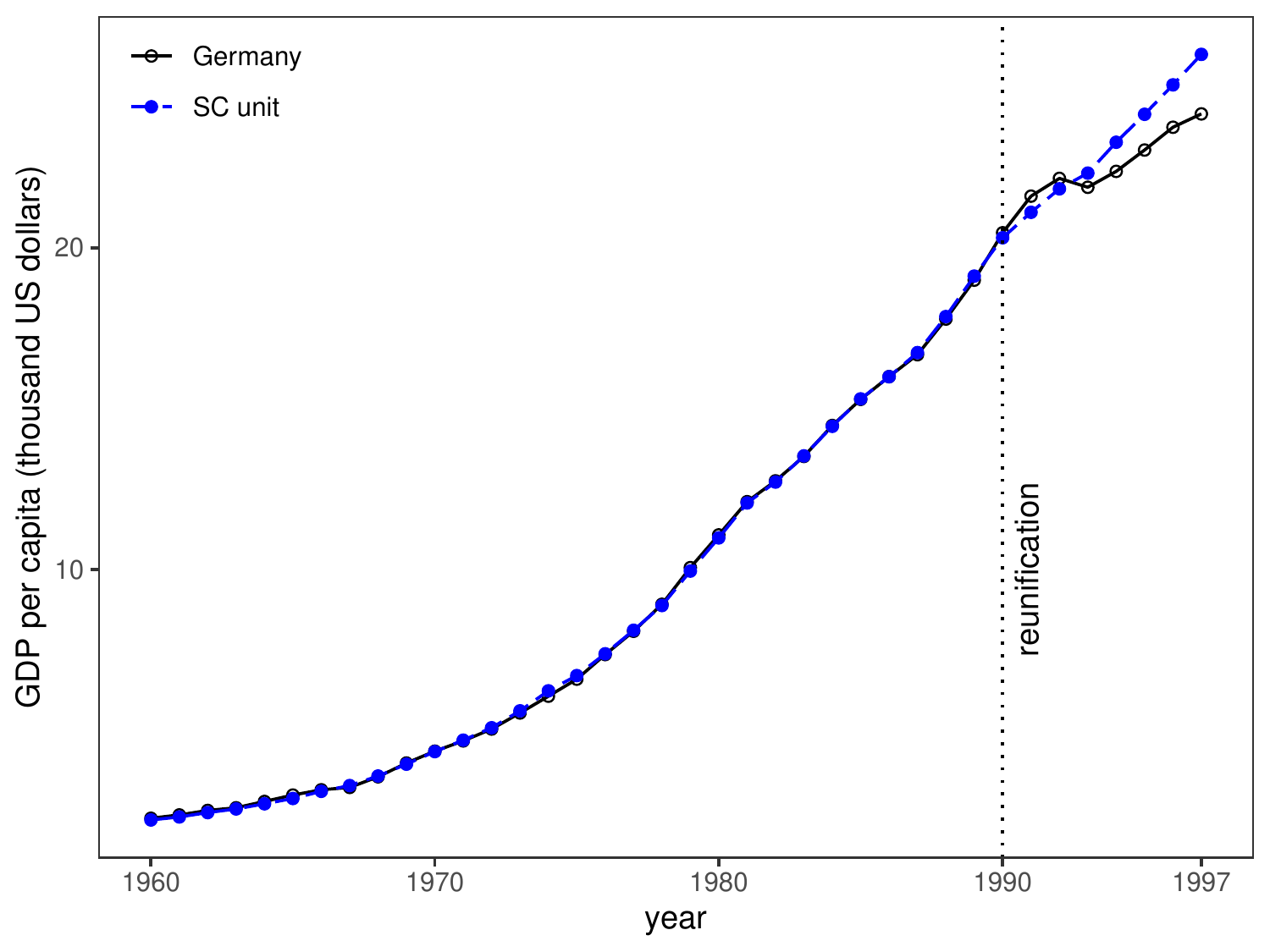}
			\caption{Synthetic West Germany}
		\end{subfigure}
		\begin{subfigure}{0.48\textwidth}
			\includegraphics[width=\textwidth]{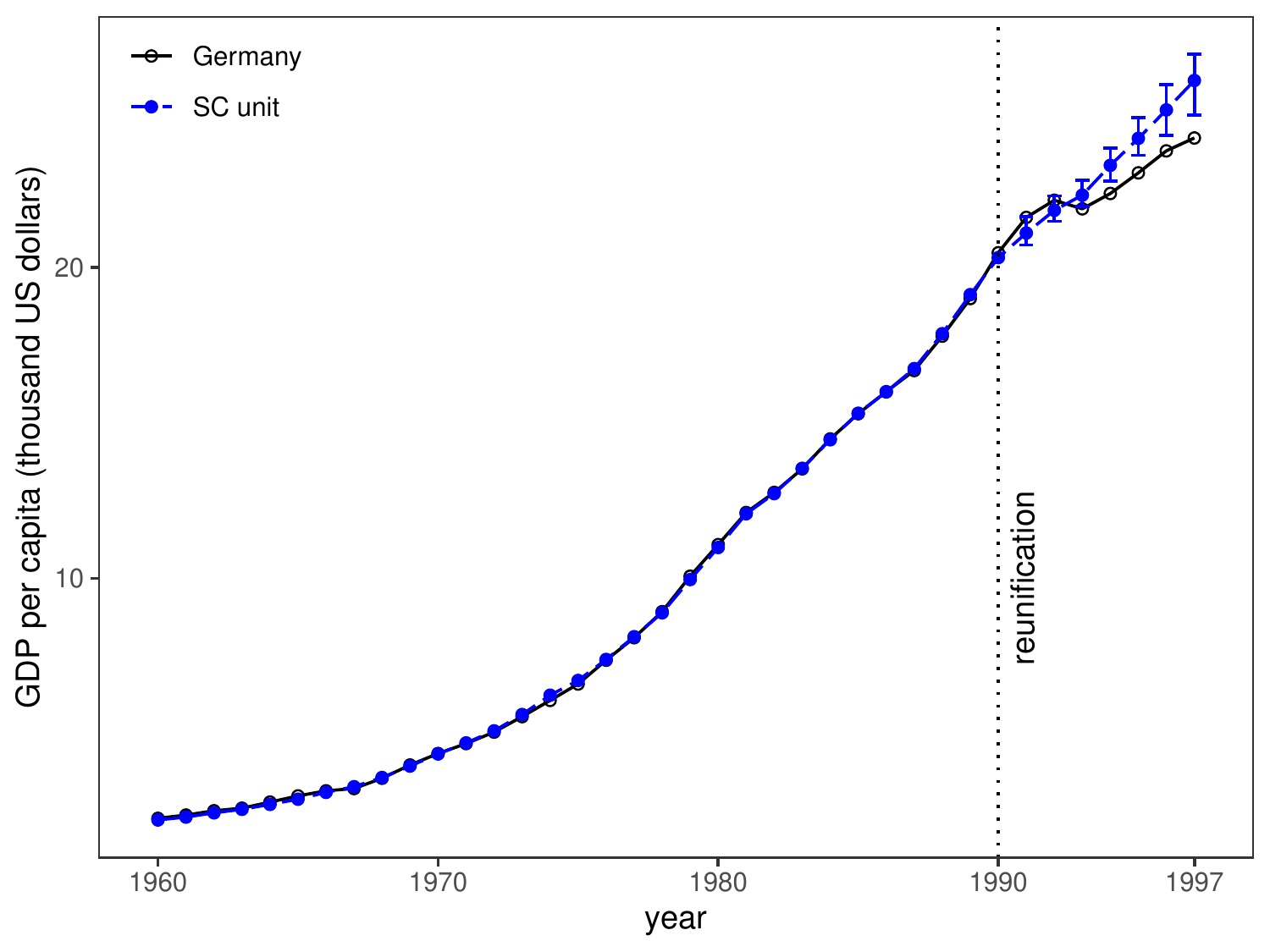}
			\caption{Prediction Interval for $\bx_T'\bw_0$}
		\end{subfigure}\\
		\begin{subfigure}{0.48\textwidth}
			\includegraphics[width=\textwidth]{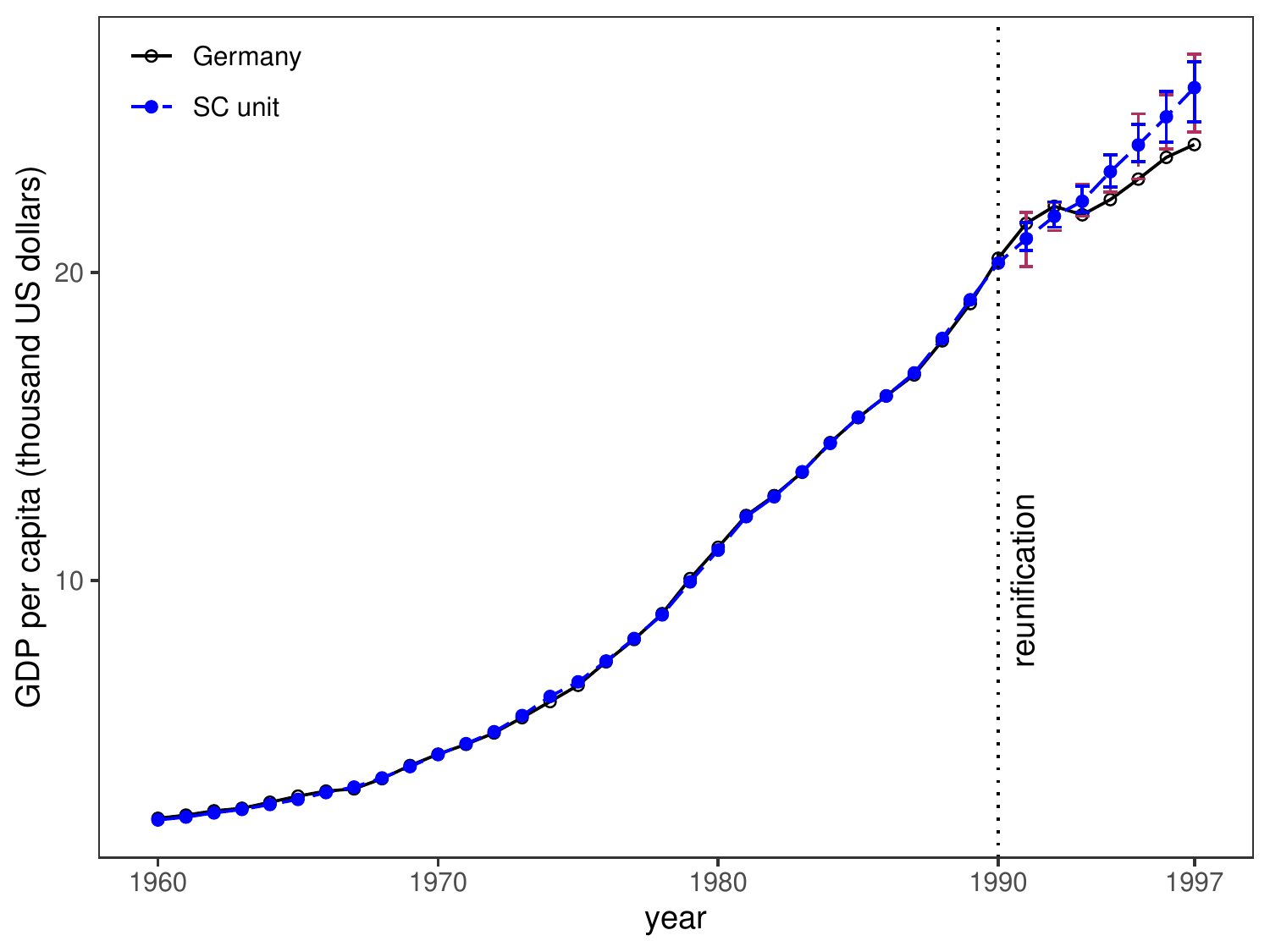}
			\caption{Prediction Interval for $Y_{1T}(0)$, approach 1}
		\end{subfigure}
		\begin{subfigure}{0.48\textwidth}
			\includegraphics[width=\textwidth]{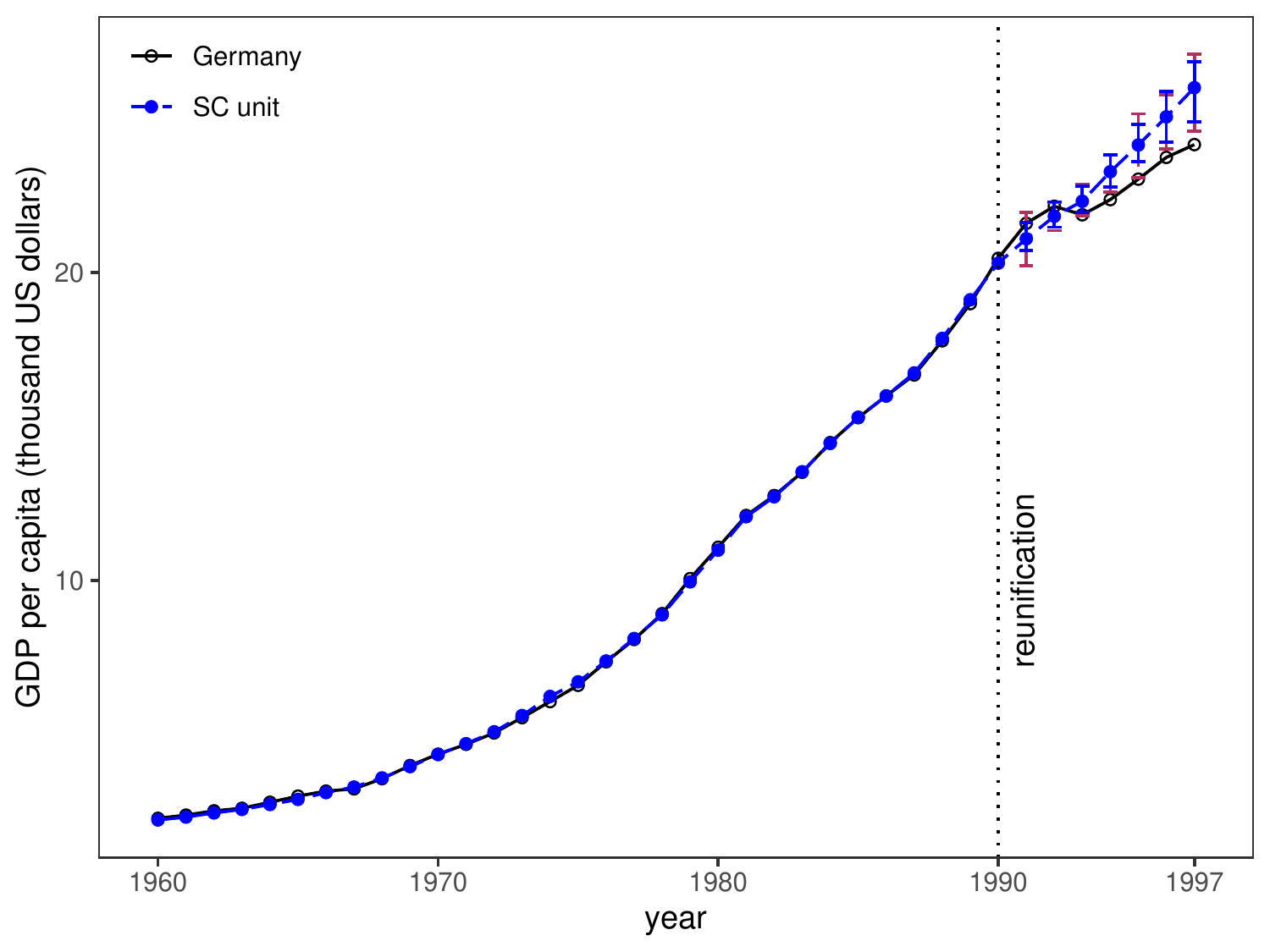}
			\caption{Prediction Interval for $Y_{1T}(0)$, approach 2}
		\end{subfigure}\\
		\begin{subfigure}{0.48\textwidth}
			\includegraphics[width=\textwidth]{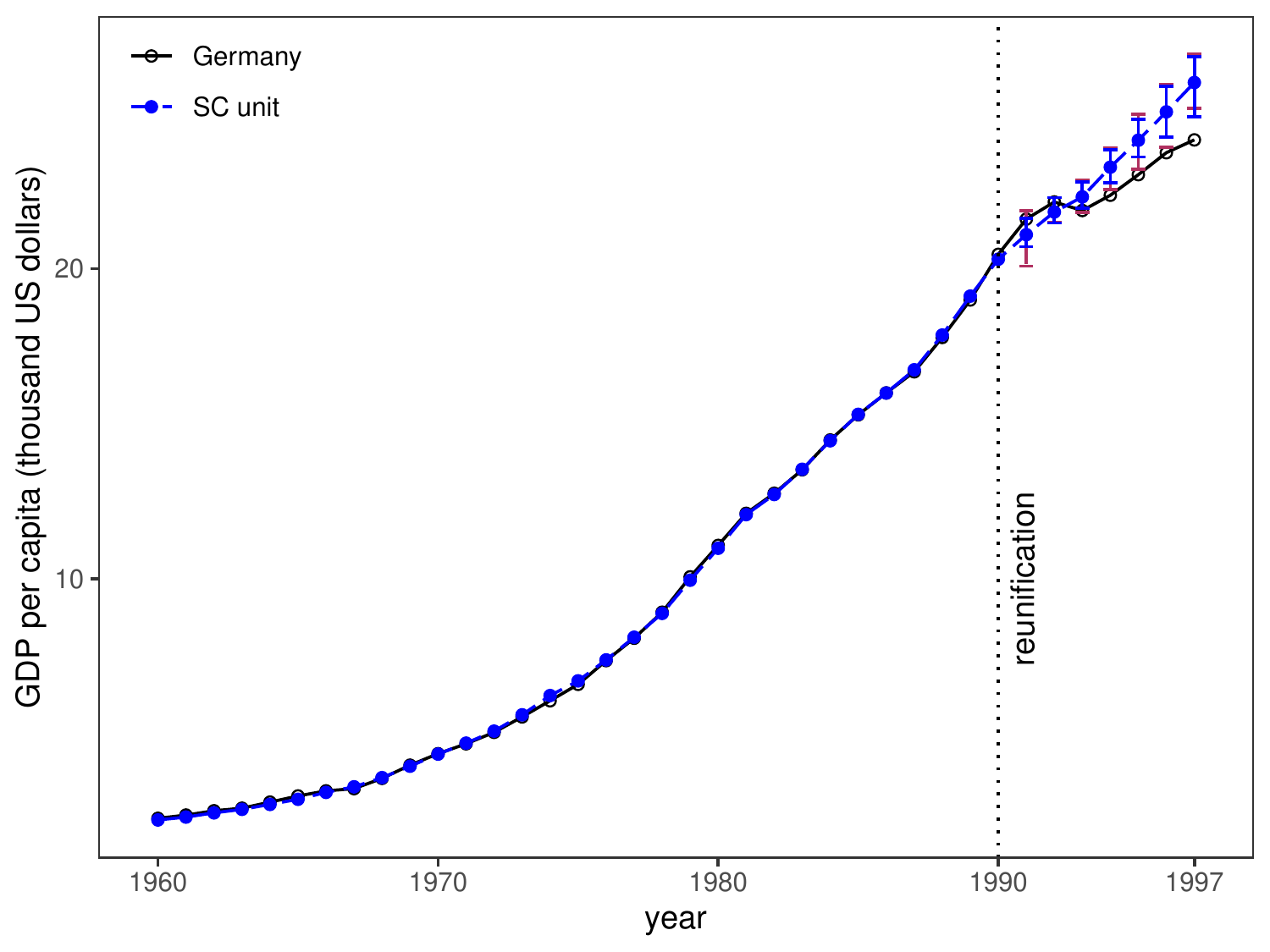}
			\caption{Prediction Interval for $Y_{1T}(0)$, approach 3}
		\end{subfigure}
		\begin{subfigure}{0.48\textwidth}
			\includegraphics[width=\textwidth]{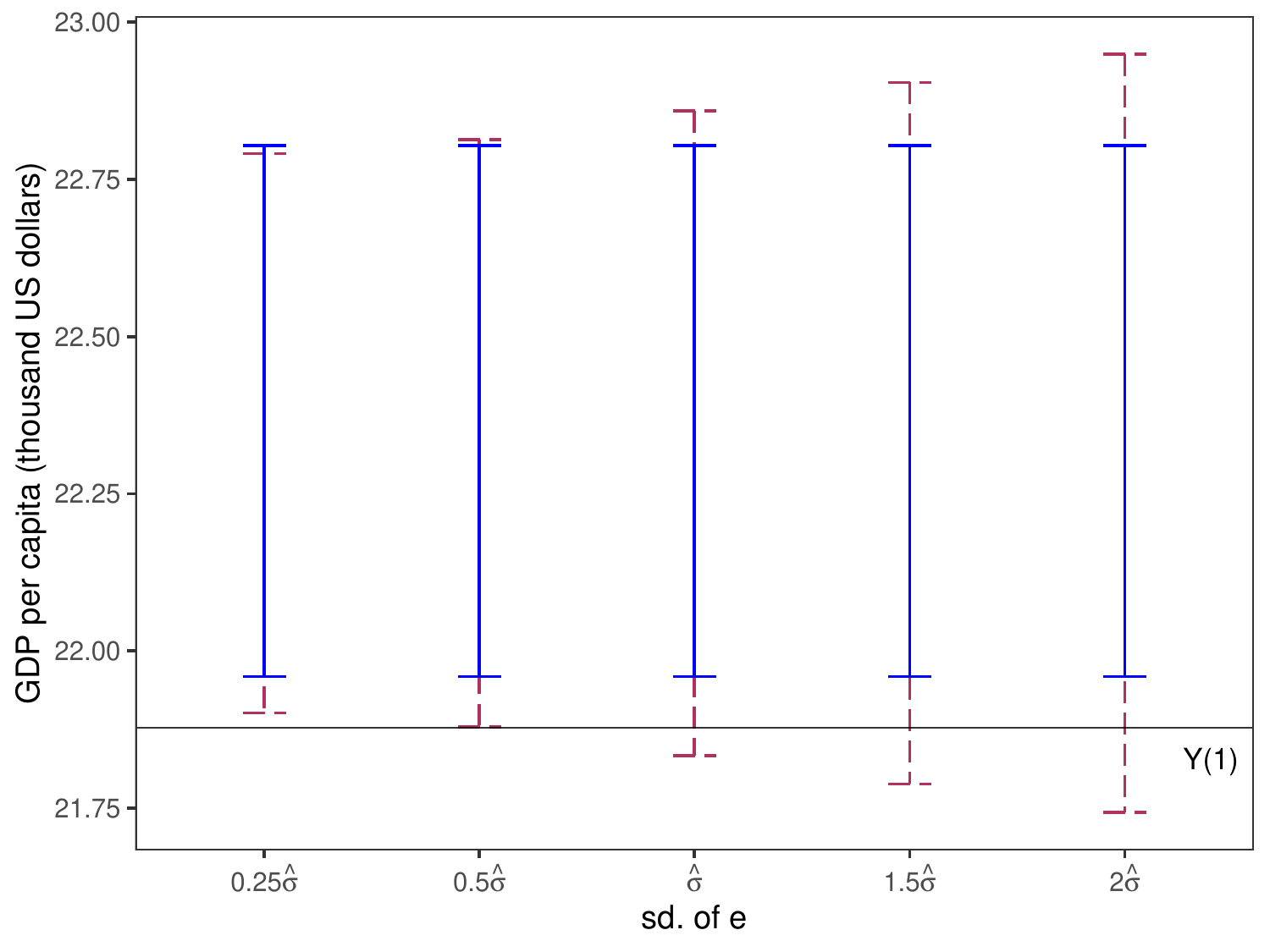}
			\caption{Sensitivity Analysis: PIs in 1993}
		\end{subfigure}
	\end{center}
	\scriptsize\textit{Notes}. Panel (a): GDP per capita of West Germany and synthetic West Germany. Panel (b): Prediction interval for synthetic West Germany with at least 95\% coverage probability. Panels (c)-(e): Prediction intervals for the counterfactual of West Germany with at least 90\% coverage probability based on three methods described in Section 5, respectively. Panel (f): Prediction intervals for the counterfactual West Germany based on approach 1, corresponding to $c\times\sigma_{\mathscr{H}}$, where $c=0.25,0.5,1,1.5,2$. The horizontal solid line represents the observed outcome for the treated.
\end{figure}

\clearpage
\begin{figure}
	\begin{center}\caption{1990 German Reunification: GDP Growth Rate.\label{fig:Germany, growth rate}}
		\vspace{-.1in}
		\begin{subfigure}{0.48\textwidth}
			\includegraphics[width=\textwidth]{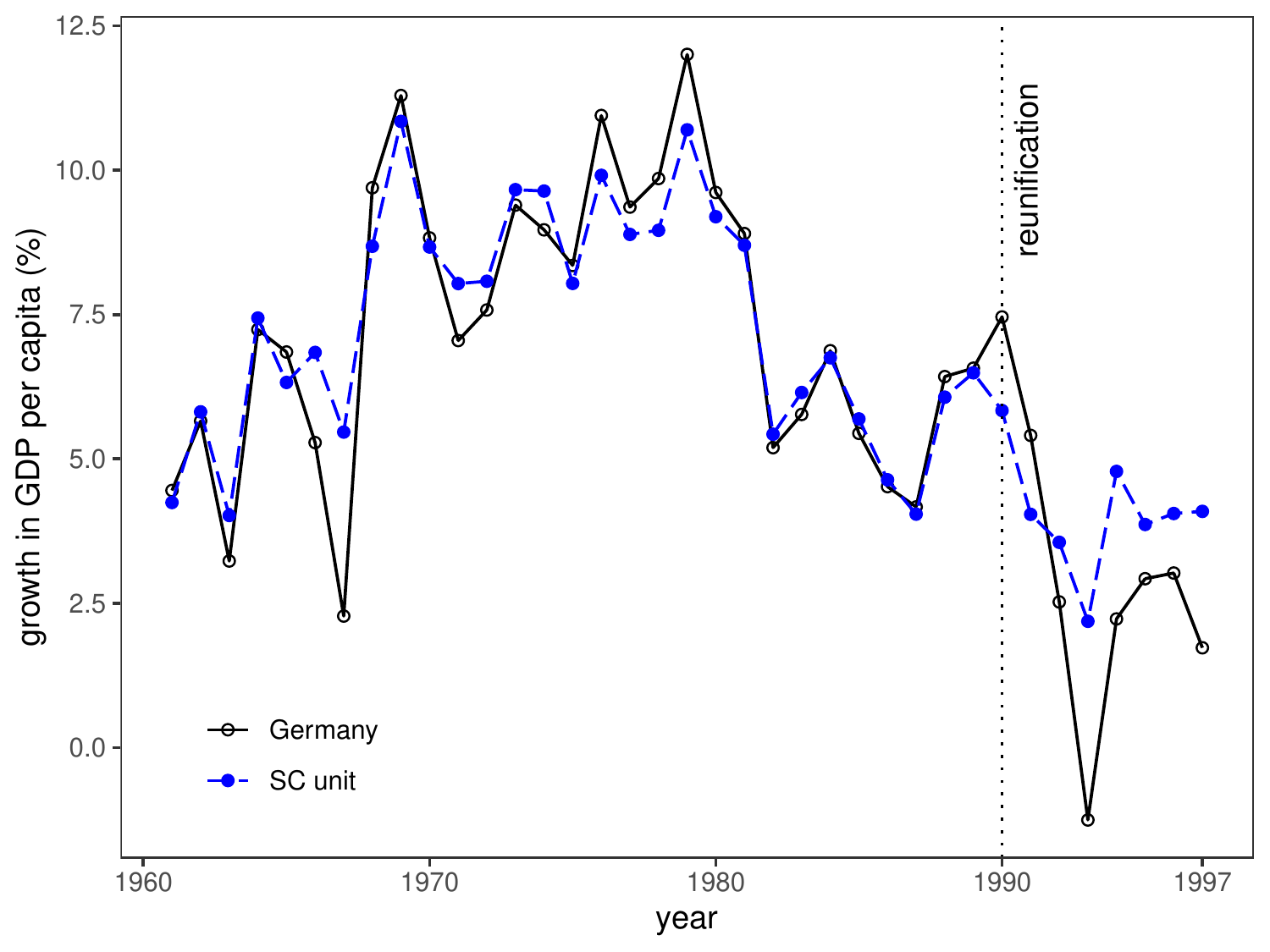}
			\caption{Synthetic West Germany}
		\end{subfigure}
		\begin{subfigure}{0.48\textwidth}
			\includegraphics[width=\textwidth]{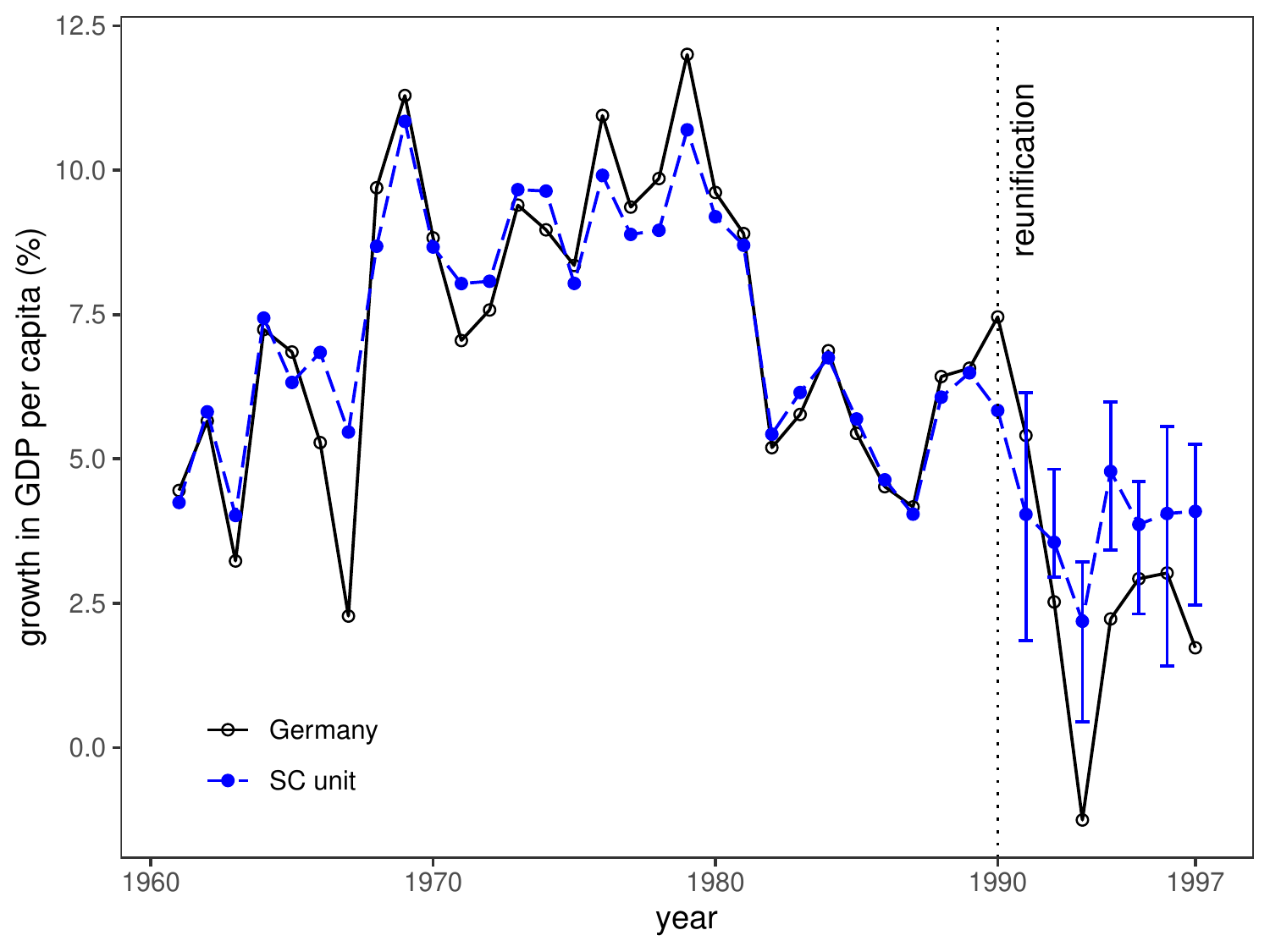}
			\caption{Prediction Interval for $\bx_T'\bw_0$}
		\end{subfigure}\\
		\begin{subfigure}{0.48\textwidth}
			\includegraphics[width=\textwidth]{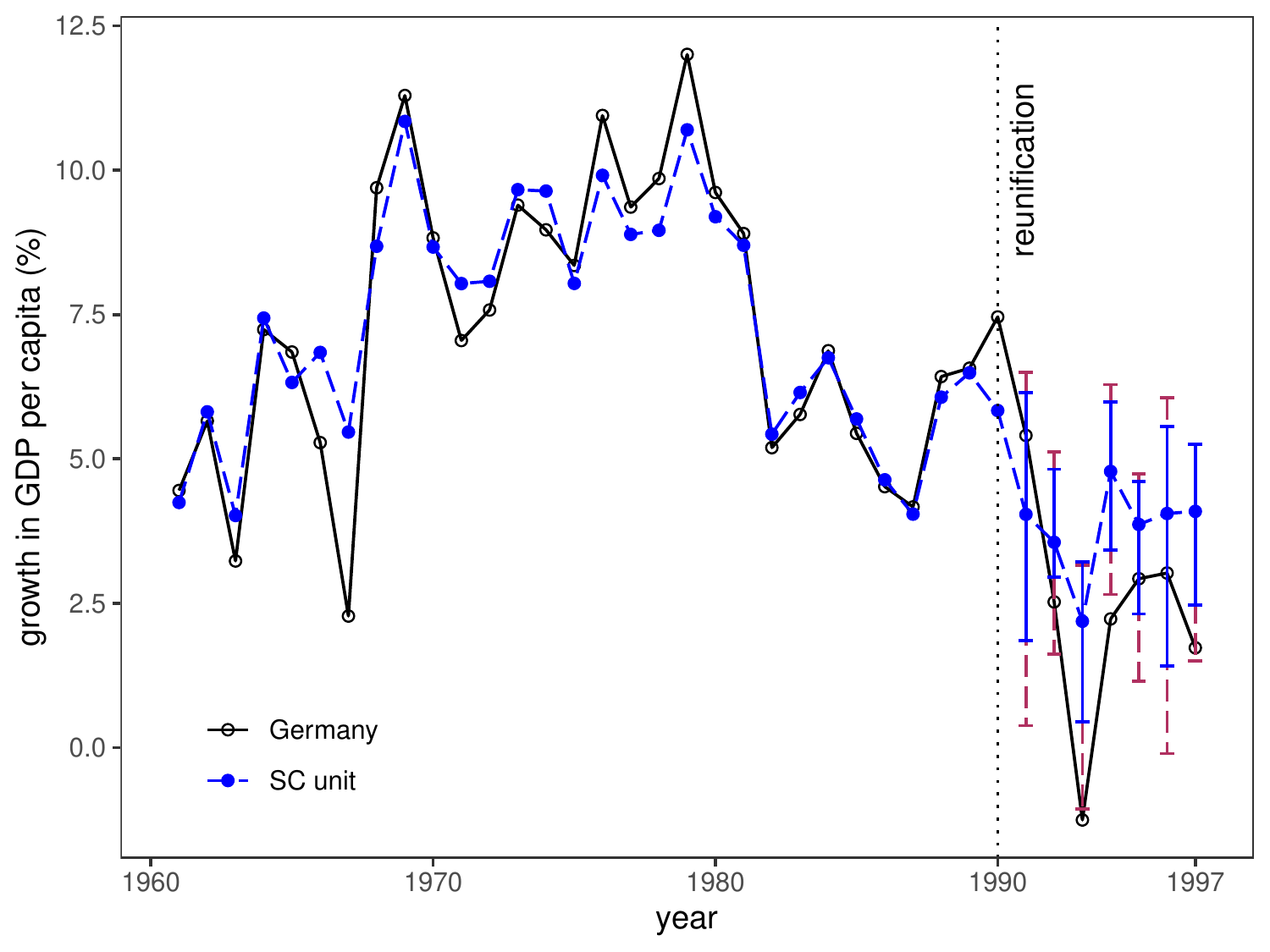}
			\caption{Prediction Interval for $Y_{1T}(0)$, approach 1}
		\end{subfigure}
		\begin{subfigure}{0.48\textwidth}
			\includegraphics[width=\textwidth]{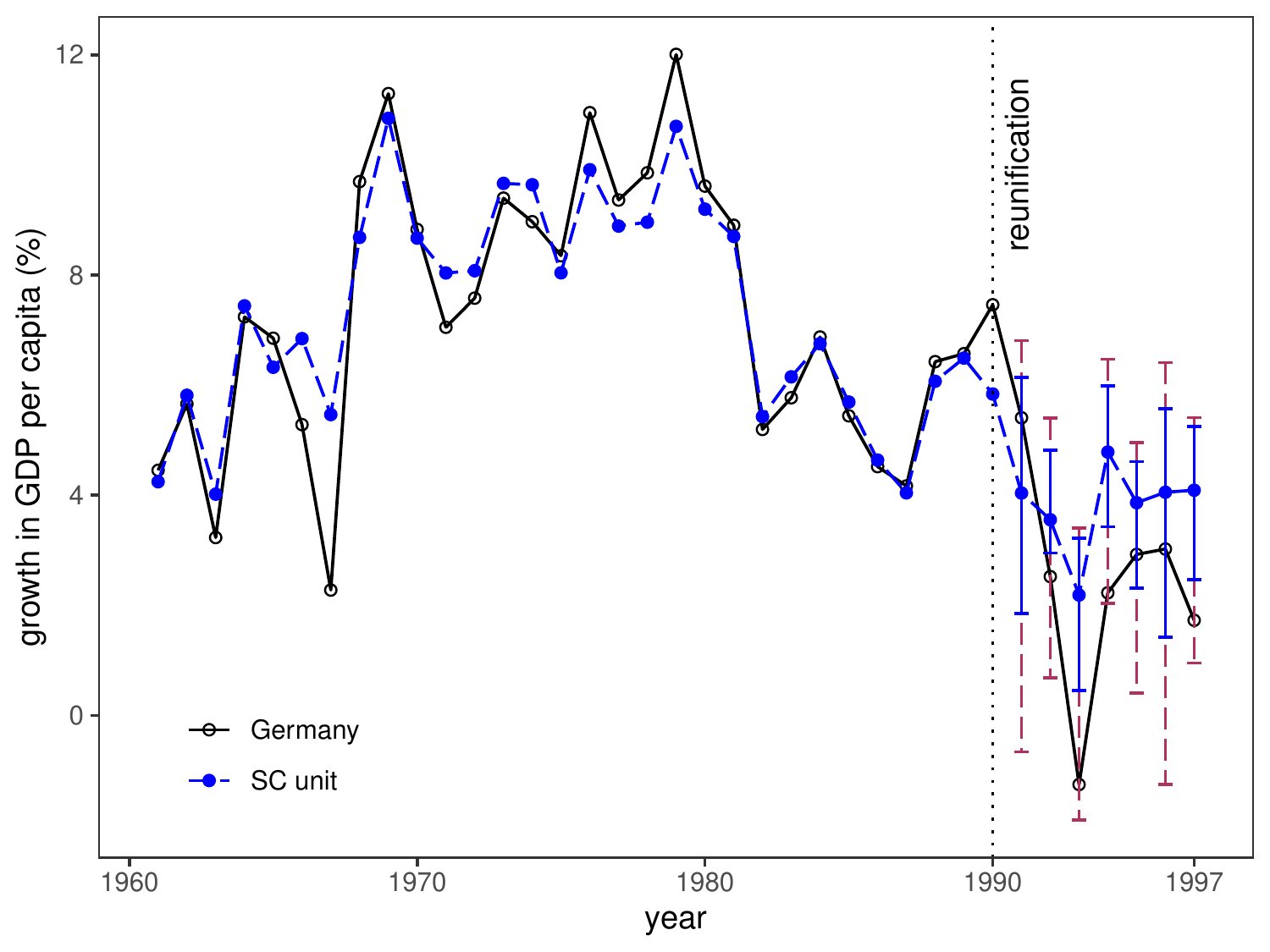}
			\caption{Prediction Interval for $Y_{1T}(0)$, approach 2}
		\end{subfigure}\\
		\begin{subfigure}{0.48\textwidth}
			\includegraphics[width=\textwidth]{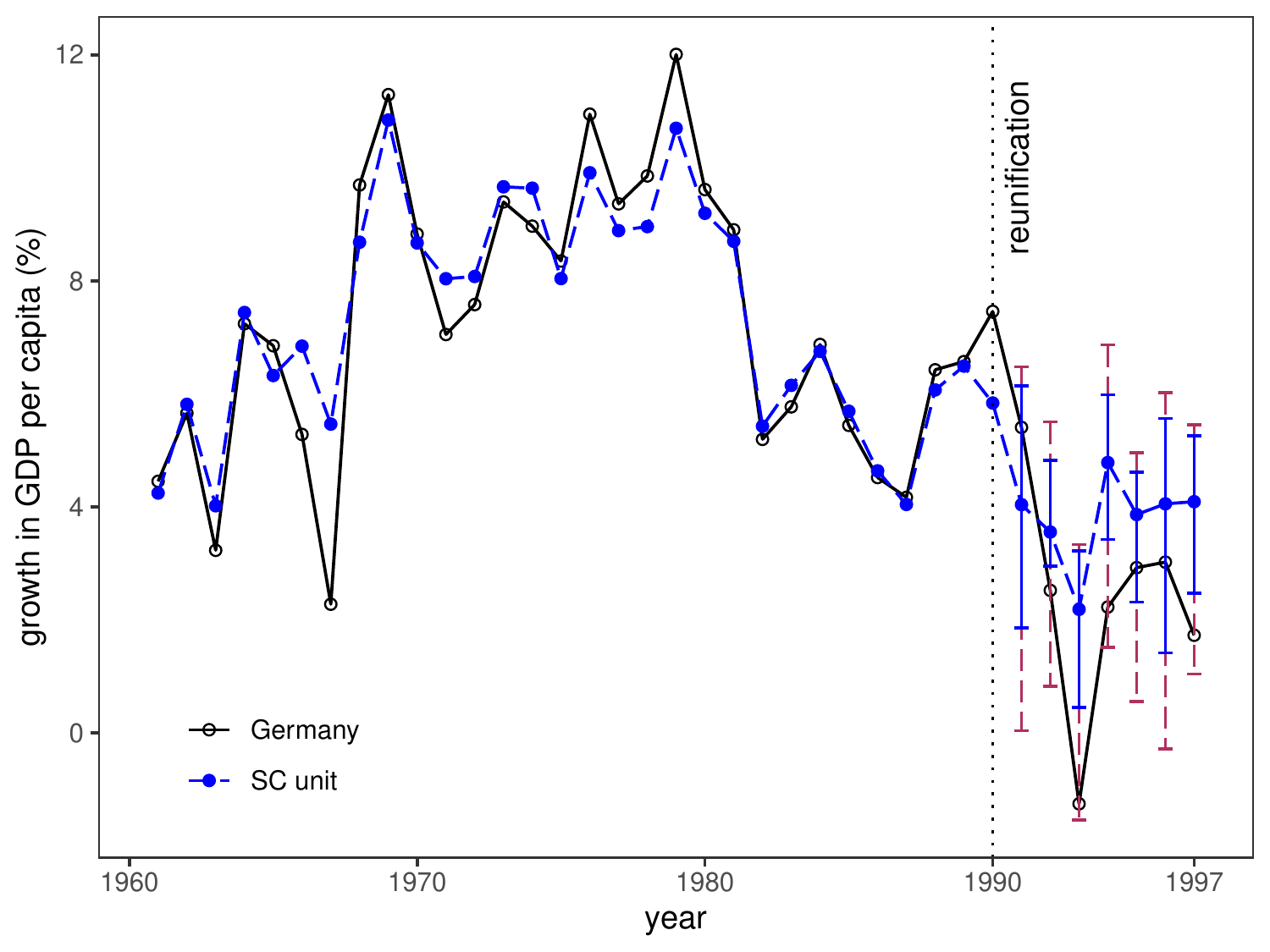}
			\caption{Prediction Interval for $Y_{1T}(0)$, approach 3}
		\end{subfigure}
		\begin{subfigure}{0.48\textwidth}
			\includegraphics[width=\textwidth]{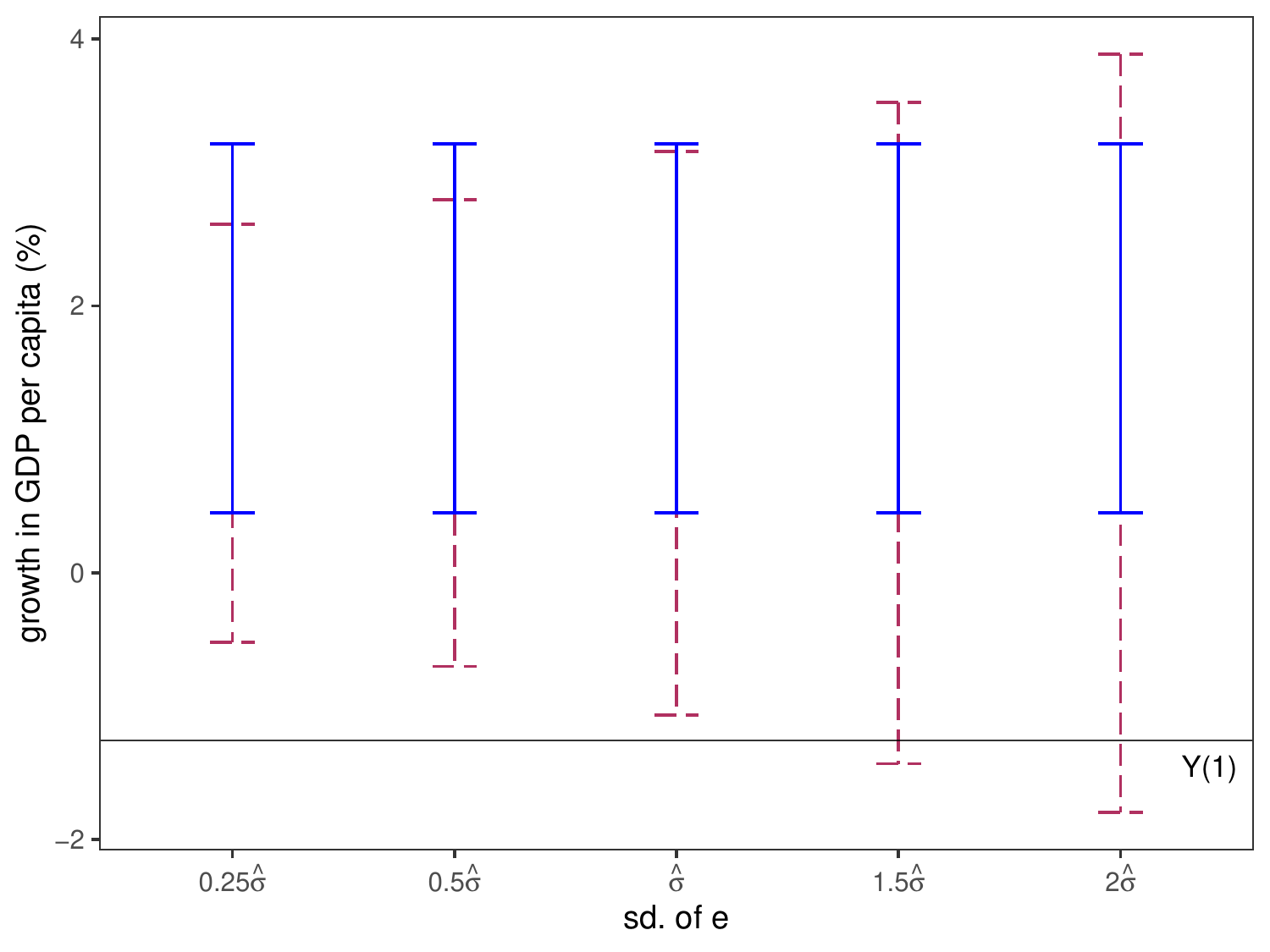}
			\caption{Sensitivity Analysis: PIs in 1993}
		\end{subfigure}
	\end{center}
	\scriptsize\textit{Notes}. Panel (a): GDP growth rate of West Germany and synthetic West Germany. Panel (b): Prediction interval for synthetic West Germany with at least 95\% coverage probability. Panels (c)-(e): Prediction intervals for the counterfactual of West Germany with at least 90\% coverage probability based on three methods described in Section 5, respectively. Panel (f): Prediction intervals for the counterfactual West Germany based on approach 1, corresponding to $c\times\sigma_{\mathscr{H}}$, where $c=0.25,0.5,1,1.5,2$. The horizontal solid line represents the observed outcome for the treated.
\end{figure}

%% file: input/simulation/Table_Simul.txt
\begin{tabular}{rllcllcllcllcll}
\hline\hline
\multicolumn{1}{c}{\bfseries }&\multicolumn{14}{c}{\bfseries Panel A: Models with Misspecification Error}\tabularnewline
\multicolumn{1}{c}{\bfseries }&\multicolumn{2}{c}{\bfseries M1}&\multicolumn{1}{c}{\bfseries }&\multicolumn{2}{c}{\bfseries M1-S}&\multicolumn{1}{c}{\bfseries }&\multicolumn{2}{c}{\bfseries M2}&\multicolumn{1}{c}{\bfseries }&\multicolumn{2}{c}{\bfseries M3}&\multicolumn{1}{c}{\bfseries }&\multicolumn{2}{c}{\bfseries CONF}\tabularnewline
\cline{2-3} \cline{5-6} \cline{8-9} \cline{11-12} \cline{14-15}
\multicolumn{1}{r}{}&\multicolumn{1}{c}{CP}&\multicolumn{1}{c}{AL}&\multicolumn{1}{c}{}&\multicolumn{1}{c}{CP}&\multicolumn{1}{c}{AL}&\multicolumn{1}{c}{}&\multicolumn{1}{c}{CP}&\multicolumn{1}{c}{AL}&\multicolumn{1}{c}{}&\multicolumn{1}{c}{CP}&\multicolumn{1}{c}{AL}&\multicolumn{1}{c}{}&\multicolumn{1}{c}{CP}&\multicolumn{1}{c}{AL}\tabularnewline
\hline
{\bfseries $\rho=0$}&&&&&&&&&&&&&&\tabularnewline
Cond. 1&0.949&2.168&&0.987&2.891&&0.979&2.625&&0.978&2.682&&0.864&1.646\tabularnewline
2&0.931&2.118&&0.982&2.844&&0.970&2.577&&0.967&2.624&&0.854&1.642\tabularnewline
3&0.922&2.136&&0.977&2.867&&0.966&2.599&&0.957&2.630&&0.842&1.643\tabularnewline
4&0.928&2.242&&0.979&2.980&&0.966&2.710&&0.956&2.719&&0.830&1.651\tabularnewline
5&0.936&2.406&&0.981&3.154&&0.971&2.881&&0.960&2.862&&0.819&1.665\tabularnewline
Uncond.&0.961&2.373&&0.991&3.094&&0.982&2.833&&0.979&2.892&&0.886&1.687\tabularnewline
\hline
{\bfseries $\rho=0.5$}&&&&&&&&&&&&&&\tabularnewline
Cond. 1&0.954&2.423&&0.986&3.140&&0.980&2.876&&0.976&2.928&&0.854&1.680\tabularnewline
2&0.965&2.488&&0.990&3.203&&0.984&2.939&&0.982&2.992&&0.865&1.691\tabularnewline
3&0.973&2.579&&0.992&3.296&&0.988&3.031&&0.986&3.078&&0.875&1.706\tabularnewline
4&0.980&2.694&&0.993&3.414&&0.990&3.149&&0.989&3.183&&0.887&1.729\tabularnewline
5&0.983&2.830&&0.995&3.556&&0.992&3.289&&0.988&3.305&&0.896&1.756\tabularnewline
Uncond.&0.962&2.387&&0.990&3.106&&0.981&2.846&&0.983&2.921&&0.882&1.695\tabularnewline
\hline
{\bfseries $\rho=1$}&&&&&&&&&&&&&&\tabularnewline
Cond. 1&0.979&2.798&&0.995&3.602&&0.991&3.308&&0.979&3.278&&0.896&3.403\tabularnewline
2&0.978&2.730&&0.995&3.554&&0.990&3.252&&0.987&3.270&&0.985&3.369\tabularnewline
3&0.970&2.756&&0.992&3.637&&0.987&3.317&&0.982&3.287&&0.170&3.345\tabularnewline
4&0.956&2.896&&0.982&3.876&&0.975&3.525&&0.951&3.331&&0.000&3.333\tabularnewline
5&0.935&3.188&&0.969&4.323&&0.961&3.926&&0.901&3.413&&0.000&3.336\tabularnewline
Uncond.&0.974&2.970&&0.994&3.733&&0.990&3.458&&0.989&3.530&&0.895&3.443\tabularnewline
\hline

\multicolumn{1}{c}{\bfseries }&\multicolumn{14}{c}{\bfseries Panel B: Models without Misspecification Error}\tabularnewline
\multicolumn{1}{c}{\bfseries }&\multicolumn{2}{c}{\bfseries M1}&\multicolumn{1}{c}{\bfseries }&\multicolumn{2}{c}{\bfseries M1-S}&\multicolumn{1}{c}{\bfseries }&\multicolumn{2}{c}{\bfseries M2}&\multicolumn{1}{c}{\bfseries }&\multicolumn{2}{c}{\bfseries M3}&\multicolumn{1}{c}{\bfseries }&\multicolumn{2}{c}{\bfseries CONF}\tabularnewline
\cline{2-3} \cline{5-6} \cline{8-9} \cline{11-12} \cline{14-15}
\multicolumn{1}{r}{}&\multicolumn{1}{c}{CP}&\multicolumn{1}{c}{AL}&\multicolumn{1}{c}{}&\multicolumn{1}{c}{CP}&\multicolumn{1}{c}{AL}&\multicolumn{1}{c}{}&\multicolumn{1}{c}{CP}&\multicolumn{1}{c}{AL}&\multicolumn{1}{c}{}&\multicolumn{1}{c}{CP}&\multicolumn{1}{c}{AL}&\multicolumn{1}{c}{}&\multicolumn{1}{c}{CP}&\multicolumn{1}{c}{AL}\tabularnewline
\hline
{\bfseries $\rho=0$}&&&&&&&&&&&&&&\tabularnewline
Cond. 1&0.943&2.154&&0.985&2.871&&0.976&2.609&&0.973&2.661&&0.879&1.619\tabularnewline
2&0.927&2.105&&0.980&2.826&&0.966&2.563&&0.962&2.604&&0.880&1.617\tabularnewline
3&0.918&2.125&&0.974&2.852&&0.960&2.587&&0.953&2.612&&0.880&1.621\tabularnewline
4&0.922&2.231&&0.975&2.966&&0.961&2.699&&0.953&2.701&&0.881&1.632\tabularnewline
5&0.937&2.394&&0.978&3.139&&0.966&2.868&&0.955&2.841&&0.881&1.649\tabularnewline
Uncond.&0.960&2.358&&0.990&3.073&&0.981&2.810&&0.981&2.878&&0.888&1.656\tabularnewline
\hline
{\bfseries $\rho=0.5$}&&&&&&&&&&&&&&\tabularnewline
Cond. 1&0.959&2.416&&0.988&3.127&&0.981&2.866&&0.978&2.918&&0.875&1.666\tabularnewline
2&0.970&2.480&&0.990&3.191&&0.985&2.929&&0.984&2.980&&0.876&1.671\tabularnewline
3&0.975&2.571&&0.993&3.283&&0.989&3.022&&0.987&3.064&&0.879&1.682\tabularnewline
4&0.981&2.685&&0.995&3.401&&0.990&3.139&&0.987&3.168&&0.885&1.699\tabularnewline
5&0.985&2.820&&0.996&3.542&&0.992&3.278&&0.989&3.287&&0.888&1.721\tabularnewline
Uncond.&0.961&2.370&&0.991&3.083&&0.981&2.825&&0.983&2.894&&0.884&1.656\tabularnewline
\hline
{\bfseries $\rho=1$}&&&&&&&&&&&&&&\tabularnewline
Cond. 1&0.964&2.533&&0.988&3.278&&0.983&3.006&&0.972&2.972&&0.860&1.561\tabularnewline
2&0.963&2.406&&0.990&3.134&&0.983&2.867&&0.979&2.892&&0.851&1.548\tabularnewline
3&0.947&2.364&&0.983&3.106&&0.972&2.834&&0.961&2.835&&0.854&1.542\tabularnewline
4&0.923&2.411&&0.965&3.200&&0.955&2.915&&0.918&2.801&&0.859&1.540\tabularnewline
5&0.887&2.571&&0.940&3.447&&0.927&3.136&&0.848&2.800&&0.847&1.547\tabularnewline
Uncond.&0.982&2.773&&0.997&3.485&&0.993&3.227&&0.991&3.287&&0.868&1.642\tabularnewline
\hline
\end{tabular}

%% file: input/simulation/Table_Main3.txt
\begin{tabular}{rllcllcllcllcllcll}
\hline\hline
\multicolumn{1}{c}{\bfseries }&\multicolumn{2}{c}{\bfseries M1}&\multicolumn{1}{c}{\bfseries }&\multicolumn{2}{c}{\bfseries M1-S}&\multicolumn{1}{c}{\bfseries }&\multicolumn{2}{c}{\bfseries M2}&\multicolumn{1}{c}{\bfseries }&\multicolumn{2}{c}{\bfseries M3}&\multicolumn{1}{c}{\bfseries }&\multicolumn{2}{c}{\bfseries PERM}&\multicolumn{1}{c}{\bfseries }&\multicolumn{2}{c}{\bfseries CONF}\tabularnewline
\cline{2-3} \cline{5-6} \cline{8-9} \cline{11-12} \cline{14-15} \cline{17-18}
\multicolumn{1}{r}{}&\multicolumn{1}{c}{CP}&\multicolumn{1}{c}{AL}&\multicolumn{1}{c}{}&\multicolumn{1}{c}{CP}&\multicolumn{1}{c}{AL}&\multicolumn{1}{c}{}&\multicolumn{1}{c}{CP}&\multicolumn{1}{c}{AL}&\multicolumn{1}{c}{}&\multicolumn{1}{c}{CP}&\multicolumn{1}{c}{AL}&\multicolumn{1}{c}{}&\multicolumn{1}{c}{CP}&\multicolumn{1}{c}{AL}&\multicolumn{1}{c}{}&\multicolumn{1}{c}{CP}&\multicolumn{1}{c}{AL}\tabularnewline
\hline
{\bfseries $\rho=0$}&&&&&&&&&&&&&&&&&\tabularnewline
Cond. 1&0.965&2.211&&0.995&2.939&&0.987&2.654&&0.987&2.654&&0.299&0.646&&0.864&1.646\tabularnewline
2&0.962&2.155&&0.994&2.883&&0.986&2.598&&0.986&2.598&&0.270&0.763&&0.854&1.642\tabularnewline
3&0.961&2.165&&0.994&2.893&&0.985&2.608&&0.985&2.608&&0.230&0.891&&0.842&1.643\tabularnewline
4&0.969&2.260&&0.996&2.989&&0.988&2.704&&0.988&2.704&&0.195&1.026&&0.830&1.651\tabularnewline
5&0.975&2.410&&0.997&3.139&&0.991&2.853&&0.991&2.853&&0.162&1.166&&0.819&1.665\tabularnewline
Uncond.&0.971&2.434&&0.995&3.164&&0.988&2.881&&0.988&2.881&&0.411&0.829&&0.886&1.687\tabularnewline
\hline
{\bfseries $\rho=0.5$}&&&&&&&&&&&&&&&&&\tabularnewline
Cond. 1&0.980&2.493&&0.996&3.221&&0.993&2.934&&0.993&2.934&&0.224&0.930&&0.854&1.680\tabularnewline
2&0.983&2.562&&0.997&3.290&&0.993&3.003&&0.993&3.003&&0.270&0.792&&0.865&1.691\tabularnewline
3&0.985&2.655&&0.998&3.383&&0.995&3.095&&0.995&3.095&&0.317&0.678&&0.875&1.706\tabularnewline
4&0.989&2.768&&0.999&3.496&&0.995&3.208&&0.995&3.208&&0.363&0.596&&0.887&1.729\tabularnewline
5&0.991&2.898&&0.999&3.626&&0.998&3.338&&0.998&3.338&&0.408&0.548&&0.896&1.756\tabularnewline
Uncond.&0.975&2.471&&0.996&3.207&&0.989&2.920&&0.989&2.920&&0.467&1.015&&0.882&1.695\tabularnewline
\hline
{\bfseries $\rho=1$}&&&&&&&&&&&&&&&&&\tabularnewline
Cond. 1&0.990&5.049&&1.000&6.474&&1.000&6.166&&1.000&6.166&&1.000&11.253&&0.896&3.403\tabularnewline
2&1.000&4.842&&1.000&6.267&&1.000&5.959&&1.000&5.959&&0.869&11.283&&0.985&3.369\tabularnewline
3&0.849&4.687&&0.989&6.112&&0.958&5.804&&0.958&5.804&&0.004&11.488&&0.170&3.345\tabularnewline
4&0.041&4.583&&0.332&6.008&&0.172&5.699&&0.172&5.699&&0.000&11.934&&0.000&3.333\tabularnewline
5&0.000&4.545&&0.001&5.970&&0.000&5.661&&0.000&5.661&&0.000&12.575&&0.000&3.336\tabularnewline
Uncond.&0.989&5.632&&0.999&7.083&&0.995&6.512&&0.995&6.512&&0.922&17.377&&0.895&3.443\tabularnewline
\hline
\end{tabular}

%% file: input/simulation/Table_Main4.txt
\begin{tabular}{rllcllcllcllcllcll}
\hline\hline
\multicolumn{1}{c}{\bfseries }&\multicolumn{2}{c}{\bfseries M1}&\multicolumn{1}{c}{\bfseries }&\multicolumn{2}{c}{\bfseries M1-S}&\multicolumn{1}{c}{\bfseries }&\multicolumn{2}{c}{\bfseries M2}&\multicolumn{1}{c}{\bfseries }&\multicolumn{2}{c}{\bfseries M3}&\multicolumn{1}{c}{\bfseries }&\multicolumn{2}{c}{\bfseries PERM}&\multicolumn{1}{c}{\bfseries }&\multicolumn{2}{c}{\bfseries CONF}\tabularnewline
\cline{2-3} \cline{5-6} \cline{8-9} \cline{11-12} \cline{14-15} \cline{17-18}
\multicolumn{1}{r}{}&\multicolumn{1}{c}{CP}&\multicolumn{1}{c}{AL}&\multicolumn{1}{c}{}&\multicolumn{1}{c}{CP}&\multicolumn{1}{c}{AL}&\multicolumn{1}{c}{}&\multicolumn{1}{c}{CP}&\multicolumn{1}{c}{AL}&\multicolumn{1}{c}{}&\multicolumn{1}{c}{CP}&\multicolumn{1}{c}{AL}&\multicolumn{1}{c}{}&\multicolumn{1}{c}{CP}&\multicolumn{1}{c}{AL}&\multicolumn{1}{c}{}&\multicolumn{1}{c}{CP}&\multicolumn{1}{c}{AL}\tabularnewline
\hline
{\bfseries $\rho=0$}&&&&&&&&&&&&&&&&&\tabularnewline
Cond. 1&0.929&2.141&&0.975&2.866&&0.963&2.600&&0.951&2.588&&0.299&0.646&&0.864&1.646\tabularnewline
2&0.909&2.083&&0.967&2.806&&0.947&2.541&&0.929&2.495&&0.270&0.763&&0.854&1.642\tabularnewline
3&0.889&2.084&&0.956&2.805&&0.934&2.542&&0.899&2.426&&0.230&0.891&&0.842&1.643\tabularnewline
4&0.881&2.175&&0.941&2.899&&0.925&2.637&&0.863&2.398&&0.195&1.026&&0.830&1.651\tabularnewline
5&0.876&2.342&&0.931&3.082&&0.914&2.818&&0.804&2.384&&0.162&1.166&&0.819&1.665\tabularnewline
Uncond.&0.947&2.350&&0.979&3.082&&0.971&2.814&&0.955&2.728&&0.411&0.829&&0.886&1.687\tabularnewline
\hline
{\bfseries $\rho=0.5$}&&&&&&&&&&&&&&&&&\tabularnewline
Cond. 1&0.926&2.347&&0.974&3.054&&0.964&2.792&&0.942&2.741&&0.224&0.930&&0.854&1.680\tabularnewline
2&0.941&2.402&&0.978&3.106&&0.968&2.845&&0.948&2.789&&0.270&0.792&&0.865&1.691\tabularnewline
3&0.945&2.488&&0.980&3.195&&0.970&2.934&&0.941&2.824&&0.317&0.678&&0.875&1.706\tabularnewline
4&0.947&2.613&&0.976&3.331&&0.967&3.068&&0.918&2.844&&0.363&0.596&&0.887&1.729\tabularnewline
5&0.942&2.788&&0.969&3.536&&0.961&3.266&&0.871&2.845&&0.408&0.548&&0.896&1.756\tabularnewline
Uncond.&0.944&2.356&&0.980&3.083&&0.968&2.822&&0.947&2.760&&0.467&1.015&&0.882&1.695\tabularnewline
\hline
{\bfseries $\rho=1$}&&&&&&&&&&&&&&&&&\tabularnewline
Cond. 1&0.914&3.071&&0.946&4.033&&0.937&3.685&&0.716&2.608&&1.000&11.253&&0.896&3.403\tabularnewline
2&0.950&2.875&&0.976&3.791&&0.969&3.453&&0.900&3.028&&0.869&11.283&&0.985&3.369\tabularnewline
3&0.880&3.284&&0.923&4.444&&0.907&4.038&&0.742&2.891&&0.004&11.488&&0.170&3.345\tabularnewline
4&0.717&6.925&&0.768&9.935&&0.751&9.157&&0.507&2.199&&0.000&11.934&&0.000&3.333\tabularnewline
5&0.581&256.704&&0.614&384.612&&0.603&392.478&&0.374&0.962&&0.000&12.575&&0.000&3.336\tabularnewline
Uncond.&0.962&2.940&&0.986&3.713&&0.979&3.433&&0.959&3.372&&0.922&17.377&&0.895&3.443\tabularnewline
\hline
\end{tabular}